\documentclass{emulateapj}
\shorttitle{M31's  inner and outer halo}
\shortauthors{Koch et al.}
\slugcomment{Accepted to the Astrophysical Journal}
\begin{document}

\title{Kinematic and Chemical constraints on the formation of M31's inner and outer 
halo\altaffilmark{$\dagger$}}

\author{Andreas Koch\altaffilmark{1},  R.~Michael Rich\altaffilmark{1}, David B. Reitzel\altaffilmark{1}, 
Nicolas F. Martin\altaffilmark{2}, Rodrigo A. Ibata\altaffilmark{3}, Scott C. Chapman\altaffilmark{4}, 
Steven R. Majewski\altaffilmark{5},  Masao Mori\altaffilmark{6}, Yeong-Shang Loh\altaffilmark{1},  
James C. Ostheimer\altaffilmark{5}, and Mikito Tanaka\altaffilmark{7}}
\email{akoch@astro.ucla.edu}

\altaffiltext{$\dagger$}{Some of the data presented herein were obtained at the W.M. Keck Observatory, which is 
operated as a scientific partnership among the California Institute of Technology, the University of 
California and the National Aeronautics and Space Administration. The Observatory was made possible 
by the generous financial support of the W.M. Keck Foundation.} 
\altaffiltext{1}{UCLA, Department of Physics and Astronomy, Los Angeles, CA, USA}
\altaffiltext{2}{Max Planck Institute for Astronomy, D-69117 Heidelberg, Germany}
\altaffiltext{3}{Observatoire de Strasbourg, F-67000 Strasbourg, France}
\altaffiltext{4}{Institute of Astronomy, Cambridge University, Cambridge, UK}
\altaffiltext{5}{Department of Astronomy, University of Virginia, Charlottesville, VA, USA}
\altaffiltext{6}{Center for Comuputational Sciences, University of Tsukuba,
Tsukuba, Ibaraki 305-8577, Japan}
\altaffiltext{7}{National Astronomical Observatory of Japan, 2-21-1 Osawa,
Mitaka, Tokyo 181-8588, Japan}

\begin{abstract}
The halo of M31 shows a wealth of substructures, some of which are 
consistent with the assembly from satellite accretion. 
Here we report on kinematic and abundance results from Keck/DEIMOS spectroscopy in the 
near-infrared calcium triplet region  
of  over 3500 
red giant star candidates along the minor axis and in off-axis  spheroid fields of M31. 
These data reach out to large radial distances of 
about  160 kpc. 
The derived radial velocity distributions show an indication of a kinematically cold substructure 
around $\sim 17$ kpc,   
which has been reported before. 
We devise a new improved method to measure spectroscopic metallicities from 
the calcium triplet in low signal-to-noise spectra 
using a weighted coaddition of the individual lines. 
The resulting distribution (accurate to $\sim$0.3 dex down to signal to noise ratios of 5) 
leads us to note an even stronger gradient in the abundance distribution 
along M31's minor axis and in particular towards the outer halo fields than previously detected. 
The mean metallicity in the outer fields reaches below $-2$ dex, with individual values 
as low as $\la -2.6$ dex. This is the first time such a metal poor halo has been detected in M31. 
In the fields towards the inner spheroid we find a sharp decline of $\sim$0.5 dex in metallicity in 
a region at  $\sim$20 kpc, which roughly coincides with the edge of an  extended disk, previously 
detected from star count maps. 
A large fraction of red giants that we detect in the most distant fields are likely members of 
M33's overlapping halo. 
A comparison of our velocities with those predicted by new N-body 
simulations argues that the event responsible for the giant Stream is most 
likely not responsible for the full population of the inner halo. 
We show further 
that the abundance distribution of the Stream is different from that of the 
inner halo, from which it becomes  evident, in turn, that the merger event that formed the 
Stream and the outer halo cannot have contributed any significant material 
to the inner spheroid. All these  
severe structure changes in the halo suggest a  high degree of infall and  
stochastic abundance accretion governing the build-up of M31's inner and outer halo. 
\end{abstract}

\section{Introduction}
With the discovery of the Giant Stream (Ibata et al. 2001), the mapping
of complex structures in the ``halo'' of M31 (Ferguson et al. 2002; Gilbert et al. 2007; Ibata et al. 2007), and
the isolation of an
extended kinematic disk structure (Ibata et al. 2005) the idea that the radially more 
distant populations of M31 originate in accretion events has become 
established.   Even a subset 
of M31 satellites might relate to the breakup
of a massive progenitor, based on the polar and planar alignment
of a number of its early-type satellite galaxies (Koch \& Grebel 2006).

Prior to this paradigm shift in the description of the M31 halo,
relatively shallow Hubble Space Telescope (HST) imagery revealed what appeared to be a mostly
metal rich M31 halo (e.g. Rich et al. 1996; Bellazini et al. 2003). 
Mould \& Kristian (1986) were the first to find a metal rich (47
Tuc-like) halo population, using ground-based imaging.
The widespread presence of this metal
rich population as well as the descending red giant branch (RGB; in contrast
to the metal poor globular clusters) was noted by Bellazzini et al.
(2003).  Ground-based minor axis star counts appear to show a smooth
$r^{1/4}$ spheroid (Pritchet \& van den Bergh 1994), and some studies
have argued that the metal rich halo population is an extension of the
metal rich bulge (e.g. Mould \& Kristian 1986; Guhathakurta et al. 2006).
But alternatively, the case has been made that the metal rich stars more closely
coincide with the perturbed regions (Ferguson et al. 2002).

Metal poor stars around M31 were already indicated in photometric studies 
of its halo at projected distances of 7 kpc (Mould \& Kristian 1986) to 20 kpc (Durrell 2001).  
Subsequent spectroscopic surveys then fully revealed a complex picture of the
halo composition:   
Using the well-established near-infrared calcium triplet (CaT) as a metallicity indicator, 
Reitzel \& Guhathakurta (2002) find metal
poor stars at 19 kpc on the minor axis.  Benefitting from much larger
samples, Chapman et al. (2006) and Kalirai et al. (2006a) argue based on
kinematics and metallicity that the
expected metal poor halo population is in fact present.
Giants with the radial velocity of M31 are claimed
as members of the halo to distances in excess of 100 kpc 
(Ostheimer 2003; Gilbert et al. 2006) and an overall metallicity
gradient of the M31 halo is proposed by Kalirai et al. (2006a).   Yet,
over much of this region, Ibata et al. (2007) find clear evidence of
density enhancements associated with accretion.   What
fraction of the outer halo is then comprised of such recently accreted material?  
In particular the chemical composition of the outer stars remains unknown, and 
depending on the adopted [$\alpha$/Fe] ratio,  Kalirai et al. (2006a) estimate a 
mean metallicity in the outermost field $\sim$0.3 dex higher than the mean Milky Way metallicity 
(for Solar scaled abundances), or  a mean that is comparable to the Milky Way halo under the 
assumption of strong $\alpha-$enhancement. 

In fact, pencil beam ultradeep imaging using the Advanced Camera for
Surveys (ACS) on board the HST, offers a complementary
picture of the complexity present in the halo populations.   The placement
of these deep imaging fields has benefitted from the starcount maps and
Keck/DEIMOS kinematic studies. 
Brown et al. (2003) first demonstrated that a minor axis field projected
at 11 kpc contains an indisputable range in age and abundance,
extending to nearly-Solar, and a predominant age
range from 6--10 Gyr.  Comparing 3 fields, in the inner spheroid (at 11\,kpc),  the M31 disk 
and on the Ibata et al. (2001) debris stream, Brown et al. (2006) find the disk to be younger and 
more metal rich, and lacking old stars. 
They find the stream and spheroid fields to be indistinguishable based on their color-magnitude diagrams (CMDs), leading to the conclusion that only one progenitor is responsible for the debris field in the inner halo region.  However, if dynamical mixing were efficient in these regions, it could also erase the signatures from different sources.
Brown et al. (2007) investigate a field at 21 kpc and find
evidence that its population  is marginally older and more metal
poor than the inner halo field. 

The present-day pencil beam surveys  have found clear
evidence of an age range in every field studied and appear to support
other evidence of a gradient in age and abundance.  What has been lacking
to date has been a survey of abundances and kinematics that ties these
fields together and provides a context for the interpretation of these
deep fields.  This is 
one aim of this paper, and a natural extension of our systematic survey of
the structure and kinematics of M31 along its southwest minor axis.

While it is attractive to seek {\it one} massive progenitor for the inner debris
field 
there are several arguments against this
position.  First, the existence of both the giant stream and the extended
disk suggest at least two very different sources for the debris field at 11 kpc. 
Second, the dramatic variation of the extended spatial structure of the debris
field as a function of metallicity and age (Ferguson et al. 2002)
is best understood by invoking  multiple events involving different accretors. 
Moreover, the distant rotating disk-like population (Ibata et
al. 2005) is superposed on other, likely unrelated, structures that are
suggestive of shells associated with the giant Stream merger (see also Ibata et al. 2007).  
It is then an intriguing question, whether a corresponding measurable abundance
change occurs at the point where this field ends.  

In the context of CDM models (e.g. Bullock \& Johnston 2005) halos
are thought, in general, to accrete from the debris of lower mass
satellites.   Yet Mouhcine et al. (2005) find a correlation between parent
galaxy luminosity and halo metallicity. Nominally,   M31 has a high metallicity halo
and falls in this relationship alongside galaxies with more prominent
bulges. There is a paradox: how can the halo of M31 be dominated by stochastic accretion
 events, yet still have an $r^{1/4}$ profile and still appear to follow trends set by luminosities 
 of the host galaxies?

We report here the culmination of an observational campaign begun in
2002; Table~1 lists the observing run details by Principal Investigator. 
This Paper is organized as follows: 
In \textsection 2 we present our observations and the standard reduction steps taken, while 
\textsection 3 describes our radial velocity measurements.  The dwarf/giant separation 
is discussed in \textsection 4 and in \textsection 5 we devise a new technique to measure 
spectroscopic metallicities from the CaT. The following sections are then dedicated to 
the analysis of kinematic (\textsection 6) and abundance (\textsection 7) substructures and gradients 
in M31's halo. Finally, \textsection 8  summarizes our findings.  

\section{Observations and reduction}
In the course of an ongoing  large Keck program (PI: R.~M. Rich) that aims at elucidating the 
formation history of M31's halo structures based on the kinematics and chemical analyses 
of red giants, we collected a vast spectroscopic data set, which covers, amongst others, 
fields on the minor axis of M31 reaching from 9 kpc out to large projected distances of $\sim$160 kpc towards the 
south-east\footnote{Throughout this work we will adopt a distance to M31 of 784 kpc (Stanek \& 
Garnavich 1998)}. These fields were originally imaged by Ostheimer (2003). 
Two additional fields at 60 kpc on the minor axis were obtained in the course of a large DEIMOS 
survey covering a wide set of fields spread across M31's full halo and disk components (PI: S. Chapman; see Chapman et al. 2006). 
 In this Paper, we will focus on the analysis of the minor axis data and those off-axis fields 
 in the southeast halo quadrant, while the 
fields located on top of the Giant Stellar Stream (Ibata et al. 2001), 
and those coinciding with the HST fields of Brown et al. (2003; 2006; 2007) will be the subject of a 
series of forthcoming papers.  
For details on the overall target selection, observation strategy and data collection for the whole 
project we refer the reader to Kalirai et al. (2006a,b) and Gilbert et al. (2006, 2007). 

\subsection{Observations}
Observations were carried out using the DEIMOS multislit spectrograph at the 
Keck\,II 10\,m telescope over a number of observing runs from 2002 through 2006 (Table~1), using 
a slit width of 1''.   
We used the 1200 line mm$^{-1}$ grating, which gives a dispersion of 0.33\,\AA\,pixel$^{-1}$ 
and a spectral resolution of 1.41\AA, as estimated from the width of the sky lines.  
The majority of the spectra were centered at a wavelength of 7800\AA, yielding  
a full spectral coverage of $\sim$6500--9200\AA, which comprises the dominant near infrared lines 
of the CaT around 8500\AA.  
Typical integration times were 1 hour per mask, while setup f109\_1 (at 9 kpc) was exposed 
for 3 hours in total. 
Fig.~1 shows the location of the slit masks discussed in this paper on an INT based star count map 
(M. Irwin, private communication; Ibata et al. 2007).
Details on these masks are given in Table~1. 

\subsection{Data reduction}
Reduction of the spectra was performed with the {\em spec2d} pipeline, which 
has been designed at the University of California, Berkeley for the DEEP2 survey\footnote{
\url{http://astron.berkeley.edu/$\sim$cooper/deep/spec1d/primer.html}}.  
The standard reduction steps comprise flat fielding, wavelength calibration via arc lamp spectra 
and sky spectrum removal. 
The total number of extracted science spectra\footnote{The pipeline also extracts additional 
point sources that serendipitously fell on the slit during the exposures. 
These objects are not considered 
in the present work.} finally amounts to 3631 (see Table~2), where   the 
signal to noise ($S/N$) ratios typically range from 2 to 60 per pixel 
(although a handful of the brightest foreground 
dwarf spectra reach as high as $\sim$120) with a median of 8.5.  
Fig.~2 displays a number of sample spectra of both high and low $S/N$ around 
the CaT and the sodium doublet region, which we will utilize to separate M31 giants from foreground dwarfs in 
Section 4. 

\subsection{Photometry}
The photometry of our targets, which will be required later on to perform a color-based foreground 
separation and to calibrate our spectroscopic metallicity measurements, was taken from two sources: 
For the fields targeted in the outer regions of M31's halo (R$\ga$25 kpc) 
we used the Washington $M$, $DDO51$ and $T_2$ photometry of 
Ostheimer  (2003), which provides a strong separation criterion for red giant selection 
(Palma et al. 2003). 
These filters were transformed from the Washington system into standard Johnson-Cousins 
V and I magnitudes by applying eqs.~1,2 of Majewski (2000). 

The photometry of targets in our inner fields  (R$\la$25 kpc) on M31's minor axis was, on the other 
hand, taken from the MegaCam/Megapipe archive (Gwyn 2008). These data are  
 available in $i'$ and either $g'$ or $r'$.   Typical exposure times range from 800
to 3757 s for $i'$, 1600 to 3200 s for $r'$ and 1445 to 3468 s for $g'$.   
Photometric errors in the catalog are below 0.1 mag for $g' < 24$ mag, and rise to 0.25 mag 
at $g = 26$ mag.  The errors in $r'$ are well below $0.1$ mag for $r' < 24$ mag and reach 0.2 mag
at $r' = 25$ mag.   Finally,  errors on the $i'$-band magnitudes are below 0.1 mag for almost the entire sample below 22.5 mag, with a maximum error of  0.15 mag at $i'=24$ mag.  
The photometric data was then  matched to our spectroscopic catalog by 
 requiring the coordinates from the two sets to match within better than 1$\arcsec$.  
 In most cases, the match was better than 0.2$\arcsec$.    

In order to determine the spectroscopic metallicity of each star, V-band magnitudes in 
the Johnson-Cousins system are 
required.  As only the $g'$ or $r'$ and $i'$ filters in the photometric system of the CFHT are available, 
a transformation to V magnitudes is determined from the latest  Padova stellar isochrones (Marigo et al.
2008),  which are available in the CHFT photometric system as well as  for Johnson V.
In practice, we obtained transformations from 
isochrones with metallicities ranging from $-$2.3 to +0.18 dex and ages of 10 and 12 Gyr.
 The transformation from $g',i'$ to V was obtained in four sections: 
\begin{equation}
\begin{array}{c@{V=g'}r@{\,(g'-i')}r@{;\qquad}r@{g'-i'}r}
& +0.39 & +0.010& & <1.25\\
& -0.34  & -0.050 &   1.25 \le &  < 1.70 \\
& -0.06 & -0.525 & 1.70 \le & < 3.85\\
& -0.22 & +0.740 &3.85 \le &
\end{array}
%V &=& g'  + 0.39\,(g'-i') + 0.0102; \qquad(g'-i' < 1.25) \\
%V &=& g'  - 0.34\,(g'-i')  - 0.05; \qquad(1.25 \le g'-i' < 1.7)\\
%V &=& g'  - 0.06\,(g'-i') - 0.525 ;\qquad(1.7 \le g'-i'< 3.85)\\
%V &=& g'  - 0.215\,(g'-i') + 0.74; \qquad(3.85 < g'-i')
\end{equation}
On the other hand, the transformation from $r,'i'$ to V is defined as follows:
\begin{equation}
\begin{array}{c@{V=r'}r@{\,(r'-i')}r@{;\qquad}r@{r'-i'}r}
& \,+\,0.98 & \,+\,0.048 & & <1.8\\
& \,+\,0.19  & \,+\,0.445 &   1.8 \le &  < 4.0 \\
& \,+\,1.03 & \,-\,1.500 & 4.0 \le & \\
%V &=& r' + 0.98\,(r'-i') + 0.048 ; r'-i' < 1.8 \\
%V &=& r' + 0.19\,(r'-i') +  0.445; 1.8 \le r'-i' < 4 \\
%V &=& r  + 1.03\,(r'-i')  - 1.5;  4 < r'-i'
\end{array}
\end{equation}

Typically, these equations are insensitive to the adopted metallicity and age of the isochrones 
and differences in V-magnitude from the different isochrones are less than 0.05 mag. 
Only for the super-solar isochrone, the overall systematic errors increase to 0.1 mag.  
To account for this we  add a 0.1 mag uncertainty in quadrature to the final error estimates 
on the V-band magnitude.

The resultant CMDs for all targeted objects are shown in Fig.~3, separately for M31 giant candidates, 
foreground dwarfs and contaminating background galaxies that were separated using the methods 
outlined in Sect.~4.

\section{Velocity measurements}
Radial velocities were measured  by cross correlating our DEIMOS spectra against 
a high $S/N$ template spectrum of the bright 
K1 red giant HD139195, which was observed using the same instrumental setup as 
our observations  (M. Geha, private communication; Simon \& Geha 2007). 
In this way we avoid any  systematic uncertainties occurring from 
potentially different spectral resolutions and dispersions. 
The correlation was performed using  {\sc 
iraf}'s\footnote{{\sc iraf} is distributed by the National Optical
Astronomy Observatories, which are operated by the Association of
Universities for Research in Astronomy, Inc., under cooperative
agreement with the National Science Foundation.} {\em fxcor} task. 
Preferentially, the entire covered spectral region was used in the correlation,    
where we rejected the wavelength regions of telluric absorption, in particular 
the prominent atmospheric A- and B-bands at $\lambda\lambda$\,7600, 6860\AA.  
For cases in which the correlation of the entire spectrum produced weak or no 
correlation peaks, i.e., for the lowest $S/N$ spectra, we restricted the windows to single narrow 
regions around prominent 
absorption features, such as the CaT from 8475--8662\AA\ and/or the Na doublet from  
8179--8200\AA.   
Choice of these band passes will minimize the contribution of potential residual telluric absorption 
lines (e.g., Schiavon et al. 1997). 

Each correlation peak was examined by eye to avoid spurious detections, which 
might lead to significantly erroneous  velocity estimates.  A Gaussian fit to the strongest 
correlation peak then yielded the final relative radial velocity value for the respective spectrum. 
Finally, heliocentric corrections were computed for each star individually to yield 
 heliocentric radial velocities, v$_{\rm HC}$, that we will use for the remainder of this work. 

The measurement errors on the radial velocities are returned by {\em fxcor} and 
are internally computed based on the Tonry-Davis 
$R$-value (Tonry \& Davis 1979) of the cross correlation. 
Thus our median random velocity error amounts to 8.0 km\,s$^{-1}$. 
Due to our interactive procedure of assessing each spectrum by eye, we were able to 
discard bad spectra and cases in which no correlation peak could be discerned 
on the spot from our sample. These are not considered for the remainder of this work. 
Moreover, red background galaxies, which significantly contaminate our target sample  
(of the order of 11\% by numbers),  were identified based on their emission and absorption lines 
and culled from the present sample (see appendix). 

A total of 110 stars were observed on adjacent masks. These repeat observations allow us 
to further assess the accuracy of our velocity measurements. We find an overall good 
agreement between the independent velocity measurements of the same stars from 
different masks: the mean deviation is 0.3 km\,s$^{-1}$ and the 1\,$\sigma$ scatter amounts to 
8.2 km\,s$^{-1}$, which is consistent with our measurement errors. 
The according reduced 
$\chi^2=1/N\,\sum_i\frac{(v_{i,1}-v_{i,2})^2}{(\sigma_{i,1}^2+\sigma_{i,2}^2)}$ 
is close to unity. Thus we conclude that our duplicate velocity measurements 
are consistent within the uncertainities.  Moreover, this shows that the formal errors returned 
by {\em fxcor} correctly reflect the accuracy of our data, also in the light of potential 
template-target mismatches in stellar type (e.g., Majewski et al. 2004), so that there is no need 
to re-scale these values (cf. Koch et al. 2007). 
As the final velocity for stars with repeat measurements we adopted the error-weighted 
mean of the individual values. 
As as result, reliable velocities could be determined for 2262 of our stars (see Table~2). 

\section{Membership separation}
In order to isolate the true sample of M31 member stars from undesired  
contamination of numerous foreground Milky Way dwarfs, we utilize the three strongest 
discriminators, viz. 
 $V-i'$ color, the equivalent width (EW) of the \ion{Na}{1} doublet at $\lambda\lambda$\,8189, 8193\AA, 
and radial velocity.  In practice, the EWs of the two Na lines were measured by 
numerically integrating the spectral flux within a bandpass from 8179\AA\ to 8200\AA\ with 
suitable continuum bandpasses (Gilbert et al. 2006), and the errors were obtained  by Monte Carlo 
simulations accounting for the continuum variance of the spectra. 
 
To achieve a separation of dwarfs and giants, we follow Gilbert et al. (2006) 
in splitting our complete observed data set into one dwarf- and one giant training sample:  
For this purpose, all stars with radial velocities below $-200$ km\,s$^{-1}$ were considered to be M31 red giants, 
whereas those exceeding $-50$ km\,s$^{-1}$ are most likely  foreground stars. 
Since it cannot be assumed {\em a priori} that the color or Na EW distributions constitute well defined 
analytical, e.g. Gaussian,  profiles, we instead  
define empirical probability distributions (PDFs) $P$ 
from the actual observed training samples in color and Na EW space. 
The PDFs were then convolved with the individual measurement errors. 
Likewise, for defining the PDF with respect to  radial velocity, we adopted a color cut at  $V-i'$=2.0, where stars above this limit 
are taken to be dwarf candidates. In parallel, 
taking advantage of the surface gravity sensitivity of the Na doublet (Schiavon et al. 1997), we 
flagged stars with EWs above  2.5\AA\ as dwarfs. These empirical limits were chosen because 
they turned out 
to optimize the separation in the  $V-i'$ vs. Na EW parameter space. 
For the M31 giant radial velocity distribution, on the other hand, we adopted a one-sided Gaussian centered 
at a systemic velocity of $-$300 km\,s$^{-1}$ with a radial velocity dispersion of 85 km\,s$^{-1}$, in concordance 
with the  values of M31's outer halo component (e.g., Reitzel \& Guhathakurta 2002).  
All stars with velocities below 
$-$300 km\,s$^{-1}$ were assigned a probability of unity  in the giant training sample. 

Although the number of dwarf stars in this low-velocity regime is expected to be well below 1\% 
and their contribution to the final velocity 
histograms may be negligible (Gilbert et al. 2006), we will turn to a detailed treatment of this component 
in  Sect.~4.1. 

Fig.~4 shows the resulting PDFs of the training samples as dashed lines. 
The final probability of any star from our sample, with its set of ($v_{\rm HC}$, $V-i'$, Na EW),  
to be either a dwarf or a giant was then determined 
from the PDFs, normalized to unity, 
in each parameter and combining these values into a total probability of being a giant: 
$P_{\rm giant}=P_{\rm giant}$($V-i'$)$\,\times\,P_{\rm giant} $(Na EW)$\,\times\,P _{\rm giant}$(v$_{\rm HC}$),  
and likewise for the probability of being a dwarf. If the resulting logarithmic likelihood 
$L=\log\left(P_{\rm giant} / P_{\rm dwarf} \right)$ is less than zero, the star is considered a dwarf, while 
$L>0$ signifies  likely giant star candidates (Gilbert et al. 2006).  
The {\em a posteriori} histograms in Fig.~4 illustrate our full data set, based  on the selection 
method described above. It is evident that the selected samples in all parameters are  fully compatible 
with the pre-defined simple training samples and that the number of remaining dwarf contaminants in 
our cleaned M31 giant sample can be expected to be negligible. 
There is nonetheless a considerable overlap of dwarf and giant stars in color space. Moreover, 
we note the presence of a population of stars, flagged as dwarfs based on the full likelihood analysis, 
which exhibit small EWs of the Na doublet. From this it becomes obvious that it is in fact necessary to 
include the entire set of available information in color, EW and velocity in the analysis to obtain an optimal separation. 
On the other hand, there will be an inevitable, though small, fraction of interlopers that cannot be 
reliably detected using the traditional separation criteria, due to potential 
covariances between each of the parameters. For instance, we note the 
presence of one star with colors and moderate Na widths representative of a dwarf star (as 
subsequently verified by its spectral features), but with  
a high negative velocity of $-$280 km\,s$^{-1}$. In this case, the velocity criterion will 
override the other discriminators and lead to classifying this star as an M31 member. 
Such cases could be mainly resolved by visual inspection of the individual spectra. 
    
We also note  that we refrain in our dwarf/giant separation from adopting 
additional secondary parameters 
such as weaker spectral features (e.g., \ion{K}{1} EWs, TiO bands, embedded within telluric absorbtion 
bands) to avoid adding further noise to 
the final PDFs. 
Furthermore, we did not weight our combined likelihood by the number 
of available diagnostics to account for potential outliers in either of the parameters, thereby yielding
 a statistically more robust and uniform rejection of dwarfs  (cf. Gilbert et al. 2006). 

To assess the accuracy and efficiency of the separation methods described above, we ran 
a comprehensive suite of {\em Monte Carlo simulations}.  To this end, each of the three indicators
was varied 10$^4$ times by its measurement uncertainty, from which subsequently new PDFs 
were built and each target's dwarf/giant status was re-determined. 
{\em As a result, 92\% of those 
stars previously classified as giants were still classified as such in 95\% (2$\sigma$) of the 
Monte Carlo realizations.  Thus we are convinced to have obtained a solid dwarf/giant 
separation of our sample.
 In particular, none of the results obtained in this work changes 
significantly, whether ``all'' giants are included or only those with secure, 2$\sigma$-classifications}. 
The final number of 
giant candidates per field is listed in Table~2. 
The ratio of giant to dwarf stars decreases
with increasing radial distance from M31 (Fig.~4, bottom right),
as expected as the M31 halo density levels off, 
until the halo of M33 contributes giants in the
outermost fields.

There is still a non-negligible fraction of stars classified as giants present  
in a transitional region around $-$150 km\,s$^{-1}$ ($\sim 1.7\,\sigma$ above the systemic velocity), which 
prevails in the outermost minor axis fields (Sect.~6.2). 
Another noteworthy outcome of our 
dwarf removal is that the addition of velocity as a membership criterion 
effectively deprives the sample of red giants above $-$100 km\,s$^{-1}$.
This limit already corresponds to removing 
stars that deviate by more than approximately 2.3\,$\sigma$ from the sample, and we do not expect the presence of 
any major population of M31 stars in this high velocity regime.  
Considering the limited S/N of the spectra in our dataset as well as that of the photometry, we have 
concluded that it is essential to use radial velocity as a dwarf/giant separation criterion. As a result, the velocities 
of our dwarf-cleaned sample span a full range from $-$570 to $-100$ km\,s$^{-1}$.  
%We further 
%suspect that the predominance of dwarfs for $V_{\rm HC}>-100$ km s$^{-1}$ 
%would render any dwarf/giant separation problematic in that velocity range.
Additionally, the constancy of the foreground Galactic dwarf sample's velocity and dispersion  (Fig.~11, 
bottom left) over the entirety of the M31 halo strengthens further our case for efficient dwarf/giant separation.

In a recent work, Sherwin et al. (2008) predicted that a total number of $\sim$5 hypervelocity stars 
with velocities below $-$420 km\,s$^{-1}$ should be identifiable in M31's halo. 
However, given M31's large overall velocity dispersion and the low number of stars with the highest negative velocities, it is  impossible to resolve, whether any 
of those stars in our sample are in fact ejected from the center of M31, 
or if they are canonical (2$\sigma$-) members of its genuine halo. 
\subsection{Comparison with the Besan\c con model}
As Fig.~3 shows, there are still a number of stars present bluewards  of the most metal poor isochrone 
that were classified as M31 giants based on all separators. However, it cannot be excluded that a subset 
of these may be blue, metal poor Galactic halo dwarfs, which typically have negligible Na doublet lines and 
a broad range in radial velocities (Fig.~5). Thus these contaminants are indistinguishable from the giant sample 
and their separation is insoluble based on the canonical membership criteria. 
It is important to assess the fraction of these blue stars, since their systematically weaker 
CaT lines will yield falsified, low metallicities. 

To this end, we queried models of the Besan\c con Galactic foreground population (Robin et al. 2003) 
using color cuts and spatial locations in analogy to the observed samples. The resulting distribution 
for the outermost field at 160 kpc is illustrated in Fig.~5 (right panel). We chose to exemplary plot  
this outermost field, because it is the one for which the Milky Way contamination is expected to be the highest. 
The first thing to note is that there is in 
fact a non-zero population of Galactic stars predicted  at M31's systemic velocity,  
with  radial velocities as low as $-$420 km\,s$^{-1}$ (although we note that the velocity dispersion in 
the Galactic model may in fact be overestimated).  Most of the contaminants are, however, 
distinguishable, either by their  high velocities or their redder colors. We thus estimate the number 
of undetectable blue stars in our sample by determining the {\em predicted fraction of dwarfs} with 
v$_{\rm HC}\la-150$ km\,s$^{-1}$ and $V-i'\la1$ to those dwarf stars in the color-velocity space that 
we are able to distinguish (Fig.~4; Sect.~4). Multiplying this fraction with the number  of our observed 
giant candidates (Table~2) then shows that there are typically no more than 0--2  blue 
dwarf stars to be expected per field in our giant sample that cannot separated by any of the observable criteria 
(see also Fig.~5, right panel), leading to a total predicted number of 35 such contaminants in the entire 
sample. If present, these will have a negligible effect on the more populous, true giant sample. 

Martin et al. (2007) model the CMD of these fields, reaching fainter than the limit of our spectroscopy,   
and find no evidence for excess star counts of Galactic members. 
In principle, dwarf members of the Andromeda-Triangulum 
stream or a potential contaminant may be present, especially in the M31 giant-poor outer fields.  
Rocha-Pinto et al. 
(2004) have measured radial velocities of Andromeda-Triangulum stream members, finding one star 
with $-245$ km\,s$^{-1}$, but with most stars at  higher velocity.  
Furthermore, the Monoceros Ring, whose main sequence stars can overlap with M31 stars near the 
Tip RGB, have a radial velocity that is high enough (at $-$75  km\,s$^{-1}$ with a dispersion of 
26  km\,s$^{-1}$; Martin et al. 2006a) so that they are not an issue in the analysis. 
Based on these studies, neither of the  systems 
poses any risk for contaminating the field with stars at the radial velocity of M31.

\section{A new method for calcium triplet metallicities}
In the frequent cases of low spectral resolution and/or low $S/N$ ratios the last resort is to 
measure gravity and/or abundance-sensitive indices in 
band passes a few times the spectral resolution,  
which are then calibrated theoretically via a grid of synthetic spectra (e.g., Jones et al. 1996). 
The latter is in general a critical endeavor for the CaT, since these lines are formed 
in the upper chromospheres of the stars, which ideally requires a full non-local thermodynamic 
equilibrium treatment  (Smith \& Drake 1990; J{\o}rgensen et al. 1992). 
Detailed model computations are however sparse so that present-day studies of stellar 
populations mostly rely on empirical calibrations (Cenarro et al. 2001; and references therein). 

Canonically, the line strength of the near-infrared CaT, $\Sigma W$, has been defined 
by a weighted sum of the three individual lines' pseudo equivalent widths, EW$_i$: 
\begin{eqnarray}
\Sigma W \, = \, \sum_{i=1}^{3}w_i\,EW_i 
\, = \, \sum_{i=1}^{3}w_i\int_{BP_i}\left(1 - \frac{F	(\lambda)}{F_{c,i}}\right)\,d\lambda,    
\end{eqnarray}
where $F$ denotes the flux in a predefined set of line band passes ($BP_i$) and 
$F_c$ is the continuum level as determined in a set of continuum bands. 
There is no physical motivation to prefer any set of the weight factors $w_i$ over the other, 
as long as a consistent definition is used between the target red giants to be calibrated and those 
in the calibrator systems, i.e., the high $S/N$ Galactic globular cluster spectra. 
The choice of the weights is mostly governed by the spectral quality and measurability of each of the 
Ca lines. 
In this vein, the most frequently used weights throughout the literature are  
 ($w_i$=1 $ \forall i$; e.g., Armandroff \& Zinn 1988),  
 ($w_1=0$,  $w_2=1$,  $w_3=1$; e.g., Armandroff \& DaCosta 1991) and 
 ($w_1=0.5$,  $w_2=1$,  $w_3=0.6$; e.g., Rutledge et al. 1997a).  
The EWs are then either determined by numerically integrating the spectral flux 
over the full bandpass or by fitting an analytical function $F(\lambda$)
to the line profile. However, both methods tend to fail at the lowest $S/N$ ratios,  
where a pure numerical integration merely reflects the noise of the spectrum rather than 
the actual EWs,Ê while the lines cannot be reliably fit anymore in the low $S/N$ regime. Typically, 
the limiting $S/N$ for which CaT based [Fe/H] measurements are given 
in the literature lies at 10--15. 

At the low $S/N$ ratios of our faintest DEIMOS targets, which reach as low as 2--5, 
a reliable determination of the CaT EWs is not feasible. Moreover, the presence 
of sky line residuals around the CaT  often leads to artificially increased widths if  
uncritically integrated over the respective band passes. 
In particular, the third of the CaT lines at 
8662\AA\ is susceptible to this increased noise component. 

Hence, in order to enhance the $S/N$ in the CaT of each individual spectrum, we define a coadded line strength $<$$\Sigma$$W$$>$  
in that we interchange the order of summation and integration in eq.~3:  
\begin{equation}
<\Sigma W>=\int_{<BP>}\sum_{i=1}^{3}w_i\left(1 - \frac{F(\lambda-\lambda_{0,i})}{F_{c,i}}\right)
\,d(\lambda-\lambda_{0,i}).
\end{equation}
That is, each line center is shifted towards a zero wavelength before performing a weighted coaddition of the 
three lines in this rest frame, using the identical weight factors as in the canonical 
definition. Mathematically, this expression is fully equivalent to the traditional 
definition in eq.~3, but it provides the advantage of integrating the {\em coadded flux}, resulting in an 
increased effective $S/N$.   
The integration is then carried out over a single common band pass. 
This procedure strictly presupposes that the individual band passes $BP_i$ all have the same 
width and location relative to the line center so that $BP_i - \lambda_i$ is the same 
for each of the lines. 
However, we will tie our measurements to a metallicity reference scale by 
using our own suite of measurements in Galactic globular clusters. Thus 
possible differences in the band passes will affect the reference spectra and our target spectra 
in the same manner and these will not introduce any systematic bias.  
 In the following, we will follow the prescription of Rutledge et al. (1997a) 
 by adopting the line weights of 0.5, 1 and 0.6. 
Our coaddition method will improve the effective $S/N$ in the CaT region by a factor 
of $\sqrt{\sum w_i}$, or 1.45, so that we will be capable of measuring metallicities 
even with $S/N$ as low as $\sim$7--10. Furthermore, the coadded line is more robust 
against potential sky residuals and noise spikes, enabling us 
to fit a line profile 
(see the illustration in Fig.~6). 
In practice, we fit the resulting line with a Penny function, i.e., a Gaussian plus a Lorentz 
component, which has proven to provide the best representation of  the line wings (Cole et al. 2004). 

In order to tie our CaT metallicity measurements in the M31 stars to a reference  
scale, we performed the identical coaddition technique as described above 
on a sample of globular clusters of known 
metallicity and thus derived the CaT line strengths of the 
globular cluster stars. 
Since no observations of globular cluster standard stars were taken for the 
present project 
using the same instrumental set up as for the M31 science observations, 
we exploited the  data set of 
Koch et al. (2006), which consists of high $S/N$ spectra of 80 red giants in four Galactic 
clusters, 
NGC\,3201, 4590, 4147 and 5904, obtained with the FLAMES spectrograph at ESO/VLT (Pasquini 
et al. 2002). Since the FLAMES spectrograph provides a higher spectral 
resolution than DEIMOS,  we  degraded the spectral resolution of the FLAMES spectra 
to match that of DEIMOS.  
The final calibration of the line strengths onto metallicity was then achieved 
accounting for the stars' magnitude above the horizontal branch (HB; e.g., Rutledge et al. 1997a,b), where  
we find the following relations 
\begin{eqnarray}
W' &=& <\Sigma W> +\,0.55\,(V-V_{\rm HB})\\
\mathrm{[Fe/H]}_{\rm CaT}  &=&-2.90 + 0.45\,W' 
\end{eqnarray}
with an r.m.s. scatter of 0.02 dex on the [Fe/H]$_{\rm CaT}$ calibration of eq.~6. 
These relations from the FLAMES data are shown in Fig.~7.
In this calibration we explicitly adopted the globular cluster metallicity 
scale of Carretta \& Gratton (1997). Furthermore, we assumed 
a HB apparent magnitude of 25.17\,mag for the M31 stars, which 
corresponds to the mean magnitude of its halo red HB population (Holland et al. 1996).  
The extent of M31's halo will inevitably lead to a spread of distances along the sight 
so that the adoption of a single HB magnitude is a simplifying assumption. 
Ibata et al. (2007) estimate that, for an extended $\rho (r) \propto r^{-2.9}$ density 
profile, the variation in the distance modulus is typically less than 0.5 mag. 
Translating this into a HB spread and by applying our calibrations (Eqs.~ 5,6) this 
leads to an uncertainty of 0.12 dex on the spectroscopic metallicities. 
Given the {\em a priori} unknown distance of individual red giants, this 
source of uncertainty  cannot be eliminated and will affect all spectroscopic metallicity 
measurements in M31. 
 Furthermore, the adoption of the HB of the oldest stars will introduce a systematic effect on 
 spectroscopic [Fe/H] estimates in the presence of a non-negligible intermediate-age population. 
Koch et al. (2006) estimate that the simplification of a single-age HB results in in a bias 
of the order of $-$0.1 dex, that is, the intermediate-age  stars may be 0.1 dex too metal poor. 
This is the same order of magnitude that Cole et al. (2004) find from their cluster sample, 
which includes a number of young open clusters.
Of course, one cannot assess which stars are affected to what extent, since their 
ages cannot be assigned {\em a priori}. 
In this context, we note that also the M31 globular clusters show a mild trend
in mean $M_v$(HB) vs. [Fe/H]; this may be lessened
by the brighter HBs of intermediate age populations.
This trend is 0.4\,mag but has a full
width of 0.6\,mag (Rich et al. 2005, their Fig.~12), with a 1$\sigma$ scatter of 
0.2 dex. 
This compares to a small formal random uncertainty of typically less than 0.01 dex 
on [Fe/H]$_{\rm CaT}$ that is introduced by the photometric errors through Eqs. 5,6.  
Even an unrealistically large error of 1mag on a star's $V$ magnitude would result in 
a metallicity error of 0.23 dex. 

We note that the coefficients of our relations in Eqs. 5,6 are slightly different from 
the standard calibration of Rutledge et al. (1997b), but for the sake of consistency between 
our own measurements and our own suite of calibration clusters, we will use these 
relations for the remainder of this work (see also the discussions in Koch et al. 2006). 

In order to estimate the random and systematic measurement uncertainties on our 
derived coadded CaT metallicity, we performed a set of Monte Carlo simulations 
(e.g., Simon \& Geha 2007), accounting for three effects. 
For this purpose, we added artificial Poisson noise to theoretical spectra, consisting of three 
Penny lines with line strengths representative of typical red giants as expected 
in our M31 sample. This noise addition accounts for the spectral quality in terms 
of the $S/N$ ratio and the variance of the continuum.
Secondly, we also added additional random 
noise peaks and troughs on top of the individual CaT lines, thus  simulating 
potentially bad sky subtraction residuals that hamper CaT measurements.  
Even if one were to assume that the individual line profiles were perfect Penny functions,  
the (weighted) sum will not necessary be a Penny anymore. 
By thus measuring the simulated coadded spectrum and comparing the 
obtained line strength with the traditional strength of the individual theoretical 
line profiles in the unperturbed spectrum according to eq.~3, we 
can finally estimate the influence of random noise, residual noise peaks and deviations 
from the assumed analytical line profile. In this vein, we adopt the standard deviation 
of the difference of the measured $<$$\Sigma$$W$$>$ in the noise added spectrum and 
the $\Sigma W$ of the input spectrum as a measure of the measurement uncertainty 
as a function of the spectral $S/N$ ratio and continuum variance. 
Thus Fig.~8 depicts the resulting relative uncertainty estimates. Typically, the widths 
measurements are accurate to the 15\% level at a $S/N$ of 10 and the relative error on 
the widths, $\sigma$$<$$\Sigma$$W$$>$$/$$<$$\Sigma$$W$$>$, generally does not exceed 
25\% at the spectral quality of our DEIMOS data. Applying our calibration from above (Eqs.~5,6), 
this translates  into typical metallicity errors of (0.32,0.25,0.2) dex at $S/N$ of  (5,10,20). 
As another potential source of uncertainty, we estimate that a conservative continuum placement 
error of 2\% will result in metallicity errors less than 0.1 dex. 

Spectra with too low a $S/N$ ratio to even measure the coadded line strengths were 
discarded from our sample on the spot so that we end up with a total number of 
1673 CaT measurements (Table~2). 
An aggravating factor for using the CaT as a proxy for metallicity is the onset of strong 
TiO absorption in the coolest stars. Increasing strength of the TiO band at 8500\AA\ 
progressively depresses the CaT lines for redder stars so that the line strength $\Sigma$$W$ 
starts to turn over towards lower values for colors redder than $V-i'\ga2$ (e.g., Garnavich et al. 
1994; their Fig.~13). This ambiguity, which also reflects in an overturn of the RGBs of metal rich globular 
clusters, 
prompted us to only include stars with $V-i'\le2$ in the present chemical analysis. 
In this way we avoid an underestimate of the metallicities of the reddest stars. 
Thus {\em 64 giant stars} with velocities compatible with the M31 mean  were removed from the CaT sample.  We note that most of the reddest, nominally most metal poor stars 
that were thus rejected, are mostly found within $\sim$40 kpc so that they do not affect our 
later conclusions about the overall large-scale radial metallicity distributions.  
\subsection{On photometric metallicities} 
As a secondary estimate, we derived photometric metallicities of our stars by following isochrone 
fits laid out in Reitzel \& Guhathakurta (2002). For this purpose we adopted 
individual corrections for reddening from the Schlegel et al. (1998) maps.   
In practice, we fit a set of isochrones to our CMD (Fig.~3), 
where we employed the 
stellar tracks without $\alpha$-enhancement  
from the Padova group (Marigo et al. 2008).
Our choice to negelect $\alpha$ enhancement is justified
by the age range in M31's halo (Brown et al. 2006): 
Stellar populations with a wide age range have had sufficient time for supernovae of type I 
to contribute iron and therefore the composition trends toward Solar.
Each star's locus in the CMD  was then fit by a surface, using a 12th order function 
in both  $g'-i'$ and $i'$, with full cross terms, to the metallicity. In practice, we generated 
a grid of isochrones from the group's web-interface\footnote{ 
\url{http://stev.oapd.inaf.it/$\sim$lgirardi/cgi-bin/cmd}},   
covering $0.0001\le Z \le 0.03$ with a spacing of 0.001. 
Moreover, we adopted the set corresponding to a population of 12.7\,Gyr age, in concordance 
with the oldest populations found in the  M31 halo (but see also Brown et al. 2003, 2006, 2007). 

As the comparison with our CaT measurements in Fig.~9 indicates, the dwarf stars clearly deviate from 
unity, while the metallicities from both methods agree better for the giant candidates, albeit with a 
broad scatter (see also Kalirai et al. 2006a). The overall agreement within this scatter is notable 
down to the lowest metallicities. 
We will exclusively rely on our {\em spectroscopic} metallicities throughout this 
work, since the photometric metallicity estimates are generally prone to a number of inconsistencies:  
Among these are the strong influence of photometric uncertainties. 
The median error on our colors is 0.05 mag, but it reaches $\sim 0.15$ mag for a number of our targets. 
For the more metal rich stars, the {\em formal} metallicity errors due to photometric errors 
are below 0.05 dex, but they exceed 0.1 dex for stars bluer than a  $g'-i'$ of $\sim$1.5 and can 
reach as high as 0.5--1 dex for the bluest giants. 
In this locus, the more metal poor isochrones become more and more degenerate and do not permit 
a reliable metallicity determination (see Fig.~3). 
In contrast, the CaT method has its greatest sensitivity at the metal poor end. 

For populations with [Fe/H]$> -1$ dex, the red giant branch curves both red and faint in a complex way, 
due to the onset of TiO absorption in the coolest stars (e.g., Garnavich et al. 1994).  
This behavior is not well modeled and 
certainly affected by $\alpha$-enhancement.  The assignment of metallicity becomes a complex function of both 
color and magnitude.  
Moreover, the systematic uncertainties on the photometric metallicities 
will be inevitably large, given the 
{\em a priori} unknown age- and abundance distribution of our stellar sample that 
is drawn from a patchwork of M31's populations.
In this context, the presence of a broad distribution in stellar ages has been confirmed within the inner spheroid fields by the 
deep HST/ACS based CMDs of  Brown et al. (2003, 2006, 2007), where 
in fact 30\% of the stars in minor axis fields at 11 and 21 kpc were found to be 
6--8\,Gyr old.  
       
Likewise, the unknown effects of varying distance modulus (Ibata et al. 2007) can easily move stars between different isochrones and pose an additional source of uncertainty.
A plausible distance modulus spread of 0.5 mag  translates into a 0.12 dex variation in the  spectroscopic metallicities. 
This uncertainty in the stellar magnitudes will have, however,  much larger effect on the photometric values.   
We will return to the issue  of photometric metallicities in Sect.~7.1.
\section{Velocity substructures}
We shall now turn to the analysis of the dwarf cleaned M31 velocity distribution. 
For this purpose, we compare our observed data to a new suite of $N$-body simulations. 
\subsection{Simulating satellite accretion into M31's halo}
In a new set of $N$-body simulations by Mori \& Rich (2008)  
a satellite galaxy is accreted onto M31, for which the self-consistent 
potential of Widrow et al. (2003; their model A) was used. This potential is essentially a three component disk/bulge/halo 
model, the parameters of which were optimized to match M31's rotation curve and its surface brightness and 
velocity dispersion profiles. 

Mori \& Rich (2008) adopt the orbit of Fardal et al. (2006), which was originally 
constrained to reproduce the stellar distribution of the observed giant stream. 
For the accreted galaxy itself a total mass of $10^9\,M_{\odot}$ distributed as a 
Plummer sphere with a scale radius of 1 kpc  was used. The progenitor mass can in general be well 
constrained by means of the observed thickness of the M31 disk (at $z_d$=300 pc; Kuijken \& Dubinski 
1995)  -- excessive masses would have led to 
an early destruction of the thin disk component as we observe it today. 
In practice, the disk was represented by 7.4$\times$10$^6$ particles with a total  mass of 
7$\times$10$^{10}M_{\sun}$ and a scale length of 5.4 kpc.  
The bulge  mass was assumed as 2.5$\times$10$^{10}M_{\sun}$ (2.6$\times$10$^6$ 
particles), whilst the halo contained 3.2$\times$10$^{11}M_{\sun}$ (30.8$\times$10$^6$ particles) 
with a tidal radius of 80 kpc. 
Yet, there is generally a 
significant lack of knowledge about the outer density profile of the
dark matter halo. Therefore, one cannot avoid ambiguities in 
estimating  the tidal radius and the total halo mass.  
In particular,  the model of Mori \& Rich (2008) has a smaller
radius and mass compared to those of earlier studies (cf. 
8.8$\times$10$^{11}$ M$_{\odot}$ and a radius of 195 kpc;  Fardal et al. 2007). 
However, since the satellite orbit is mainly followed within 50 kpc from the center of M31,
the outer structure of the dark matter halo is of little
relevance to the dynamics of the satellite. 

In fact, the Mori \& Rich (2008) simulation gives results similar to those of Fardal et al. (2007) 
who have a more massive dark matter halo, but which otherwise have the
 same satellite orbit and position (by construction) and similar disk and bulge mass.

Apart from the vastly increased number of particles in the work of Mori \& Rich (2008), the novelty of these  simulations is the use of a self-gravitating, {\em live} disk and bulge that respond to 
the actual infall by  particle motions of the underlying components. Thus we cannot only trace stars that were 
torn from the satellite in the course of the accretion, but also follow the fate of mutually removed disk and bulge stars.  
Hence we may assess how ejected disk stars might contribute to the halo in comparison to stars originating in the disrupted satellite.  Another important feature of including the live disk and halo is that an accounting can be made of energy input into these populations from the collision event.  This aspect affects the kinematics of the satellite's stars following the collision.

Projections of the simulation at $t=1$ Gyr in the standard M31 coordinate system are shown 
in Fig.~10, separately for disk, halo and satellite particles, as well as the combined simulation data.  
As these plots show, the main features observed in surface density maps of M31's halo, such as 
the Giant Stellar Stream (Ibata et al. 2001) 
and the bubble-like feature in extension of the stream (the Eastern and Western shelves; 
Ferguson et al. 2002, 2005; Irwin et al. 2005; Ibata et al. 2007) 
are well reproduced by the simulations.

\subsection{Minor axis fields}
The results from our simulations are shown in  Fig.~11, in comparison with the 
observed distribution of 
stars on the minor axis in the analogous velocity vs. location space. It is evident that the simulations yield 
a characteristic triangular shape of the satellite particles' velocity as a function of projected 
distance, indicative of its disruption during the accretion (e.g., Merrifield \& Kuijken 1998). 
This feature is also discernible in our observed data (top panel), though to a lesser extent, 
and is consistent with 
the findings of  kinematically cold, i.e., low-radial velocity dispersion, 
localized substructure in Gilbert et al. (2007). While the simulations 
predict the maximum density of this shell-like feature to occur at a projected distance of $\sim 1.4\degr$ 
(19 kpc), the observations indicate  an onset of this cold structure slightly inwards, at $\sim 1.1\degr$ (15--16 kpc). 
Using the Kayes Mixture Modeling algorithm of Ashman et al. 
(1994), we decomposed the (unbinned) velocity data at this location into two Gaussian components, 
thereby verifying the likely presence of a considerably colder substructure with a radial velocity 
dispersion of (29$\pm$22) km\,s$^{-1} $ plus an underlying canonical halo component with 
a dispersion of (110$\pm$10) km\,s$^{-1}$. The population ratios, by number, of these 
components are 24\% vs. 76\%, in good agreement with Gilbert et al. (2007). 
It has been suggested by Gilbert et al. (2007) and  
Fardal et al. (2007) that such kinematic substructure is probably a realization of spatially localized 
shelves, which may provide the forward continuation of the giant stream. 
Other kinematic 
substructures other than the distinct V-shape in our data 
towards the inner spheroid are less conspicuous 
and the inner regions in our observations appear smoother
compared to both the simulations and
in the sample of Gilbert et al. (2007).  

\subsection{Outer halo fields}
Although the radial scale in Figs.~11 and 12  (top panel) was chosen so 
as to emphasize the substructure and  build-up of the inner halo within $\sim 35$ kpc, 
we note that we detect stars that were flagged 
as genuine M31 red giants out to large radial distances of 
160 kpc (see Fig.~12, bottom panel), thus strengthening the claim for the 
large radial extent of M31's stellar halo by Guhathakurta et al. (2006) and Kalirai et al. (2006a). 
Nevertheless, the number  density of confirmed members at large radii is sparse (these are 
addressed in further detail in Fig.~19). 
The outermost field at 160 kpc (m11) contains 16 giant candidates with a mean velocity of $-$180 km\,s$^{-1}$, 
nine of which lie above $-200$ km\,s$^{-1}$ and seven with velocities in excess of $-150$ km\,s$^{-1}$. 
In the adjacent, second most remote field (m8), 
seven stars survived the criteria for being selected as giants,  although only one exhibits a 
radial velocity close to M31's systemic mean, while the remainder shows velocities between 
$-180$ and $-120$ km\,s$^{-1}$, thus about of the order of 1.2--1.8$\sigma$ above 
M31's mean velocity. While we cannot exclude the possibility that these stars are in fact 
members of M31's outermost, extended halo, we labeled their mean velocities as 
upper limits in Fig.~12 (bottom). 
All in all, there is a progressively larger  relative contribution of stars at higher velocities with respect to  
M31's mean towards outer fields. 
Based on their star count maps, Ibata et al. (2007) estimate that the field covered by our 
four m11 masks should contain 
0--2 M31 red giants. In fact Kalirai et al. (2006a) report 
on the presence of 3 bona fide giants in this field. Given the location of this field at a separation of 
4$\degr$ ($\sim$ 50 kpc) of M33, it is then conceivable that the 
stars towards the higher velocity tail are part of the overlapping, extended halo component of M33, as 
suggested by Ibata et al. (2007). 
With its distance from the Sun of 849 kpc (Galleti et al. 2004) , 
M33 is located at a distance  of 220 kpc from M31 (e.g., Koch \& Grebel 2006) and its 
systemic velocity lies at $-180$ km\,s$^{-1}$ (e.g., McConnachie et al. 2006).  
Judging by the surface brightness profiles of Ibata et al. (2007; e.g. their Fig.~28), the contribution 
of M33 stars (coupled with the inevitable foreground component)  appears to set in at radial 
distances $\ga 11\degr$ ($\sim$150 kpc).  

\subsection{The halo's merger origin}
It is then intriguing to ask, to what extent the full halo of M31 has been assembled via the accretion 
of  one or more accretion events such as the one simulated in Mori \& Rich (2008). 
Is there a necessity to invoke more  such 
mergers or is there evidence of several disruptive events involving many smaller satellite systems?
In fact, Ibata et al. (2007) have revealed a wealth of substructures and stellar streams that 
haunt the full extent of M31's halo, thereby complicating the interpretation of 
this halo as a single, smooth entity. 
Therefore, we show in Fig.~13 our observed radial velocity distributions against the simulated ones 
as a function of radial distance. The simulation particles shown were selected 
from locations that coincide with our observed fields, but we inflated the respective selection boxes,  where necessary, so as to ensure the same number of observed and simulated stars for the 
comparison. 

The first thing to note is the 
presence of two major velocity peaks in the simulated data, which reflects the wrap of the 
material stripped from the disrupted satellite galaxy around the M31 disk, as is in fact observed 
in the form of the Giant Stellar Stream.  The regions of the aforementioned cold substructure 
are contained within the $R=17$ kpc histogram, yet there is no clear resemblance 
between observations 
and simulations in this regard. 
While the simulations still show the apparent kinematic bifurcation with 
pronounced peaks at $-400$ and $-200$ km\,s$^{-1}$,  the observations indicate the narrow 
population peaking at M31's systemic velocity of $-300$ km\,s$^{-1}$, with an underlying, smoother and broader genuine halo population. 
To this end, we also plot in Fig.~13 (dashed lines) the contribution of original M31 halo stars from our simulation (again selected from the same spatial locations as above), i.e., 
those, which are not related to the disrupting satellite in any way.  
It is striking that a major part of our velocity histograms at all radii closely resembles this genuine halo 
component, and that the strongest deviations of the velocity distributions from the simulations 
occur towards the fields at approximately 
13--17 kpc.  
Furthermore, there is a  clear discrepancy  
between the model predictions and the observed distribution in the innermost spheroid ($\lesssim$13 kpc).  
This kinematic evidence suggests that the inner spheroid of M31 cannot entirely consist of debris 
from one collision like the one responsible for the Giant Stellar Stream. In fact most of the observations at all radii appear to strongly overlap with the prevalent
 M31 halo component, again showing that the accretion of a massive satellite galaxy 
did presumably not  single-handedly drive the build-up of the inner halo of  M31.

Moreover, the discovery of at least 4 major streams perpendicular to M31's minor axis, which also 
intersect our observed fields (Ibata et al. 2007) clearly shows that essentially all of 
our targeted fields are contaminated by a more or less  significant fraction of 
stars originating in the mergers of the systems responsible for these streams (Chapman et al. 2008). 
Although the exact shape of the simulations' velocity distribution depends on the 
extensive parameter space of progenitor structure, its kinematics and orbit, 
the overall trends and 
discrepancies between model and observations discussed above will not sensitively rely 
on such parameter variations and are expected to persist, in particular, since the Mori \& Rich (2008)  
model reproduces the structural features in the halo observed in the surface brightness maps 
remarkably well. 

The observed velocity distributions (Figs.~12,13) 
in the Giant Stream fields H13s and a3 
(at 21 and 32 kpc) show the clear signatures at $-520$ and $-400$ km\,s$^{-1}$ (H13s) and 
$-450$ km\,s$^{-1}$ (a3). In accordance with Kalirai et al. (2006b) 
we find the H13s peak at $-520$ km\,s$^{-1}$ 
to be more prominent relative to the one at higher velocities. 
It is then noteworthy that, while well predicting the Stream component at $-$400 km\,s$^{-1}$ 
at $\sim$17 kpc (bottom left panel of Fig.~13), our simulations and the models of Fardal et al. 
(2006, 2007), on which our orbit is based, fail to reproduce the primary, low-velocity   
peak prominently seen in the 22 kpc histogram, as well as the outer stream field in 
the data 30 kpc. This feature is, however, well reproduced by the simple orbit model by Ibata et al. 
(2004). 
 
Interestingly, neither disk, nor bulge stars, ejected 
during the simulated merger event contribute any considerable fraction to any of 
the potential substructures in the velocity histograms. 
Only a few of these ejected disk stars are to be found within the innermost 10 kpc: 
The number ratios of model particles 
within a region corresponding to the observed field f109 are 0:2:280:1 (bulge/disk/halo/satellite). 
None of the disk particles ventures any further out. 
At a total of 158 giant stars in this field, we would expect no more than 1--2 stars ejected 
from the M31 disk. 

\section{Abundance substructures -- a strong gradient}
In Fig.~14 we show our metallicity results from the new coaddition measurements (Sect.~5) 
as a function of radial velocity and radial distance. 
It is worth noticing that the distribution of these spectroscopic metallicities with velocity (left panels) 
appears to be a powerful dwarf/giant discriminator that can be efficiently employed in cases, where 
not all of the traditional indicators are available (cf. Sect.~4). 
In this representation, the dwarf stars occupy a narrow range above $\ga -150$ km\,s$^{-1}$ 
around ``[Fe/H]''$\sim -2$ dex. Given the lack of knowledge of their distances, their assignment 
relative to a HB magnitude becomes meaningless. It is also far from self-evident that the 
CaT linestrength correlates with metallicity in the dwarf stars as in the giants. 
For this reason, the traditional calibrations (Rutledge et al. 1997a, 1997b; eqs. 5,6) are not valid 
for dwarfs anymore and their application leads  to their clear separation in Fig.~14. 

Contrary to the strong clumping 
in velocity space (Figs.~11,12), there is no apparent population substructure discernible within the 
inner fields of $R\la$20 kpc, nor in the Stream fields at 21 and 32 kpc. More striking is  the sudden decrease in the mean metallicity  
towards larger projected radii that becomes already visible in this representation. 
It is at these distances, where a transition from the dominant, more metal rich bulge population 
towards a more metal poor halo component may occur (Ostheimer 2003; Kalirai et al. 2006a). 
This view would also conform with the claim of a break in M31's surface brightness profile
between an inner $R^{1/4}$ profile (de Vaucouleurs 1958) and a gradual transition to 
an $R^{-2}$ power-law surface brightness profile that defines an outer halo 
(Irwin et al. 2005; Guhathakurta et al. 2006; Chapman et al. 2006). 
Interestingly, the run of the strength of the TiO band\footnote{The TiO$_{7100}$ band pass is a classical 
discriminator of spectral types (O'Connell 1973) and we measure its strength by a straight integration 
of the line band pass from 7055--7245\AA\ with respective continuum bands. Larger values of 
TiO$_{7100}$ indicate the presence of cool, metal rich giants.}  
at 7100\AA\ (top left panel of Fig.~15)  lends strong support to this scenario. While a number of TiO-strong, thus more metal rich, giants are predominant in the bulge and inner halo regions, these are clearly missing in the outer parts beyond $\sim$50 kpc.

In Fig.~15 (top right panel) we separate our full  data set into three radial bins, comprising 
the inner spheroid ($R<$20 kpc), the transitional region (20--40 kpc) and the outer halo fields from 
40 kpc out to our last data at 160 kpc. There is a clear indication that the outer fields are distinctly more 
metal poor by more than 1 dex compared to the inner spheroid: while the inner 20 kpc's metallicity 
distribution function (MDF) peaks at $\sim -0.6$ dex, the mean metallicity shifts progressively towards 
$\sim -1.2$ dex  for 20 kpc$<R<$40 kpc. This region already contains a considerable component  
below $-2$ dex, which becomes the characteristic metallicity regime in the outermost fields. 	
A K-S test reveals that the MDFs from all three regions are unlikely to originate from the 
same parent distribution, with all probabilities being consistent with zero. 
This suggests that there is in fact a mixture of populations, possibly 
from the inner spheroid and an outer halo component. Nevertheless, we note that the 
dispersion of the MDF (as derived from an iterative  Gaussian likelihood estimator accounting for 
measurement errors; Koch et al. 2007) does not increase considerably due to the 
potential overlap of two populations of separate peak metallicities: at  
(0.47$\pm$0.02) and 
(0.46$\pm$0.03) the dispersions are practically indistinguishable. 
However, it is perilous  to compare these numbers, when 
splitting the sample in only these two regions along a broad spatial range. 

Thus we plot in the middle panel of  Fig.~16 the dispersion in metallicities 
versus radius in smaller radial bins that were, for R$<$40 kpc, chosen such as to maintain 
the same number  of stars (viz. 100). 
This way, we guarantee a proper statistical sampling; furthermore, since 
the data were analyzed in a homogeneous manner, no systematic biasses will be introduced 
when averaging data 
across adjacent masks. 
Only for the  three outermost fields 
beyond 85 kpc do we group the sparse data by field (11, 5, and 10 giants in m6, m8, and m11) 
 to avoid averaging across the large radial 
gap.  These fields are separated by 40 kpc (already the size 
of the entire inner spheroid) and it is further likely that these fields are contaminated by both 
substructure and the halo of M33.

The  bin  at 16 kpc shows a metallicity dispersion that 
is higher by 0.24 dex than the average, while the adjacent bin at $\sim$18 kpc 
again exhibits an average dispersion. It is worth noticing that the dispersion in our data 
around 13--14 kpc shows a remarkably smaller internal (i.e., accounting for measurement errors)  dispersion in [Fe/H]$_{\rm CaT}$ of 
$\sigma = 0.21\pm 0.03$ dex (at a mean metallicity of $-0.84\pm 0.07$ dex). In fact, this region corresponds to the approximate location 
of the kinematic substructure discussed in the previous section and by Gilbert et al. (2007). 
Without applying any further velocity cuts to extract substructure stars by isolating 
their triangular distribution in velocity space, these regions are expected to contain an overlap of 
giants from the intrinsic inner spheroid population and those forming the substructure itself. 
As Gilbert et al. (2007) argue, the resulting, intermixed MDF should be expected to 
be slightly more metal rich than adjacent other fields. The radial bin in question 
is slightly more metal rich on average by $\sim$0.11 dex than the next inner data (at a significance of 
0.9$\sigma$). Hence, our mean abundances in the inner spheroid and the substructure do not lend strong  support to the cold structure being significantly more metal rich than its surroundings. 

However, the mean [Fe/H]$_{\rm CaT}$ significantly drops by about 0.4 dex 
(2.5$\sigma$) between approximately 16 and 20 kpc (Fig.~16), while this trend even proceeds further outwards. 
We note that 
this sharp drop coincides with the edge of disturbed region  visible in the Ferguson et al. (2002) maps. 
As Ibata et al. (2005) show, the majority of the  photometric contrast
in the Ferguson et al. (2002) maps, leaving the impression of an edge, 
is due to an extended, rotating disk component (see also Fig.~1, top panel).  
As this structure has a high [Fe/H], 
the transition that we map in the metallicities is mostly a result of 
moving off the rotating component and into the underlying halo dominance, where the halo falls off 
much more slowly than the exponential disk. 

Beyond 40 kpc, the MDF contains 108 stars (Fig.~15, top right), 
but it is evident that the majority of these 
M31 red giant candidates are more metal poor than the crossover region by $\sim -1$ dex. 
In particular, the 60 kpc fields provide a transition from the metal poor outermost fields 
and the more metal rich inner parts, which is to be expected as they are dominated by the more metal rich tangential streams they lie on (see Fig.~1, bottom panel; Chapman et al. 2008). 
All in all, there is a pronounced metallicity gradient seen throughout our fields within M31's spheroid, 
which becomes even more striking in the plot of individual metallicities as a function of distance from 
M31 (bottom panel of Fig.~15).

\subsection{Comparison with previous detection of a gradient}
A distinct metallicity  gradient in M31's halo has already been proposed 
by Kalirai et al. (2006a),  who 
find a more gradual leveling of their MDFs,  based on {\em photometric}  metallicities, 
from $-0.47$ dex within 20 kpc to $-0.94$ dex 
around 30 kpc down to $-1.26$ dex beyond 60 kpc (open squares in the top and bottom panels of Fig.~16).  
Thence their data imply a smooth decline as ($-0.77\pm$0.09) dex\,(100 kpc)$^{-1}$. 
The analysis of our data, on the other hand, yields a steeper   
radial metallicity gradient of the order of ($-1.50\pm0.08$) dex\,(100 kpc)$^{-1}$. 
Our finding of a clear abundance gradient is underscored by the disappearance of 
TiO strong, metal rich giants in the outermost fields (Fig.~15, top left panel). In fact, the outermost radii appear to be 
dominated by a purely metal poor population. 
Another significant difference between the gradients derived in this work and 
suggested by Kalirai et al. (2006a)  is the overall lower metallicity 
at almost any given radius in our analysis. On average, our  {\em mean} abundances in those fields in common with their data 
are  more metal poor by 0.75 dex. Given the quoted measurement errors from both sources, 
this discrepancy is significant at the $3.4\,\sigma$ level and the  
reason for such a deviation merits careful investigation.  

In the top panel of Fig.~16, we only include the minor axis data, which were radially binned 
such as to guarantee the same, statistically significant number ($\sim$100) of stars per bin. 
Our  data and those of Kalirai et al. (2006a) additionally  includes off-axis fields, 
which we highlight with the encircled points in the top panel  of Fig.~16. 
Two of these are located directly on the Giant  stream (H13s and a3), while also 
a13 and b15 fall towards the edge of the stream feature at a projected 
distance  of $\sim 50$ kpc. 
 It is known from previous spectroscopic measurements (Guhathakurta et al. 2006; Kalirai et al. 2006b) 
that the stream is intrinsically more metal rich than the halo. 
One other field in Kalirai et al. (2006a) lies on  the dwarf spheroidal (dSph) galaxy And\,III 
(d3), which exhibits a population representative of M31's moderately metal poor satellite galaxies 
(with a mean abundance of  $\sim -1.7$ dex; McConnachie et al. 2005). 
As the field may well be contaminated with dSph members, we do not consider it a representative 
choice for a study of the pure M31 halo. 
Due to our inclusion of the inner spheroid fields, we have significantly greater numbers 
within 20 kpc. 
In order to evaluate the influence of the off-axis fields on the metallicity gradient, we 
include in the bottom panel of Fig.~16 all our measurements throughout M31's  spheroid, separated 
by field (see also the color coded map in Fig.~1, bottom panel). 
These fields include H13s, a3, a13, a19 and b15, which were also targeted by Kalirai et al. (2006a). 
Nevertheless, it is obvious that the strong character of the gradient persists disregarding whether 
we focus on the minor axis or on the whole spheroid. Thus the metallicity gradient is likely a 
characteristics associated with M31's full halo. 

One major source of uncertainty is in general the use of photometric metallicities. Larger photometric errors, 
the choice of the adopted set of isochrones, the general failure of stellar 
evolutionary tracks to simultaneously reproduce the major features of CMDs 
(Gallart et al. 2005; and references therein) and undesired age-metallicity degeneracies on the RGB  
render photometry the less reliable metallicity indicator as compared to spectroscopic estimates. 
Isochrone fits produce values for the actual stellar metallicity, $Z$, which incorporates 
the admixture of heavy elements, in particular the $\alpha$-elements. 
In this vein, any unknown $\alpha$-enhancement can considerably 
alter the derived photometric metallicities. Kalirai et al. (2006a) demonstrate that 
an [$\alpha$/Fe] ratio of +0.3, as found in the Milky Way halo, yields results  
more metal poor by 0.22 dex  compared to those derived from scaled-solar isochrones 
(but see also Koch et al. 2006 for a discussion of the systematics of 
$\alpha$-variations on CaT metallicities). In order to reconcile the spectroscopic (this work) and photometric  (Kalirai et al. 2006a) metallicities a strong enhancement 
in these elements of at least 1 dex would be required. This seems an unreasonably high 
value, even if the formation of the outer halo regions were dominated 
by early star formation bursts.

Photometric metallicities are especially susceptible to overestimating the metallicities of metal poor stars.  
For stars with  [Fe/H]$<-1.5$\, dex, there is relatively little difference in the metal line blanketing in the V  band:  
In color-magnitude diagrams such as Fig.~3, 
 a 1 dex decrease in metallicity causes little discernable change in color.  
Given that for many distant systems, the only accessible metallicity estimate is photometric metallicity, we would thus urge the use of bluer filters, and greater caution in the interpretation of results.
It is reassuring, though, that despite the large overall discrepancy between the 
Kalirai et al. (2006a) and our measurements, there is a significant agreement  
between the two studies in  field m8 (at 120 kpc).  In this  field, we
 find 7 giant candidates, for 5 of which we could measure an [Fe/H]$_{\rm CaT}$. 
If we discarded those two stars with the largest velocities that are in the transition 
higher-velocity tail of our distribution ($\sim$1.2--1.5\,$\sigma$ above the systemic mean), 
the mean abundance would even drop by another 0.1 dex, confirming the 
dominance of metal poor of stars in these outer fields. 
We note in passing that m8 constitutes one of the fields for which Ibata et al. (2007) 
exclude the occurrence of any substructure, so that it may in fact be 
a contender for representing M31's genuine underlying metal poor halo (see also Chapman et al. 
2006, 2008).

Finally, we add to  Fig.~16  the mean spectroscopic [Fe/H]
 in a minor axis field at 19 kpc, based on the 29 confirmed red giant members 
of Reitzel \& Guhathakurta (2002; their Fig.~17). These authors not only find that their spectroscopic 
metallicities are systematically lower than the photometric counterparts, but they also 
detect a distinct metal poor tail in this MDF, reaching as low as   [Fe/H]$_{\rm CaT}=-2.85$ 
on the scale of Carretta \& Gratton (1997). By comparing their MDF to those of 
Galactic and M31 globular clusters and of Local Group dSph galaxies,  
they attributed its shape and the presence of such metal poor stars to the build-up of M31's 
halo from the accretion of many a small subsystems. As this comparison with our data 
shows, the mean metallicity in the Reitzel \& Guhathakurta (2002) field is 
in good agreement with our measurement in the respective radial bin. 
The most recently discovered faint dSph satellites around M31 are predominantly 
characterized by mean metallicities between $-1.3$ and $-1.7$ dex (Martin et al. 2006b; Majewski 
et al. 2007), which agrees well with  the field star population's abundance over a radial range from 
$\sim$20--50 kpc.

\subsection{Andromeda's metal poor outer halo}
In contrast, the significant discrepancy of the mean metallicity estimates in the outermost 
field, m11, is an issue of concern: While Kalirai et al. (2006a) state a value of $\sim -0.92$ dex 
from their three confirmed red giant members, we find a value as low as $\sim -2.6$ dex 
from 10 red giant candidates stars in total. It is then fair to ask {\em how reliable are our low
 metallicities} in this highly foreground contaminated field and how trustworthy is our  
 detection of metal poor stars at all radii? 
First, if those marginal stars with potentially too high velocities were removed from m11, 
the mean value in this field would essentially remain unaltered -- {\em the metal poor nature and 
strong gradient do persist}. All in all, there are 56 stars with [Fe/H]$_{\rm CaT} < -2.3$ found 
at all radii from 9 to 160 kpc. Based on their higher radial velocities, no more than 6--8 of these 
could be potential remaining foreground contaminants. 
Furthermore, it was verified by visual inspection that none of the stars is a potential mismatch 
in spectral type or exhibits spurious noise peaks. 
Also, we note that the median log-likelihood $L$ (Sect.~4) amounts to 0.7 for these stars so that it is 
5 times more likely that they are M31 red giants than Milky Way stars based on 
the adopted discriminant indicators. 
As we have shown in Sect.~4.1, there may be $\sim$35 undetectable blue dwarfs in our whole 
sample (Table~2), based on the Besan\c con predictions. Furthermore, the model predicts that 
only 15\% of these stars have nominal metallicities below $-2.3$ dex,  which translates into 
$\sim$5 such undetectable contaminants in our sample. 

The principal caveat against deriving a metal poor tail in stellar populations is that, 
generally, no calibrations of the CaT strength exist below $-2.1$ dex. 
Neither the original sample of Rutledge et al. (1997a, 1997b) 
 included any system more metal poor than $-2.02$ dex (on the scale of Carretta \& Gratton 
1997), nor did our calibration clusters (Fig.~7). 
Nevertheless, the CaT technique is widely applied throughout the literature of metal poor 
stellar populations and can 
still maintain its prime role in at least a relative ranking of stars towards the metal poor 
extrapolations (Koch et al. 2006; Simon \& Geha 2007; Battaglia et al. 2008)\footnote{The same holds for   
the calibration of metal rich stars:  the most metal rich globular cluster in our sample, M5, has [Fe/H]$_{\rm CG97}=-1.12$ 
dex), while a small fraction of our  measured M31 stars nominally reach above solar values. 
That the CaT calibration is still valid   
up to +0.47 dex has recently been demonstrated by Carrera et al.  (2007).}. 
We plot in Fig.~17 a series of spectra that were grouped and coadded in 
various metallicity bins.  It becomes obvious that the expected trend of increasing 
CaT line strength with increasing [Fe/H] is also visible in our data so that our 
metallicity scale and ranking derived from the CaT coaddition did not introduce any grossly falsified 
results, in particular towards the metal poor spectra. This coaddition of the spectra also 
emphasizes a number of \ion{Fe}{1} and \ion{Ti}{1} lines that clearly scale with metallicity  
in our spectra and will allow us to estimate  chemical abundance ratios in future works 
(see also Kirby et al. 2008). 

As a further comparison, we generated synthetic spectra of red giants, using Kurucz model atmospheres\footnote{\url{http://kurucz.harvard.edu} }
with representative stellar parameters, i.e., 
(T$_{\rm eff}$=4000 K, log\,$g$=1.0, $\xi$=1.5 km\,s$^{-1}$), and solar-scaled opacity distributions\footnote{\url{http://wwwuser.oat.ts.astro.it/castelli}} (e.g., Koch et al. 2008). In Fig.~18, we show the 
resulting syntheses for different metallicities of the atmospheres. These spectra have been degraded to 
match the spectral resolution of DEIMOS and, in the bottom panel, convolved with an additional noise 
component the mimic a representative spectrum with $S/N$=10. 
Apart from the expected weakening of the CaT, there is a visible decrease in the strength of the 
weak \ion{Fe}{1} absorption features (e.g., at $\lambda\lambda 8514.1,8468.4,8621.6,8674.7$\AA) adjacent to the CaT (see also Bosler et al. 2007) towards lower metallicities. 
This decrease is even discernible in the low $S/N$ spectra and in fact observable in our 
coadded spectra (Fig.~17). 
 
One might then argue that the metal poor stars could be an artifact of the new coaddition technique and its calibration devised in Sect.~5. To test this, we grouped the spectra of progressively 
metal poor stars, and applied the same method as before on the coadded spectra. 
It transpires that the metal poor character  of these spectra persist and there is 
a tight, close-to-unity relation between the metallicity bin of the individual stars to be grouped, and 
the final coadded estimate from the coadded spectra.  
Moreover, the use of our own consistent 
calibrations (eqs.~5,6) does not alter the qualitative detection of metal poor stars: 
if we were instead to use  the canonical calibration of Rutledge et al. (1997a, 1997b), 
we would observe a shift of the measurements towards [Fe/H] lower by 0.2 dex at $-$3 -- 
if anything, our calibration would overestimate the metallicities so that 
the metal poor character of the distribution remains.  
Further tests that we employed to ascertain the reality of the gradient and the metal poor 
stars verified that there is no unusual trend of the [Fe/H] with magnitude discernible, nor 
does the gradient change when we restrict the analyses only to the high- or low-$S/N$ spectra, respectively. 
Moreover, to test the influence of potential remaining dwarf contamination, we 
constructed another test sample by only including stars with radial velocities below $-$300 km\,s$^{-1}$. 
Apart from increased statistical uncertainties due to the decreased sample size, the 
outer regions do remain metal deficient and the gradient perseveres. On average, the ``pure'', 
velocity restricted  giant sample yields mean  metallicities in each field 
that are  more metal poor by 0.04 dex (r.m.s. scatter of 0.50 dex) on average than the full giant data. 
Essentially, the same holds if we inflict a strict color cut to select a ``pure'' giant sample: 
a gradient persists even for a, say, $V-i'>1$ subset. Neither are any significant changes found if we 
restrict the analysis only to giants classified as such in $>95$\% of the Monte Carlo runs of the 
dwarf/giant separation  (Sect.~4). The median difference between the mean metallicities from 
``all'' giants and the $2\sigma$ cases is $<0.01$ dex (r.m.s. 0.08 dex). 
Moreover, our dwarf sample does not exhibit any significant sign of a gradient 
(at 0.06$\pm$0.04 dex\,(100 kpc)$^{-1}$), 
which is expected, since the CaT is no metallicity indicator for these stars so that 
their [Fe/H] are randomly distributed. 

The color coded CMD in Fig.~3 then verifies that the trend of metallicity with location on the RGB 
is in fact as expected, with the more metal rich stars (red points) 
exhibiting progressively redder colors, and the most metal poor stars (blue points) being 
predominantly located towards blue colors. 
We also note the presence of $\sim$40 giants that fall bluewards of the most metal poor isochrone 
in Fig.~3.  Given our discussion in Sect.~4.1, it appears unlikely that these are Galactic contaminants. 
Although their spectroscopic metallicities indicate them to be metal poor objects, 
their remarkably blue colors  are surprising. If one were to assign these colors to erroneous photometry, 
and thus redden these objects towards the [Fe/H]=$-2.3$ isochrone, this color difference translates 
(Eqs. 5,6) into a spectroscopic metallicity uncertainty of less than 0.30 dex 
with a median of 0.03 dex  (r.m.s. scatter 0.09 dex). 
All in all,  we are left to believe that  there is in fact a detectable, real population of 
considerably metal poor red giants present in M31's inner halo that becomes yet more 
prominent in its outer halo.

This fact then confirms 
the view of an accretion origin of the inner halo, presumably by several events, 
and argues strongly in favor of the same mechanism governing the formation 
of the inner and the outer halo. 
Moreover, there is additional evidence (Chapman et al. 2008, in prep.) that halo fields at 
110 kpc are not affected by any of the substructures in the Ibata et al. (2007) maps and are consequently 
a plausible true halo component; they are found to have [Fe/H]$\sim-2$.

Given that the halo of M33 itself is also metal poor (at around $-$1.5 dex; McConnachie et al. 2006) 
it is again feasible that a major fraction of the red giants in the outermost fields,  are members of 
M33, the more so, since their velocities appear to be similar to the systemic velocity of M33. At present, there 
is no compelling evidence against the hypothesis that  {\it all} the
candidate giants in these distant fields $\ga 120$ kpc are genuine M33 members. 
Fig.~19 (left panel) shows the metallicities and velocities of stars in the three most 
distant fields,  where we schematically overplot each galaxy's velocity distribution, using 
the stellar halo parameters from McConnachie et al. (2006) for M33 and those derived in this work for 
M31. Under the simplifying assumption that these fields contain an equal mix of M33 and M31 
stars, we can estimate, based on our targets' velocities,   that 85\% of the red giant members 
in field m11 and m8, and 37\% of those in m6, might actually belong to M33's halo.  
Statistical removal of this contribution  would yield mean metallicities of 
($-1.94\pm 0.52$, $-2.00\pm 0.60$, $-1.60\pm 0.38$) dex in m6, m8, and m11. However, there 
is no way of reliably separating the mutual overflow of giants into each other galaxy's halo 
at present. 
 
The more metal rich stars at [Fe/H]$\ga-1.2$ dex 
and velocities around $-325$ km\,s$^{-1}$, on the other hand, 
might plausibly be considered to have originated
in a radial collision,  presumably like the one that bore 
the ancient Giant Stream (see Mori \& Rich 2008).  
This hypothesis is also consistent with them being close to the systemic velocity of M31. 
In particular, there are no stars found with very high negative velocities relative to M31 --  
all giants in these fields have velocities well within $\sim1.5\sigma$ of the M31 mean in 
the negative velocity tail (which does account for M31's radially decreasing velocity dispersion).   

\section{Conclusions}
From a spectroscopic analysis of 1316 confirmed red giant stars along the 
minor axis on M31 and in spheroid fields out to 160 kpc 
we  find and confirm the following structural specific features in M31's inner 
and outer halo:  
\begin{enumerate}
\item 
There is evidence of an abundance substructure in the sense that  the mean stellar metallicity 
strongly declines at $\sim$20 kpc, where the abundance range within this radius is typically 
$-0.5$ to $-1$ dex and falls towards $-1.4$ dex at 20--40 kpc. The latter values are consistent 
with those detected within a smooth underlying M31 halo (Ostheimer 2003; Guhathakurta et al. 2006; 
Kalirai et al. 2006a; Chapman et al. 2006).  
Interestingly, the location of the break in the metallicity profile coincides with the edge of the metal rich,  
extended rotating disk reported by  Ibata et al. (2005), but it cannot be entirely ruled out that 
there is also a contribution from the kinematic substructure (Gilbert et al. 2007), which would bias  the 
inner regions towards marginally higher metallicities.   
A metallicity gradient has been detected by 
Kalirai et al. (2006a), but here we show that the decline in our spectroscopic 
measurements appears to proceed even stronger towards the outermost fields at 160 kpc, 
where we find mean values around and even below $-2$ dex.  
\item In particular, there is a considerable fraction of metal poor stars below [Fe/H]$_{\rm CaT}\la -2$ 
found at almost all radii.
Their presence is not utterly surprising, and red giants that metal poor have been claimed to exist  
in M31's halo before (e.g., Reitzel \& Guhathakurta 2002). Furthermore, their established prominence  
in the Milky Way halo (Carney et al. 1996; 
Chiba \& Beers 2000; Carollo et al. 2007) raises the question, why there 
should not be a comparable distribution present in the M31 halo, if both systems had experienced 
a similar formation and accretion history.  
\item 
A considerable fraction of stars in the outermost fields beyond $\sim$100 kpc 
(which corresponds to about 100 kpc projected distance to M33) exhibit velocities and metallicities 
consistent with those of M33 halo stars. This confirms earlier findings from star count maps (Ibata et al.  
2007), according to which the stellar halos of these major Local Group spirals overlap to a large extent. 
\item 
We confirm the earlier detection of a kinematically cold  substructure (Gilbert et al. 2007), 
located at a radial distance of  15--20 kpc. By comparison of new N-body simulations (Mori \& Rich 
2008), we show that 
such a substructure is consistent with having originated in the merger event that 
produced M31's Giant Stellar Stream (Ibata et al. 2001, 2004). Nevertheless, the full radial 
velocity distribution along the minor axis is difficult to reconcile with a single collisional  
event of this kind, and it is more likely that a wealth of accretions occurred and formed 
the halo, which is concordant with the progressive detections of stream-like substructures 
in star count maps (Ibata et al. 2007). 
\item 
There is a considerable contribution of stars from a  genuine, ancient M31 halo, 
potentially ejected during 
such merger events,  to the velocity distribution at any radius. 
Moreover, neither stars from the pristine bulge or disk components are likely to be found 
in the outer or even inner spheroids. 
\end{enumerate}
This leaves us with a picture in which the progenitor that produced the giant Stream 
and thus presumably donated a major part of M31's halo  cannot be single-handedly 
responsible for all the substructures seen in our and previous studies -- outside the 
present-day Stream's sphere of influence,  one sees a predominant occurrence of 
minor substructures or streams from many a past mergers. 
A large number of  accretion events is also thought to have contributed to  the formation of M31's disk (e.g., Pe\~narrubia et al. 2006), as this would produce the proper, observed mix of metallicities. 
Ibata et al. (2007) note that the radial metallicity gradient is a mere reflection of Stream 
debris and numerous  other substructures 
in the {\em inner} regions, which is in concordance with model predictions.  In this context the 
metal rich, genuine Stream material is more centrally concentrated, leaving the impression 
of a more extended underlying metal poor halo (cf. the stream distribution in Fig.~13).  
It is unclear at present, whether the gradient is intrinsic to the inner substructures becoming 
more metal poor with radius, or whether the underlying halo, with its larger dominance 
in the outer regions, is becoming more metal poor. 
Note that Brown et al. (2007) find that an HST field at 25 kpc ($\sim 1.5\degr$) is 
both more metal poor and older than the innermost HST field projected at 11 kpc.
Hence, also the HST pointings appear not to be dominated by the canonical Stream's 
debris.
Moreover,  we find that the innermost deep HST field (Brown et al. 2003) is situated in a region that has some potential contribution from tidal debris, but appears to be dominated by a more metal rich "inner halo" population with a {\it smooth}  velocity distribution, free of the  structures 
at the velocity extremes that are predicted to be present from interaction models (see left middle panel 
of Fig.~13). 
  The Brown et al. (2003) 11 kpc field appears to be free of major contamination from an infall event.  The 21 kpc field lies  just outside of the metal rich inner region; Brown et al. (2007) suggest a lower metallicity for this field, which we confirm from our CaT analysis.  The field at 35 kpc apears to be yet more metal poor and may be genuinely representative of the outer halo.  
We note that all the HST fields at 11, 21 and 35 kpc have very different circular orbital periods 
(derived from a simple mass model and the assumption of circular orbits) of $\sim$0.2, 0.5 and 1\,Gyr, respectively.  
The characteristics of the HST deep fields will be discussed in a forthcoming paper (Rich et al.  in prep).

Using the same arguments, we find large 
orbital periods of $\sim$6\,Gyr at 160 kpc, which presupposes them to be bound to M31 (see also 
Majewski et al. 2007). 
It is hard to reconcile these time scales with a scenario, in which these distant  
M31 members are  pressure supported. 
It is rather an attractive notion that these distant stars might be ejecta 
of collisions, not yet having completed a single orbit around M31.  

The fact that the new metallicity scale we framed in this work is in good agreement 
with the moderately metal poor  character of the dSph satellites of M31 (Reitzel \& Guhathakurta 2002; McConnachie 
et al. 2005; Martin et al. 2006b; Majewski et al. 2007) further support 
the idea that the outer halo might plausibly emanate from a population similar to the dSphs. 
Moreover, the presence of the more metal rich stars beyond 100 kpc  
suggest an origin of these stars in collisions rather than in a primordial halo. Yet it is challenging to reconcile this scenario with 
the number count maps of Ibata et al. (2007), 
which are smooth to within the sensitivity of the MegaCam survey   
 in these regions and with the presence of very metal poor stars:  
there are no very metal poor stars detected in the Galactic dSphs (Koch et al. 2006; Helmi et al. 2006), 
nor in the faintest M31 satellites (Martin et al. 2006b, 2007),  
so that it remains unclear whether systems like the  present-day observed dwarf satellites are responsible for the bulk of the halo.  

While our observed radial metallicity gradient stretches 1.5--2 dex over the full extent of our data 
of 160 kpc, simulations of the hierarchical assembly of Galactic halos do not reproduce 
any considerable gradients. In this context, the study of Font et al. (2006) predicts gradients in 
[Fe/H] {\em averaged} over each full simulated halo of at most 0.5 dex over a few tens of kpc. 

Our findings lend striking support to paralleling the M31 and Galactic halos: 
In a recent work  Carollo et al. (2007) detected a dichotomy 
in the Milky Way's stellar halo.  In this sense, there exists a clear kinematic and 
chemical separation into an inner halo (with stars on more eccentric orbits, 
metallicities around $-1.6$ dex, a flattened density distribution and no considerable rotation) 
and an outer, independent halo component, which is more spherical in shape, shows 
evidence of rotation {\em and is more metal poor} on average. 
In particular, the outer Galactic halo exhibits a peak metallicity of $-2.2$ dex.  
The presence of two distinct halos 
is well explicable in the context of  cosmological structure formation models 
and incorporates the continuous, chaotic accretion of distinct subhalos that follow 
an earlier stage of the dissipative merging of massive, yet sub-galactic fragments 
(see Carollo et al. 2007 for a  discussion of the detailed formation scenario).  
If structure formation proceeded in analogy to form M31 -- 
and in the light of $\Lambda$CDM this is likely -- 
then the clear distinctness of at least two halo components visible in 
the MDFs in our M31 study is a natural outcome of the accompanying stochastic abundance 
accretion.

As a concluding, independent remark, we note that a further benefit of the present study is the 
derivation of a new improved method to measure metallicities from the calcium triplet, which 
allowed us, and should encourage future works,  to extract reasonably accurate information
from low $S/N$ spectra.  

\acknowledgments
Support was provided by NSF (AST-0307931, AST-074979),  
HST (GO-10265, 10816) and by R.M. Rich. 
The authors thank Stephen Gwyn for assistance with obtaining the
photometry from the CFHT archive.
We gratefully acknowledge P. Guhathakurta and J. Kalirai for assisting in the Keck observations.
H. Ferguson is thanked for helpful comments. 
We are also grateful to Marc Davis and the DEEP team, and to Phil Choi and George Helou, for obtaining some of the observations used in this study. 
We are grateful to Sandy Faber and the DEIMOS team for building an outstanding instrument
and for extensive help and guidance throughout the observing runs.  The spec2d data reduction 
pipeline for DEIMOS was developed at UC Berkeley with support from NSF grant AST-0071048. 
The staff of the W.M. Keck Observatory, in particular Greg Wirth, is thanked 
for assistance with observations and data recovery. Data presented herein were 
obtained using the W. M. Keck Observatory, which is operated 
as a scientific partnership among Caltech, the University of 
California, and NASA. The Observatory was made possible 
by the generous financial support of the W. M. Keck 
Foundation. 
This research used the facilities of the
Canadian Astronomy Data Centre operated by
the National Research Council of Canada
with the support of the Canadian Space Agency.

\begin{appendix}

\section{Background galaxies}

In the full sample of DEIMOS targets taken for the entire project, we identified a still large number of 
$\sim$400 galaxies background galaxies, which corresponds to a striking 
contamination fraction of $\sim$11\%. 
It is possible that more galaxies were missed during the classification of the spectra and 
rather flagged as bad quality data, although we verified that no ``true'' dwarf or giant star 
was mistaken as a galaxy spectrum and vice versa. 
These galaxies cover the full color  range from $\sim$0.5 to 3.5 in  $V-i'$, with  the 
majority falling in the interval  between 1 and 2 (see the CMD in Fig.~3). 
Despite the thorough pre-selection of potential M31 giant candidates on the upper RGB 
and, for some of the inner fields, using Washington photometry (e.g., Gilbert et al. 2006), the partially poor seeing conditions  during the photometric runs (obtained at the KPNO; see Ostheimer 2003; 
Gilbert et al. 2006) hampered an appropriate {\em a priori} 
rejection of non-stellar extragalactic point sources. 

For those galaxies identified here, redshifts were measured from the Doppler shifts of generally 
2--3 major emission features, such as the Balmer 
H$\alpha$, H$\beta$, H$\gamma$ lines, the [\ion{O}{2}] 3727\AA\ doublet and/or the 
 [\ion{O}{3}] 5007\AA\ line.  For rare occasions ($\sim$5 of the galaxies) we could also detect the strong 
 Ca\,H,K features in absorption. 
The final redshift distribution is displayed in Fig.~20. The first thing to note is that our sample 
includes a fairly broad range in redshifts, reaching from $z\sim$0.04 to $z\sim$1.35.

This distribution should, however, not be taken as representative of the true galaxy distribution
in the line of sight towards M31. This is due to our identification criteria based on only a small 
number of spectral features, which are then redshifted into the limited spectral range of DEIMOS 
(see also Kirby et al. 2007), biassing the distribution towards only selected redshift intervals.  
Hence,  a detailed derivation of their physical properties and 
star formation rates has to be beyond the scope of this work and is left for a future 
paper (Koch et al., in prep.). 

\end{appendix}

\clearpage
\begin{deluxetable}{rcccccr}
\tabletypesize{\scriptsize}
\tablecaption{DEIMOS mask observation details}
\tablewidth{0pt}
\tablehead{
\colhead{} &
\colhead{Date of} &
\colhead{} & 
\colhead{Projected} & 
\multicolumn{2}{c}{Mask center} &
\colhead{P.A.} \\
\raisebox{1.5ex}[-1.5ex]{Mask} &
\colhead{observation} &
\raisebox{1.5ex}[-1.5ex]{P.I.} & 
\colhead{radius\tablenotemark{a} [kpc]} & 
\multicolumn{2}{c}{(J2000.0)} &
\colhead{[degree]}  
}

\startdata
\multicolumn{7}{c}{}\\
\multicolumn{7}{c}{\raisebox{1.5ex}[-1.5ex]{Minor axis fields}}\\
\hline
f109\_1     & 2005 Aug 29  & Rich & 	      9 (194) & 00 45 47.06 & +40 56 52.3  &	23.9 \\
H11\_1      & 2004 Sep 20  & Rich & 	     12 (190) & 00 46 21.94 & +40 41 42.7  &	21.0 \\
H11\_2      & 2004 Sep 20  & Rich & 	     12 (190) & 00 46 20.57 & +40 41 41.9  & $-$21.0 \\
f116\_1     & 2005 Aug 28  & Rich & 	     13 (189) & 00 46 54.63 & +40 41 33.3  &	22.6 \\
f115\_1     & 2005 Aug 28  & Rich & 	     14 (188) & 00 47 32.51 & +40 42 05.3  & $-$20.0 \\
f123\_1     & 2005 Aug 28  & Rich & 	     17 (185) & 00 48 05.79 & +40 29 35.5  & $-$20.0 \\
f135\_1     & 2005 Aug 29  & Rich & 	     17 (186) & 00 46 25.22 & +40 11 45.1  & $-$27.0 \\
f130\_3     & 2006 Nov 22  & Guhathakurta & 	     19 (182) & 00 48 35.01 & +40 16 01.2  &	90.0  \\
f130\_1     & 2005 Aug 28  & Rich & 	     22 (180) & 00 49 11.59 & +40 11 51.9  & $-$20.0  \\
f130\_2     & 2006 Nov 21  & Guhathakurta & 	     22 (180) & 00 49 38.44 & +40 16 04.8  &	90.0   \\
a0\_3       & 2004 Jun 17  & Helou & 	     29 (175) & 00 51 50.65 & +40 07 05.9  &	 0.0  \\
a0\_1       & 2002 Aug 16  & Rich & 	     31 (172) & 00 51 51.04 & +39 50 24.9  & $-$17.9  \\
a0\_2       & 2002 Oct 12  & Rich & 	     31 (172) & 00 51 30.49 & +39 44 02.3  &	90.0  \\
mask4       & 2006 Nov 22  & Guhathakurta &  37 (166) & 00 54 08.84 & +39 41 47.0  &   172.0 \\
123Glo & 2005 Oct 01 & Chapman & 60 (143) & 00 58 20.30 & +38 01 46.0 & 147.0 \\
124Glo & 2005 Oct 02 & Chapman & 60 (143) & 00 58 07.47 & +38 05 00.9 & 147.0  \\
m6\_1       & 2003 Oct 01  & Rich &  	     87 (119) & 01 09 51.88 & +37 47 04.4  &	 0.0  \\
m6\_2       & 2005 Jun 09  & Helou &  	     87 (118) & 01 08 36.22 & +37 29 04.6  &	 0.0 \\
m8\_1       & 2005 Jul 07  & Davis &  	    118 (90) & 01 18 12.17 & +36 16 13.1  &	 0.0 \\
m8\_2       & 2005 Jul 07  & Davis &  	    119 (89) & 01 18 35.63 & +36 14 31.4  &	 0.0 \\
m11\_2      & 2003 Sep 30  & Rich &  	    159 (54) & 01 29 33.57 & +34 27 56.5  &	 0.0 \\
m11\_1      & 2003 Oct 01  & Rich &  	    162 (51) & 01 29 34.48 & +34 13 49.8  &	 0.0 \\
m11\_3      & 2005 Jul 08  & Davis &  	    162 (51) & 01 30 02.38 & +34 13 31.7  &	 0.0 \\
m11\_4      & 2005 Jul 08  & Davis &  	    164 (50) & 01 30 37.34 & +34 13 17.1  &	 0.0 \\
\hline
\multicolumn{7}{c}{}\\
\multicolumn{7}{c}{\raisebox{1.5ex}[-1.5ex]{Off axis spheroid fields}}\\
\hline
H13s\_1     &  2004 Sept 20 & Rich &  21 (186) &   00 44 15.14   &  +39 44 23.5  &    21.0  \\
H13s\_2     &  2004 Sept 20 & Rich &  21 (186) &   00 44 14.72   &  +39 44 23.4  & $-$21.0  \\
a3\_2       &  2002 Oct 11  & Rich &  32 (173) &   00 47 47.30   &  +39 06 03.4  &   178.2 \\
a3\_3       &  2002 Oct 26  & Rich &  32 (173) &   00 48 22.81   &  +39 12 38.4  &   270.0 \\
a3\_1       &  2002 Aug 16  & Rich &  34 (171) &   00 48 21.69   &  +39 02 38.0  &    64.2 \\
a13\_3      &  2005 Nov 05  & Guhathakurta &  57 (171) &   00 42 25.86   &  +37 08 28.8  &     0.0   \\
a13\_4      &  2005 Nov 05  & Guhathakurta &  57 (173) &   00 41 32.55   &  +37 08 44.6  &     0.0  \\
a13\_1      &  2003 Sept 30 & Rich &  59 (168) &   00 42 58.20   &  +36 59 04.7  &     0.0  \\
a13\_2      &  2003 Sept 30 & Rich &  61 (171) &   00 41 30.07   &  +36 50 15.7  &     0.0  \\
a19\_1       & 2005 Aug 29 & Rich &                81 (173) & 00 38 15.77 & +35 28 07.7 & 90 \\
b15\_3      &  2005 Sept 07 & Guhathakurta &  94 (129) &   00 53 36.91   &  +34 50 13.6  & $-$90.0  \\
b15\_1	    &  2005 Sept 07 & Guhathakurta &  96 (129) &   00 53 23.41   &  +34 37 17.1  & $-$90.0 
\enddata
\tablenotetext{a}{Numbers in parentheses are projected distances from M33.}
\end{deluxetable}
\begin{deluxetable}{rcccccc}
\tabletypesize{\scriptsize}
\tablecaption{Number of reliable measurements per mask}
\tablewidth{0pt}
\tablehead{
\colhead{} &
\colhead{\# Targets} &
\colhead{\# Reliable} &
\colhead{\# M31 giant} &
\multicolumn{2}{c}{\# M31 giants with Metallicities} &
\colhead{\# Expected} \\
\raisebox{1.5ex}[-1.5ex]{Mask} &
\colhead{on  mask} &
\colhead{velocities\tablenotemark{a}} &
\colhead{candidates} &
\colhead{($V-i')\le2$} &
\colhead{($V-i')>2$} &
\colhead{blue dwarfs\tablenotemark{b}} 
}
\startdata
f109\_1     &  204 & 186 & 158   &  105 & 4  &  1 \\	  
H11\_1      &  139 & 102 &  85   &   58 & 5  &  1 \\ 
H11\_2      &  140 & 115 &  93   &   53 & 5  &  1 \\ 
f116\_1     &  139 & 116 &  98   &   59 & 5  &  1 \\ 
f115\_1     &  187 & 162 & 121   &   87 & 4  &  1 \\ 
f123\_1     &  138 & 112 &  83   &   62 & 4  &  1 \\ 
f135\_1     &  146 & 123 &  92   &   73 & 2  &  1 \\ 
f130\_3     &   70 &  41 &  29   &   38 & 1  &  0 \\ 
f130\_1     &  112 &  88 &  44   &   24 & 2  &  1 \\ 
f130\_2     &  109 &  75 &  34   &   11 & 1  &  1 \\ 
a0\_3       &   93 &  65 &  36   &   24 & 1  &  1 \\ 
a0\_1       &   90 &  44 &  28   &   17 & 3  &  1 \\ 
a0\_2       &   93 &  47 &  31   &   15 & 0  &  1 \\ 
mask4       &  101 &  60 &  18   &   15 & 0  &  2 \\ 
123Glo      &  103 &  91 &  13   &   5	& 4   &  3 \\ 
124Glo      &  105 & 102 &  16   &   8	& 6   &  3 \\ 
m6\_1       &   79 &  39 &  12   &    5 & 1  &  1 \\ 
m6\_2       &   75 &  27 &   7   &    5 & 0  &  1 \\ 
m8\_1       &   62 &  20 &   2   &    2 & 0  &  1 \\ 
m8\_2       &   65 &  25 &   5   &    3 & 0  &  1 \\ 
m11\_2      &   72 &  22 &   2   &    1 & 0  &  1 \\ 
m11\_1      &   74 &  35 &   8   &    5 & 0  &  1 \\ 
m11\_3      &   85 &  27 &   1   &    1 & 0  &  1 \\ 
m11\_4      &   82 &  31 &   5   &    3 & 0  &  1 \\ 
\hline
H13s\_1     & 134 & 106& 92   &   52	&    5 &  0 \\
H13s\_2     & 100 & 75 & 52   &   30	&    4 &  1 \\
a3\_2       &  80 & 30 & 18   &    9	&    1 &  1 \\
a3\_3       &  87 & 47 & 32   &   13	&    4 &  1 \\
a3\_1       &  86 & 30 & 24   &    7	&    1 &  0 \\
a13\_3      & 113 & 37 & 16   &   11	&    1 &  1 \\
a13\_4      &  90 & 34 & 12   &    7	&    0 &  1 \\
a13\_1      &  84 & 31 & 16   &   11	&    0 &  0 \\
a13\_2      &  74 & 27 & 14   &   11	&    0 &  0 \\
a19\_1      &  76 & 37 &  6   &    3	&    0 &  1 \\
b15\_3      &  76 & 32 & 10   &    6	&    0 &  1 \\
b15\_1	    &  68 & 21 &  4   &    4	&    0 &  0 
\enddata
\tablecomments{The listed number of velocity and metallicity measurements excludes ``serendipitous'' extractions 
and  background galaxies.}
\tablenotetext{a}{Prior to the membership separation.}
\tablenotetext{b}{Based on the comparison with the Besan\c con model. See text for details.}
\end{deluxetable}

\clearpage
\thispagestyle{empty}
\setlength{\voffset}{-18mm}
\begin{figure}
\begin{center}
\vspace{-10mm}
\includegraphics[angle=0,width=0.5\hsize]{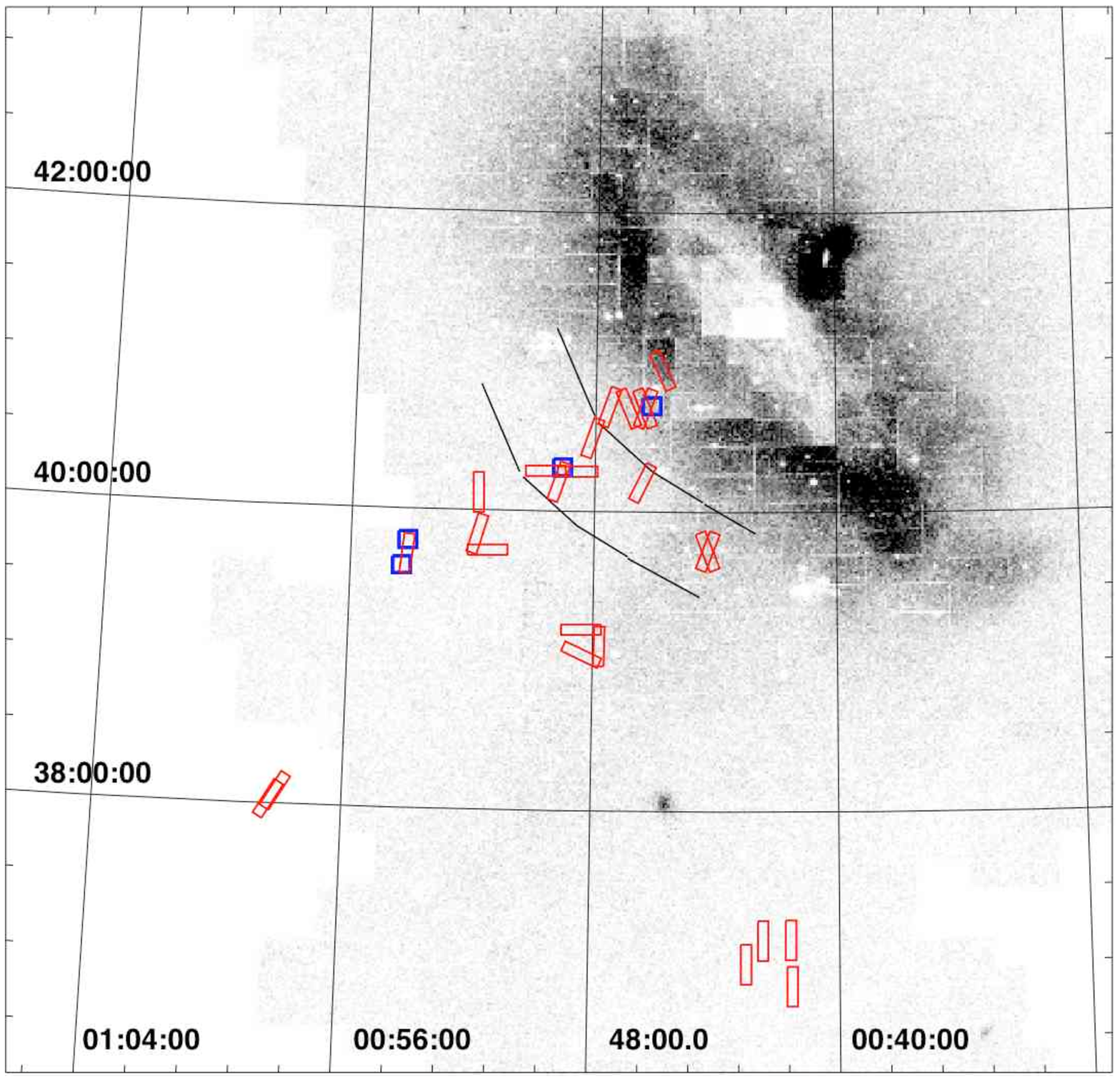}\vspace{-10mm}
\includegraphics[angle=0,width=0.75\hsize]{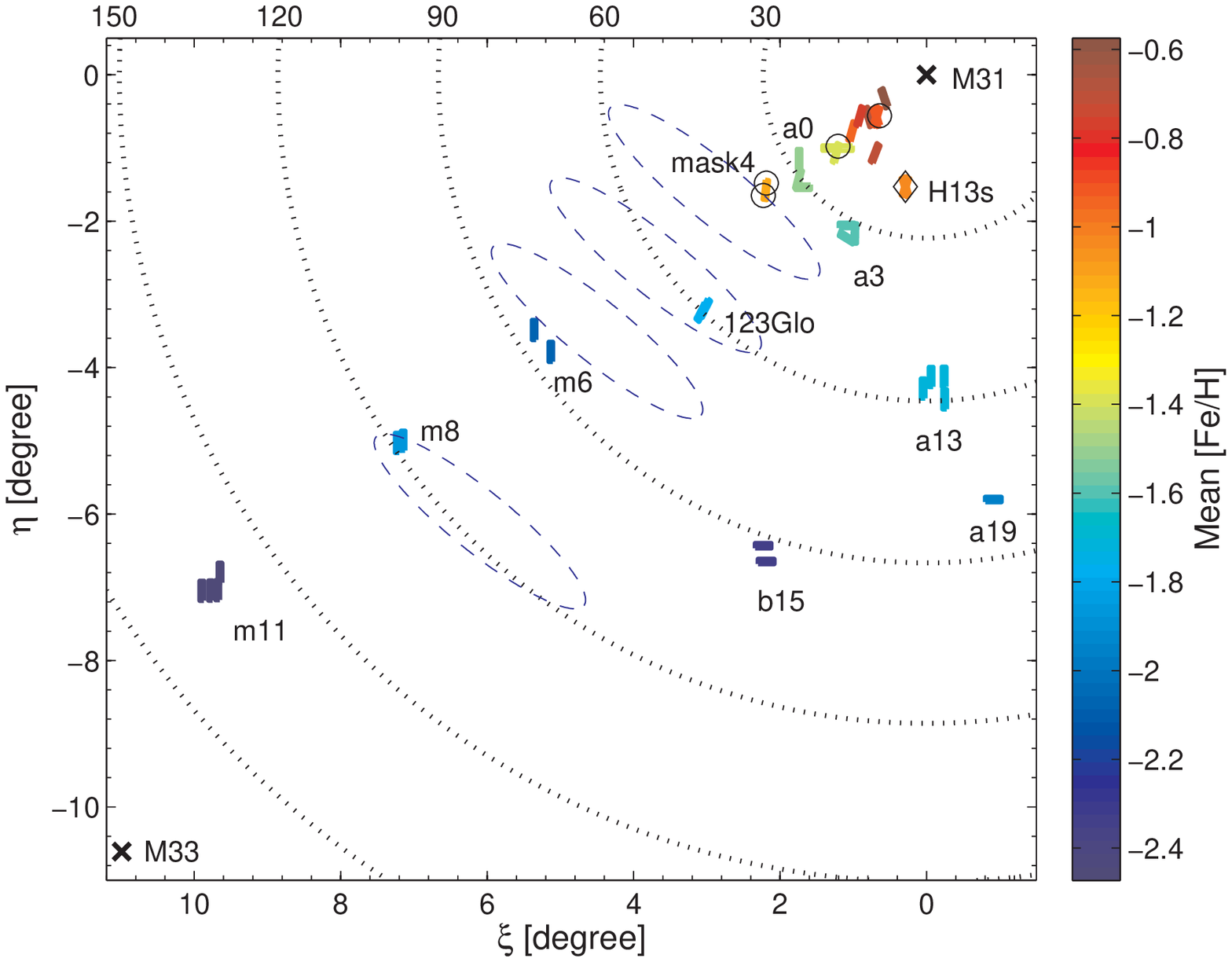}
\end{center}
\vspace*{-5mm}
\caption{Top panel: Location of the targeted inner  fields (red rectangles) on a star count map of M31 
(courtesy of M.~Irwin).  Shown as blue squares are the HST fields of Brown et al. (2003, 2006, 2007).  
The solid black lines delineate the region in which we detect a break in the abundance 
profile (see Figs.~14--16) -- this transition occurs remarkably close to the edge of the perturbed  
disk component in the star count maps. 
Bottom panel: Representation of the full data set in standard coordinates. Numbers at the top 
denote projected distances from M31 in kpc, also indicated by dashed circles. The individual 
masks are color coded by their metallicity (see Sect.~7).  
The black diamond represents the stream field H13s (Kalirai et al. 2006b), black circles are the HST fields, and M31 and M33 are 
indicated as crosses. Dashed ellipses illustrate the approximate location of the tangential streams 
found in the Ibata et al. (2007) maps.}
\end{figure}
\clearpage
\setlength{\voffset}{0mm}
\begin{figure}
\begin{center}
\includegraphics[angle=0,width=1\hsize]{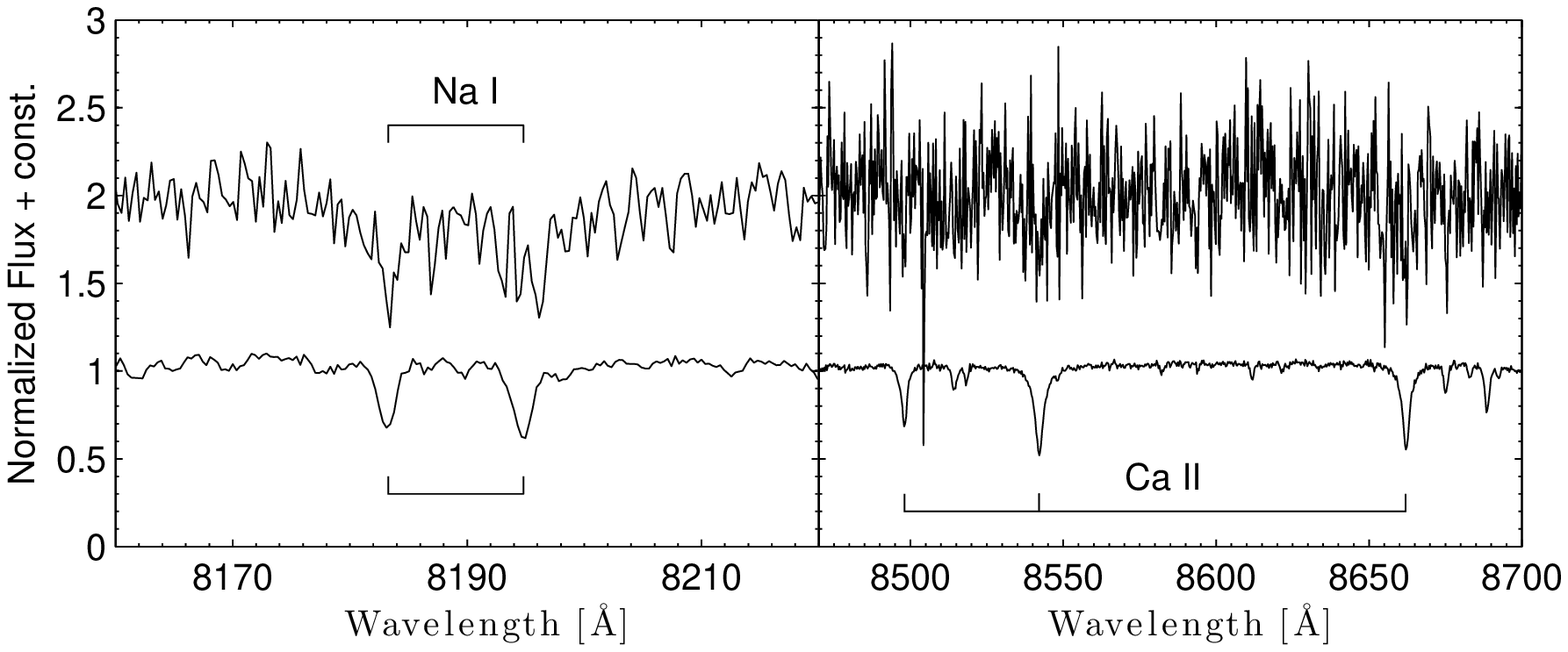}
\includegraphics[angle=0,width=1\hsize]{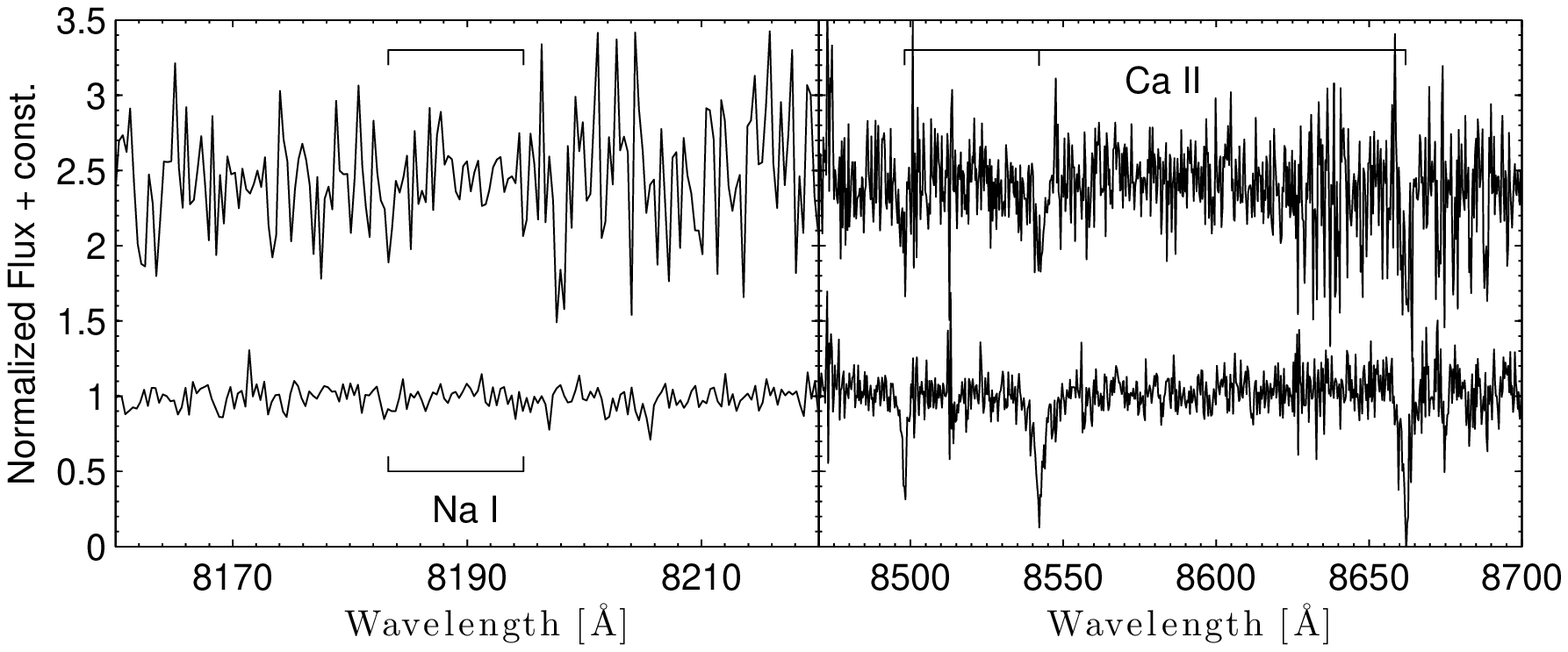}
\end{center}
\caption{Sample spectra of foreground dwarfs (top panel)  and M31 red giants (bottom panel) 
with low and higher $S/N$ ratios. The low $S/N$ data ($S/N$=6 and 8) represent the approximate limit for which velocities 
and CaT metallicities could be barely determined; the $S/N$ for the higher quality spectra 
are $\sim$120 (top) and  $\sim$20 (bottom). 
Indicated are the wavelength regions of the surface gravity sensitive 
sodium doublet and the CaT.}
\end{figure}
\begin{figure}
\begin{center}
\includegraphics[angle=0,width=1\hsize]{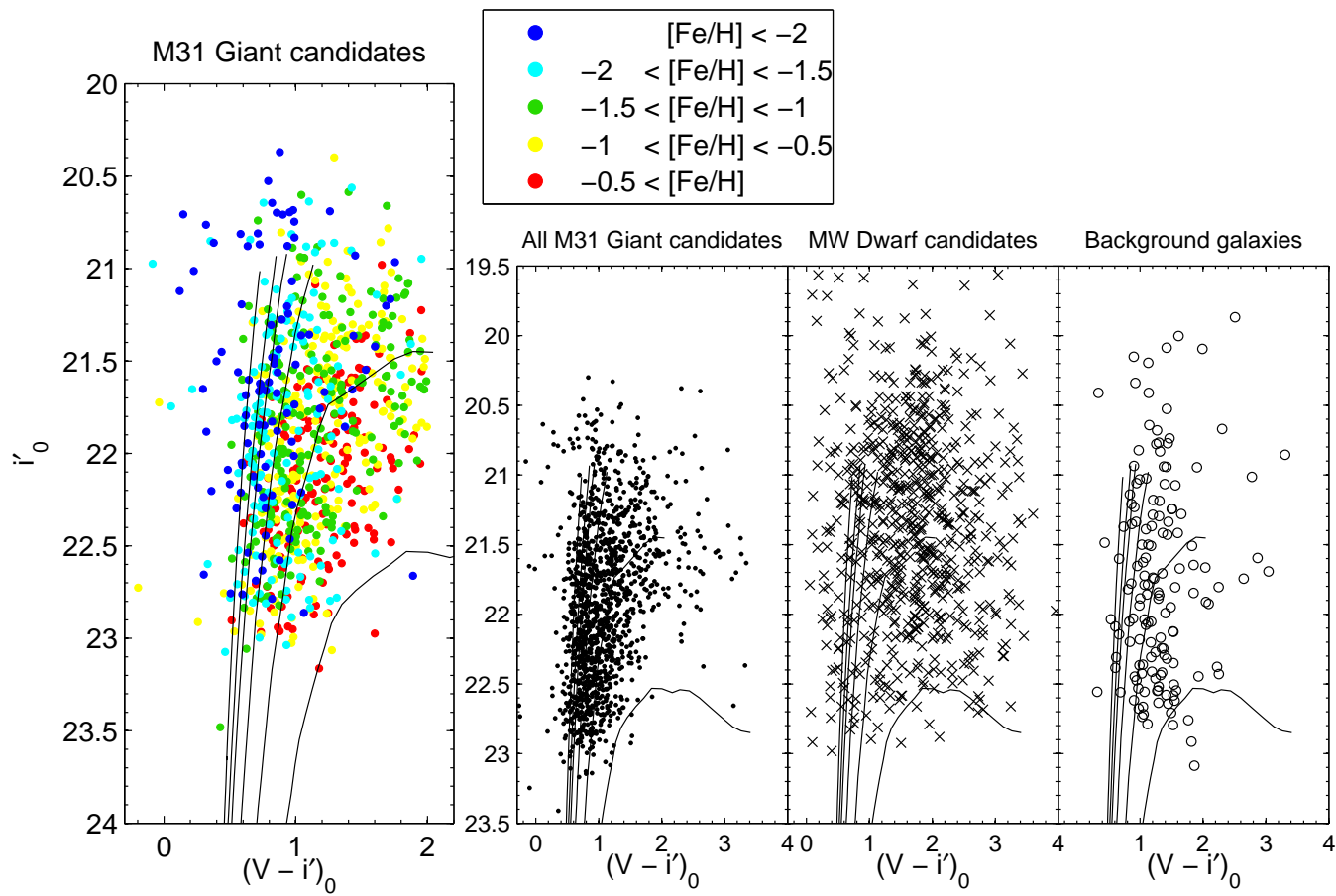}
\end{center}
\caption{Color magnitude diagrams of target objects, separated into M31 giants, foreground 
dwarfs  and background galaxy contamination. Giant candidates for which we could determine 
spectroscopic metallicities are color coded by their [Fe/H]$_{\rm CaT}$ in the leftmost panel.  
The small numbers of metal poor giants that
are plotted to lie to the blue of the bluest
isochrone were carefully vetted and pass 
as likely M31 giants. If these stars are excluded from the sample, our
conclusions are not affected. 
The solid lines are isochrone sets of Girardi et al. (2002) for an age of 
12.7\,Gyr and metallicities [$M$/H] of (left to right) $-$2.3, $-$1.7, $-$1.4, $-$1.0, $-$0.5, and 0.0}
\end{figure}
\begin{figure}
\includegraphics[angle=0,width=0.5\hsize]{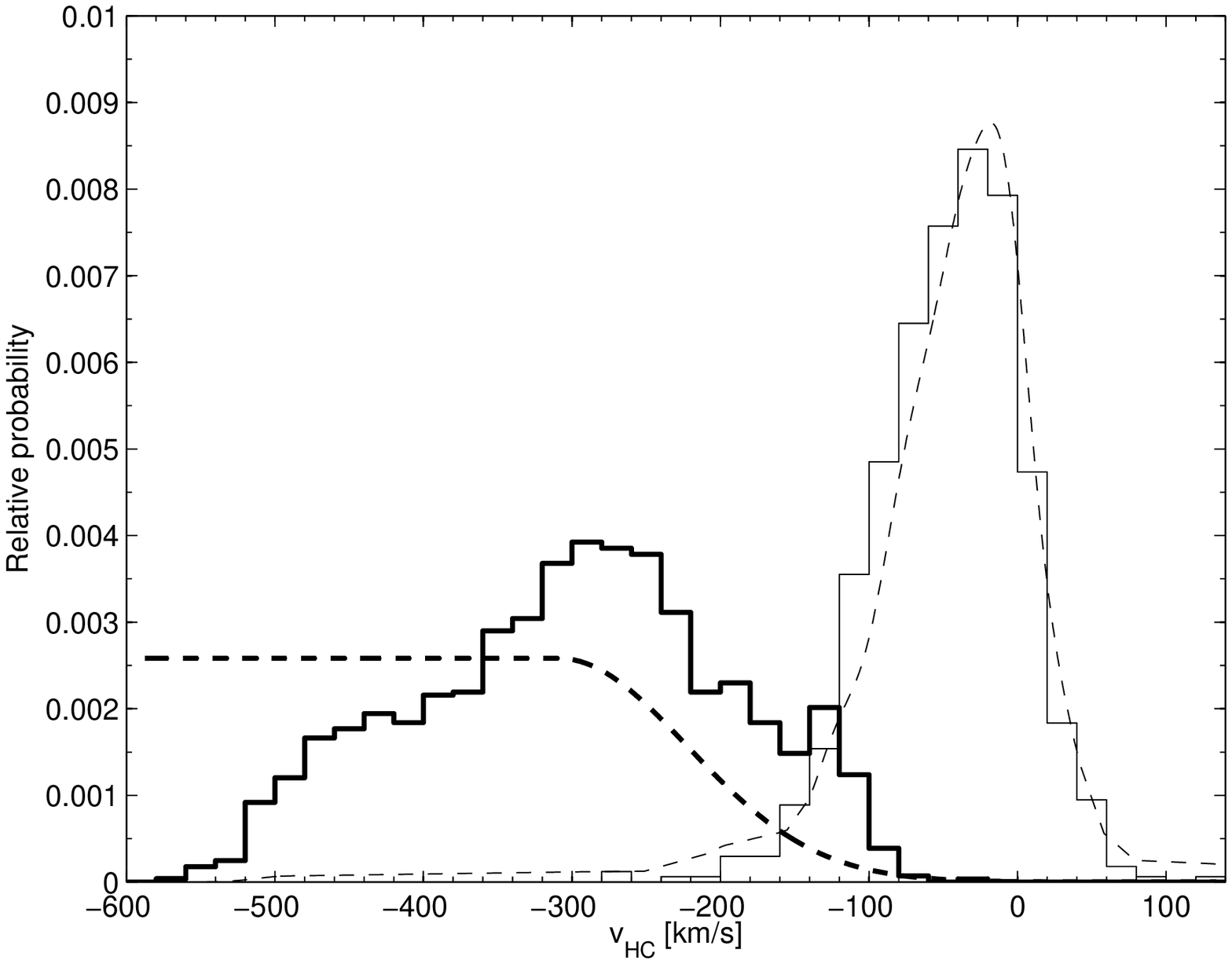}
\includegraphics[angle=0,width=0.5\hsize]{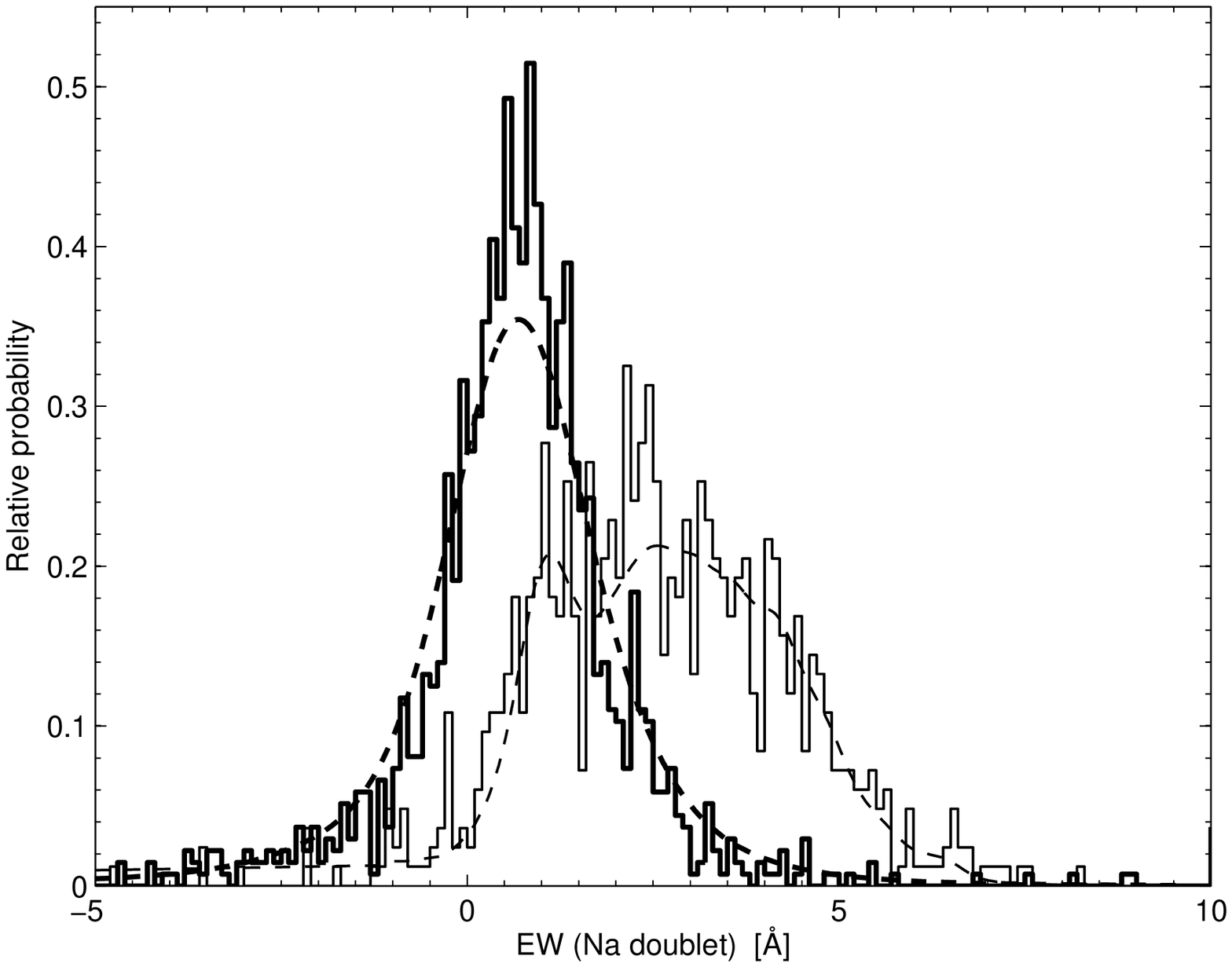}\\
\includegraphics[angle=0,width=0.5\hsize]{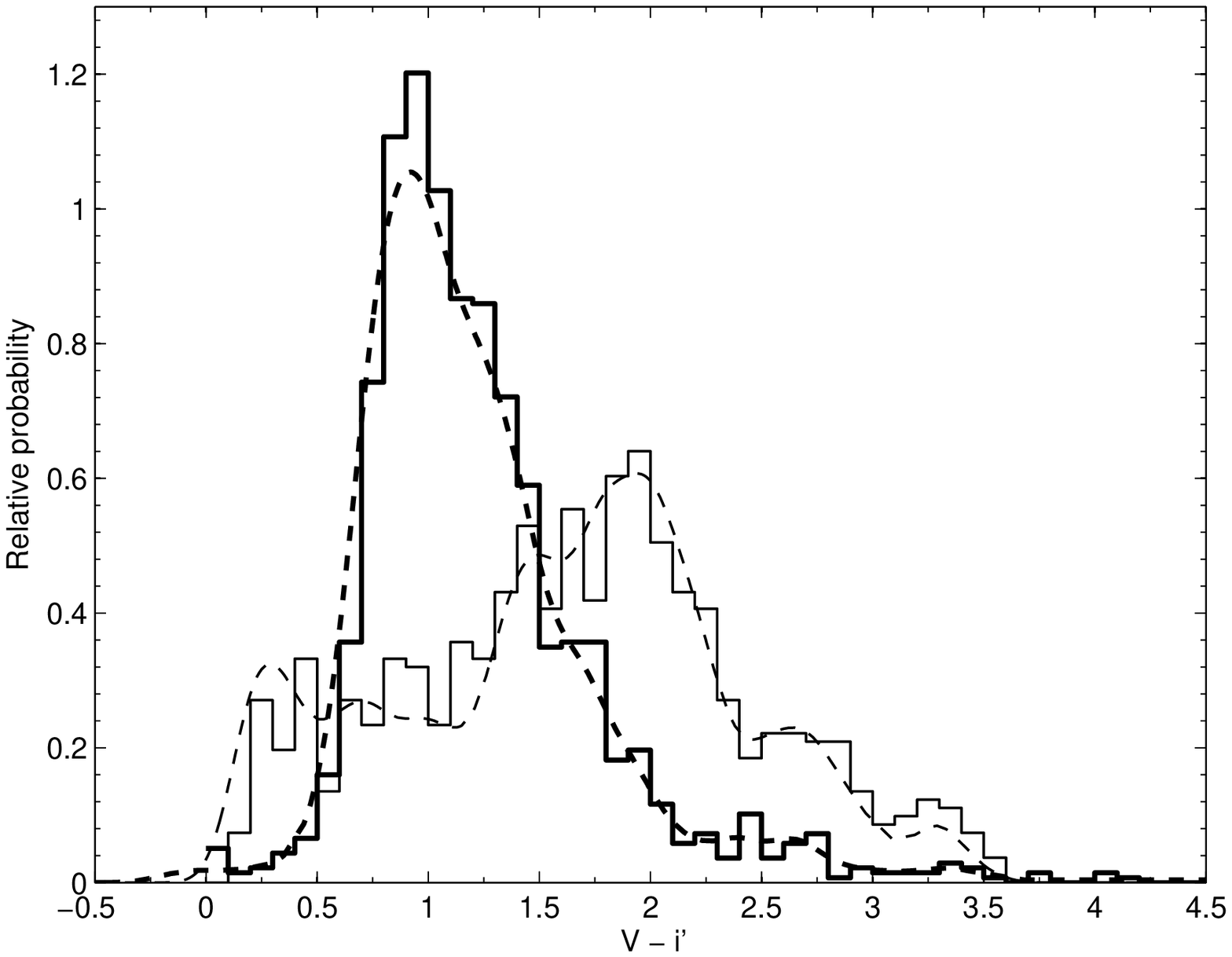}
\includegraphics[angle=0,width=0.49\hsize]{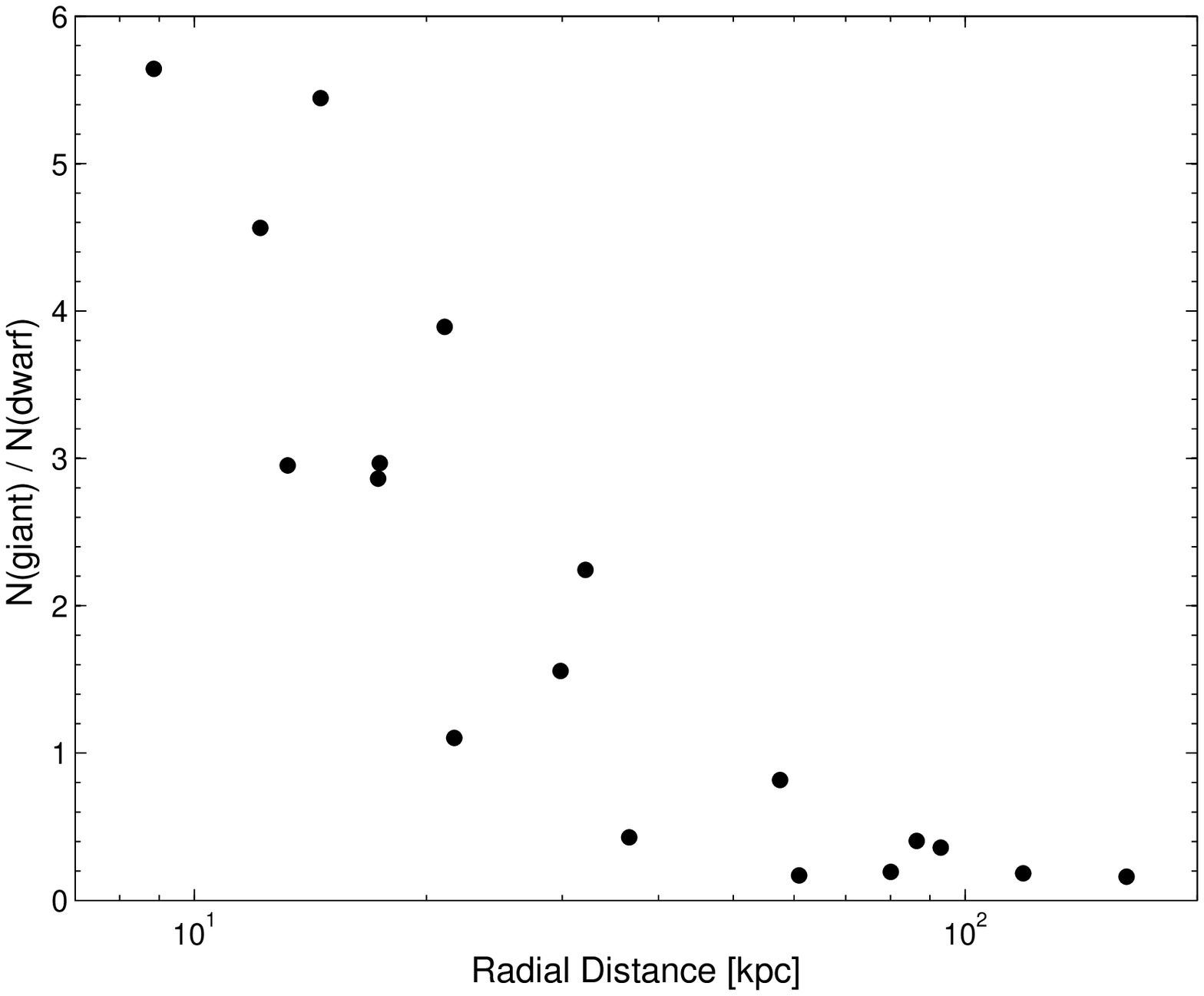}
\caption{Probability distributions (PDFs) for those parameters that we used as criteria for dwarf/giant separation, i.e., 
Na doublet EW,  $V-i'$ color and radial velocity. The dashed lines each indicate the empirical PDFs defined by 
the full training samples, while the solid histograms were separated by applying the respective criteria to our full data. 
Thin and thick lines discriminate between dwarf and giant samples, respectively.  
Despite a considerable overlap in each individual indicator, the statistical combination of all three 
pieces of information allows for an efficient dwarf removal. The bottom right panel shows the ratio of thus 
separated giant to dwarf stars, separately for each field. 
See text for details.}
\end{figure}
\clearpage
\begin{figure}
\includegraphics[angle=0,width=0.5\hsize]{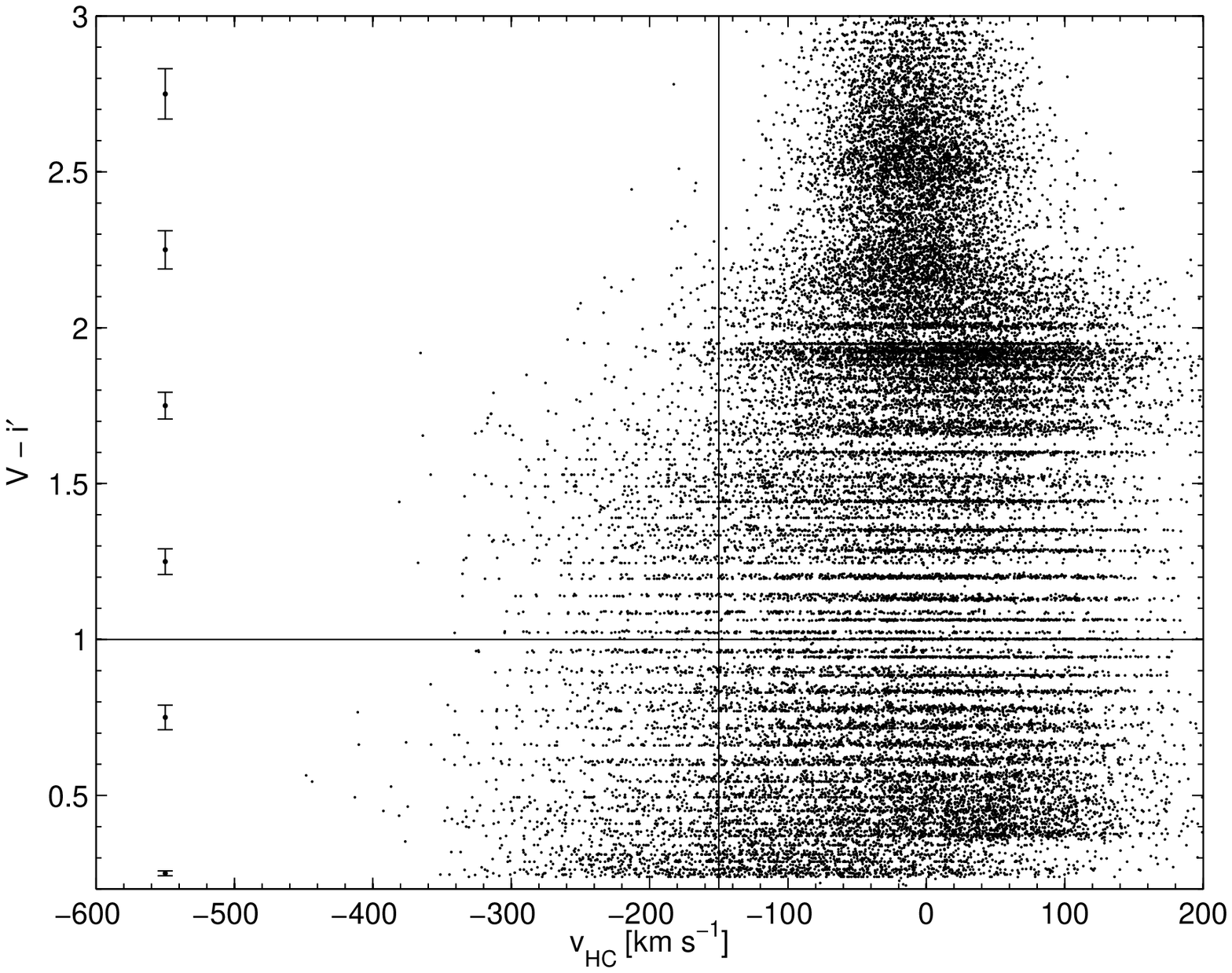}
\includegraphics[angle=0,width=0.5\hsize]{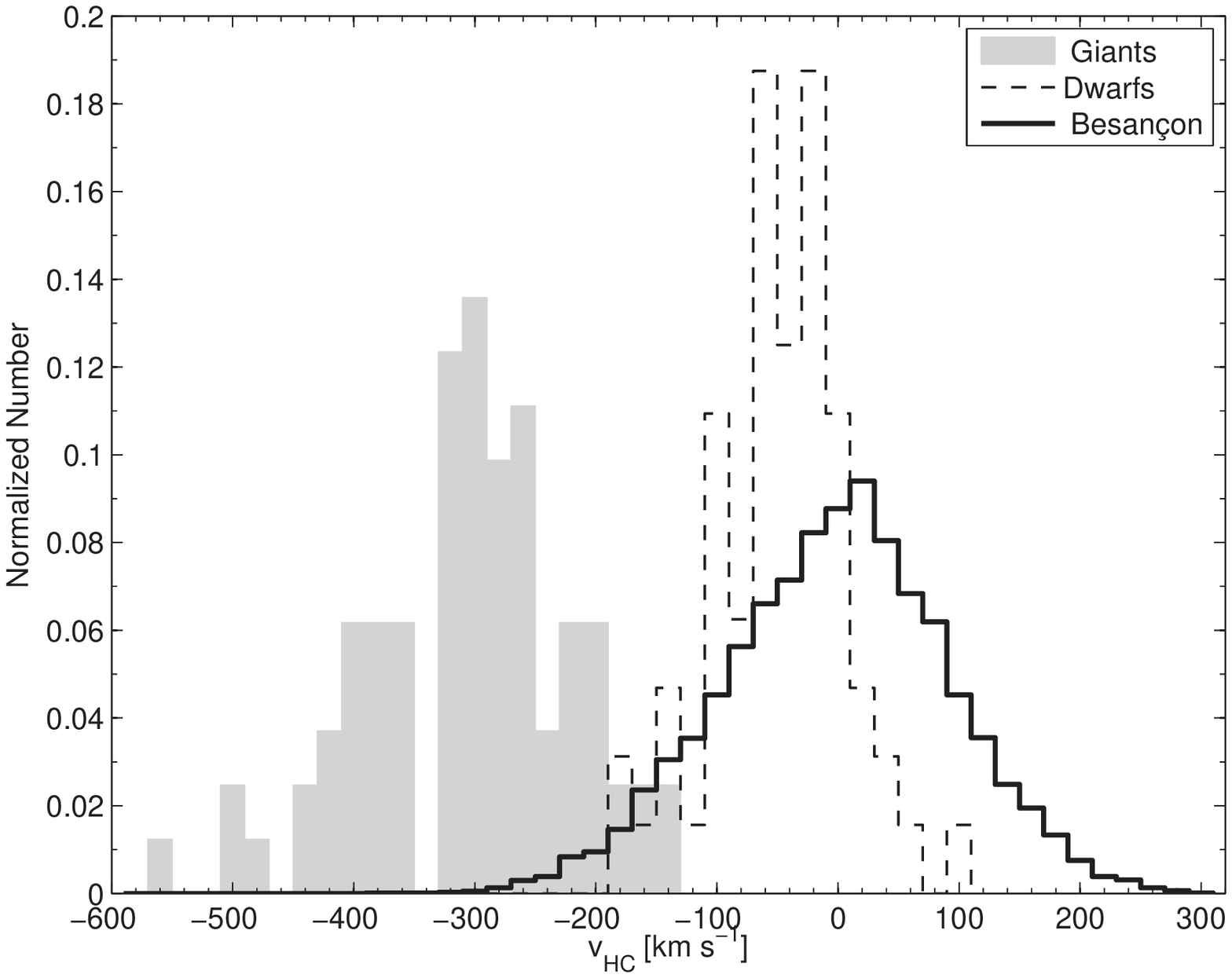}
\caption{Left panel: Color-velocity distribution of foreground dwarfs 
from the Galactic Besan\c con model (Robin et al. 2003),  at a projected distance of 160 kpc to M31. 
The solid lines illustrate the cuts we used to assess the number of undetectable blue dwarfs 
in our M31 giant sample. Also indicated are the observed photometric errors. The right panel 
shows velocity distributions of our observed stars 
(shaded and dashed histograms) with $V-i'\le1$ and the expected blue dwarfs from the 
Besan\c con model (solid histogram). The fraction of unresolvable blue dwarfs in our giant 
sample is well below 3\% and negligible.}
\end{figure}
\begin{figure}
\includegraphics[angle=0,width=0.5\hsize]{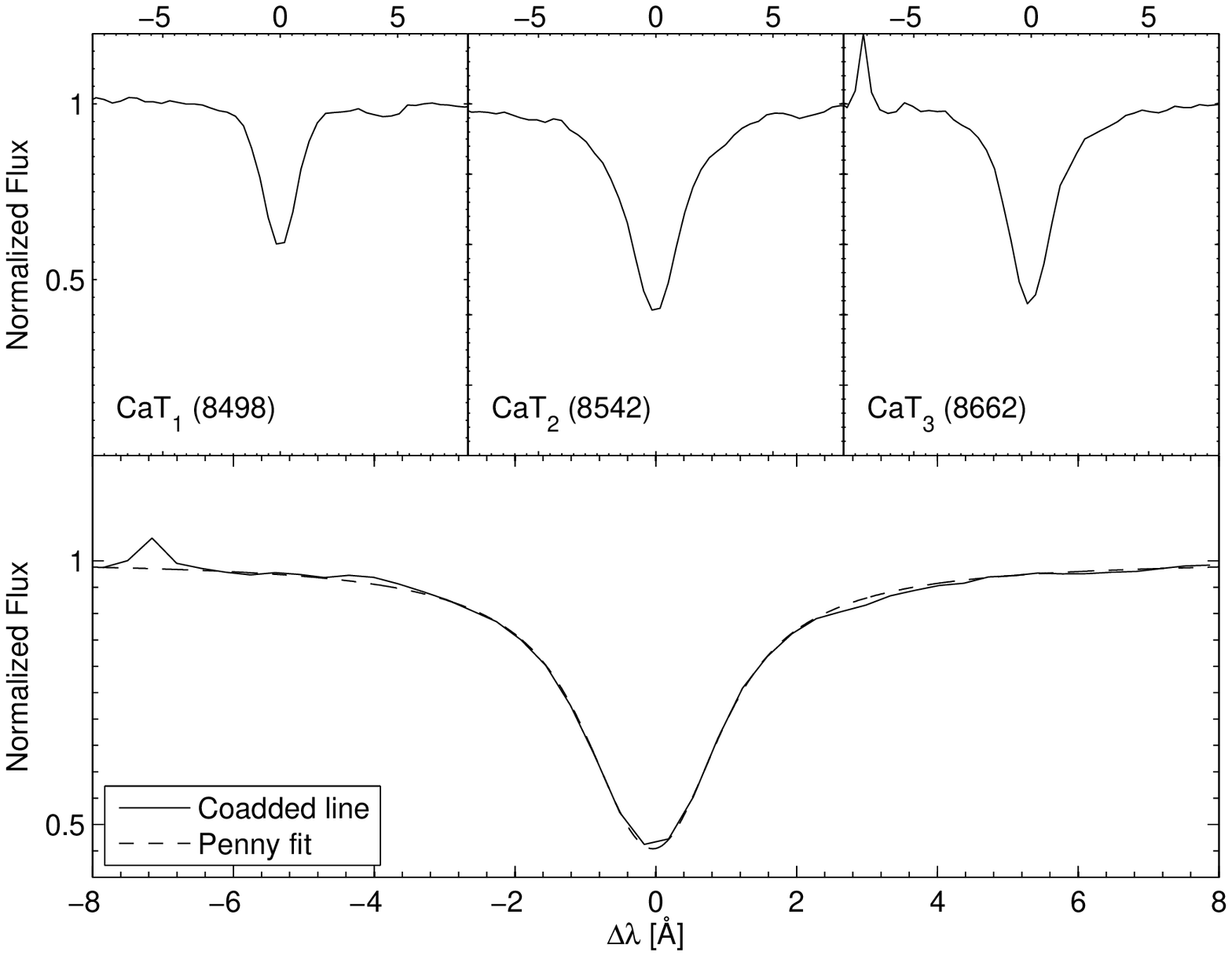}
\includegraphics[angle=0,width=0.5\hsize]{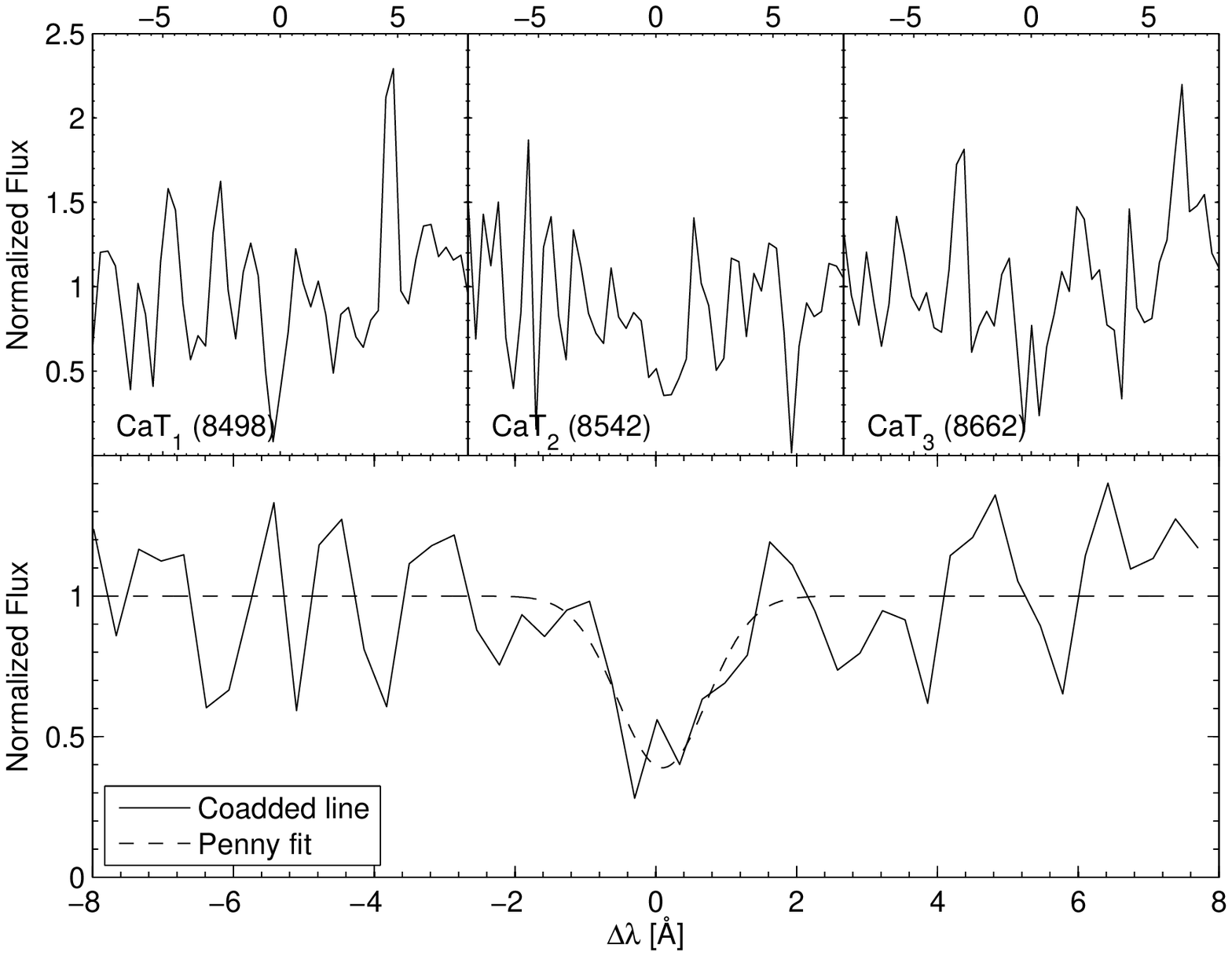}
\caption{Illustration of the CaT coaddition in a high $S/N$ spectrum of a red giant in a calibration 
globular cluster 
(star N3201-S11; $S/N$$\sim$110; V=14.5\,mag; v$_{\rm HC}=485$ km\,s$^{-1}$; 
$<$$\Sigma W$$>$$=3.8$\AA) 
and a low $S/N$ DEIMOS spectrum of a M31 red giant 
(star 5004266; $S/N$$\sim$4; V=21.7\,mag; v$_{\rm HC}=-338$ km\,s$^{-1}$;
$<$$\Sigma W$$>$$=2.4$\AA). 
The top panels display each individual Ca line, shifted to a common line center,  
and the bottom panels show the weighted coadded line (solid line), from which we measure 
the CaT line strengths using a Penny (i.e., Gaussian plus Lorentz profile) fit (dashed line).}
\end{figure}
\begin{figure}
\includegraphics[angle=0,width=0.45\hsize]{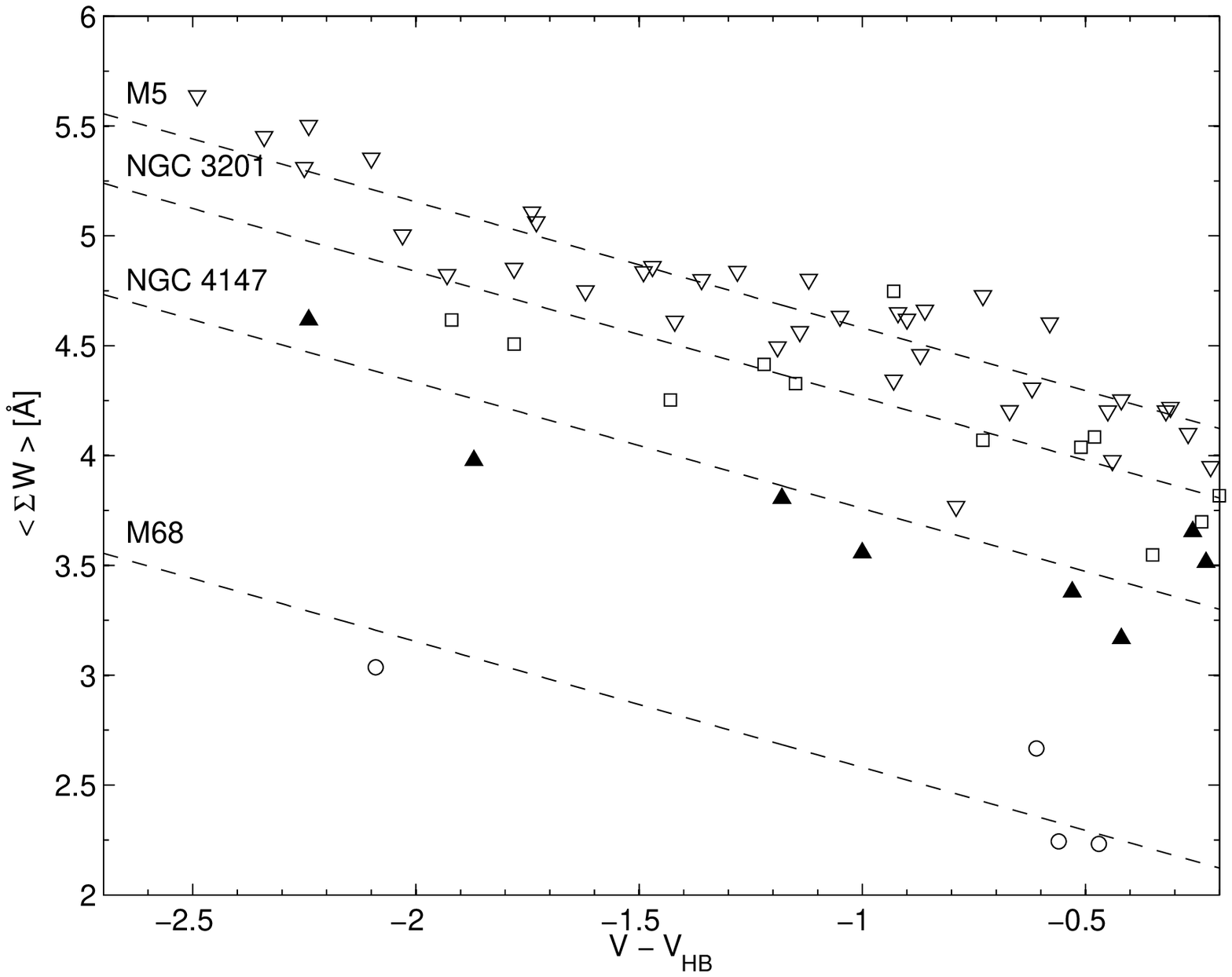}
\includegraphics[angle=0,width=0.5\hsize]{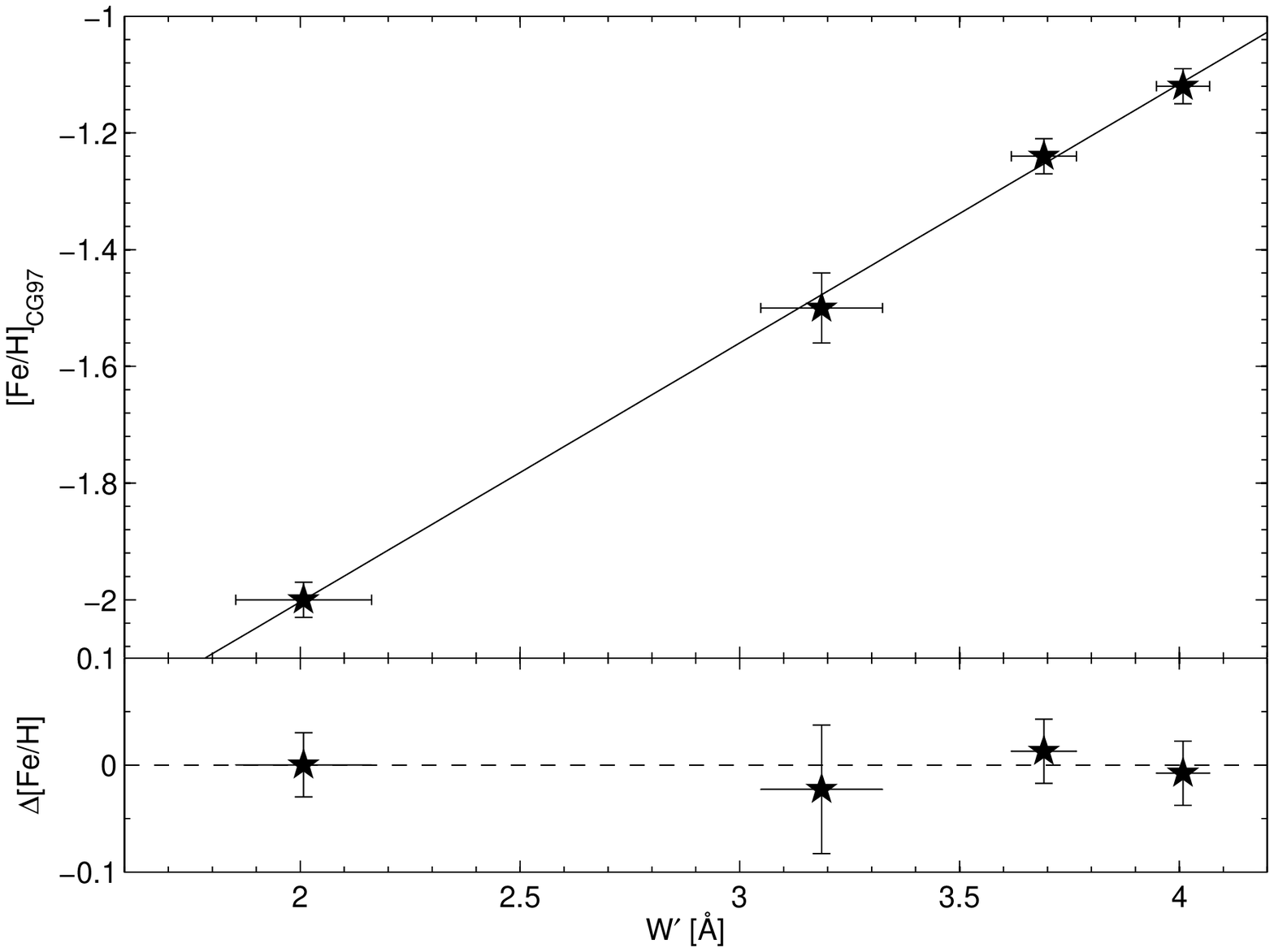}
\caption{Left Panel: Linestrengths of red giants in the calibration clusters from FLAMES data (Koch et al. 2006) vs. magnitude above the horizontal branch. The right panel shows the metallicities of these 
globular clusters on the scale of Carretta \& Gratton (1997), with the residuals of the best-fit relation 
(eq. 6) at the bottom. Solid and dashed lines indicate these best-fit calibrations (eqs. 5,6). 
All these measurements are based on the coadded CaT. }
\end{figure}
\begin{figure}
\begin{center}
\includegraphics[angle=0,width=0.7\hsize]{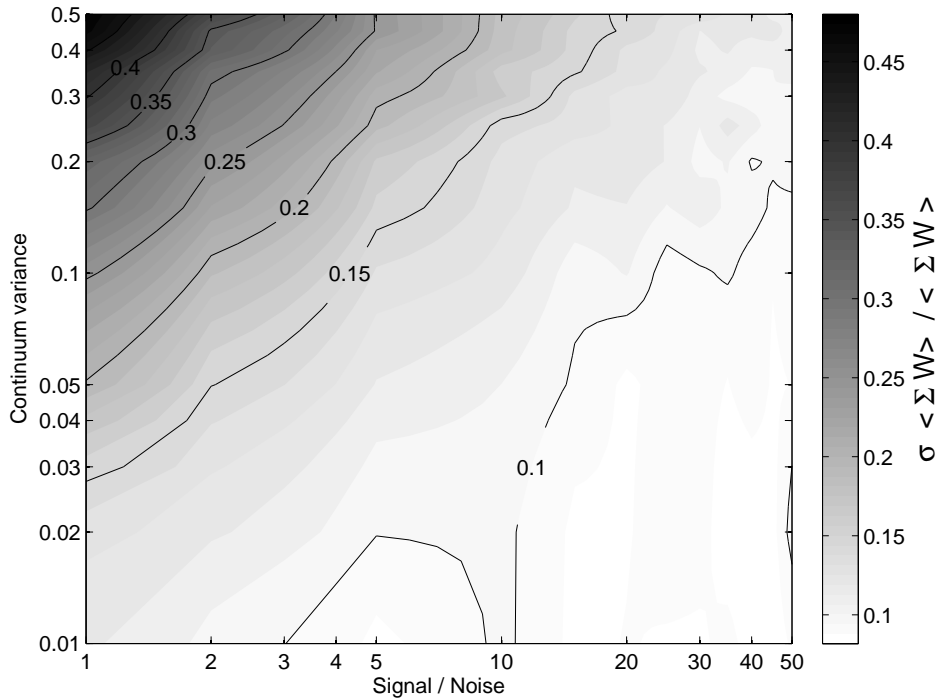}
\end{center}
\caption{Distribution of the relative error on the measured coadded line strength $<$$\Sigma W$$>$, determined from Monte Carlo simulations, as a function of spectral quality. For our observed spectra, 
these values are propagated through Eqs. 5,6 to obtain the spectroscopic metallicity uncertainty.}
\end{figure}
\begin{figure}
\begin{center}
\includegraphics[angle=0,width=0.7\hsize]{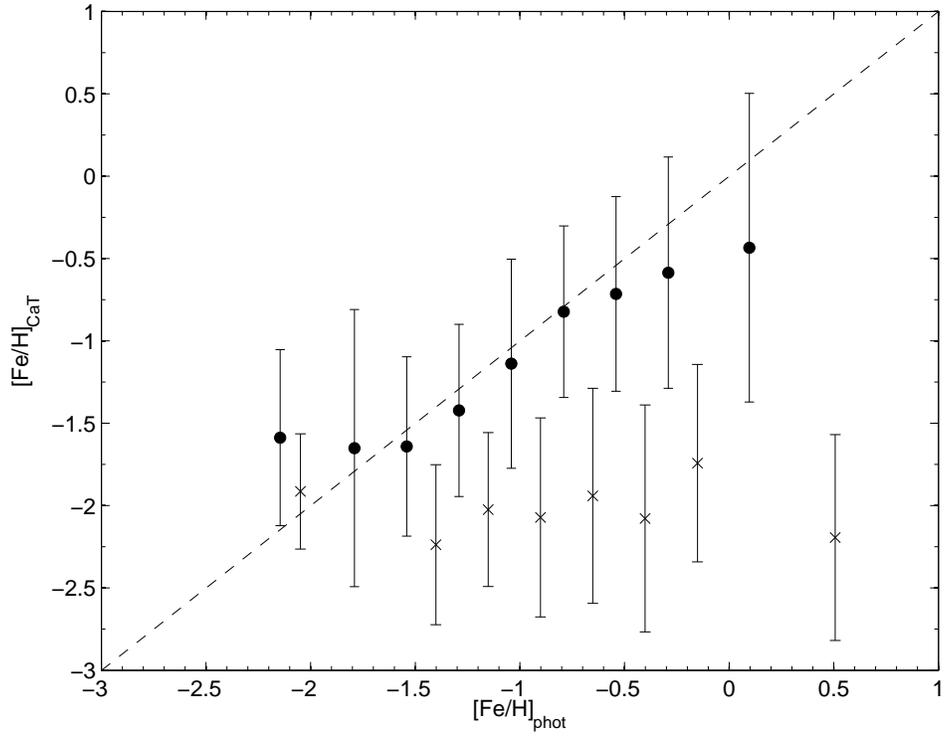}
\end{center}
\caption{Spectroscopic versus photometric metallicities for dwarf (crosses) and giant (points) 
candidates. Shown each are mean and 1$\sigma$ scatter. The dwarf stars clearly deviate from unity (solid line). Data were binned by 0.25 dex except for the most metal poor and metal rich bin, where 
we chose to include 20 stars for better sampling.}
\end{figure}
\begin{figure}
\begin{center}
\includegraphics[angle=0,width=1\hsize]{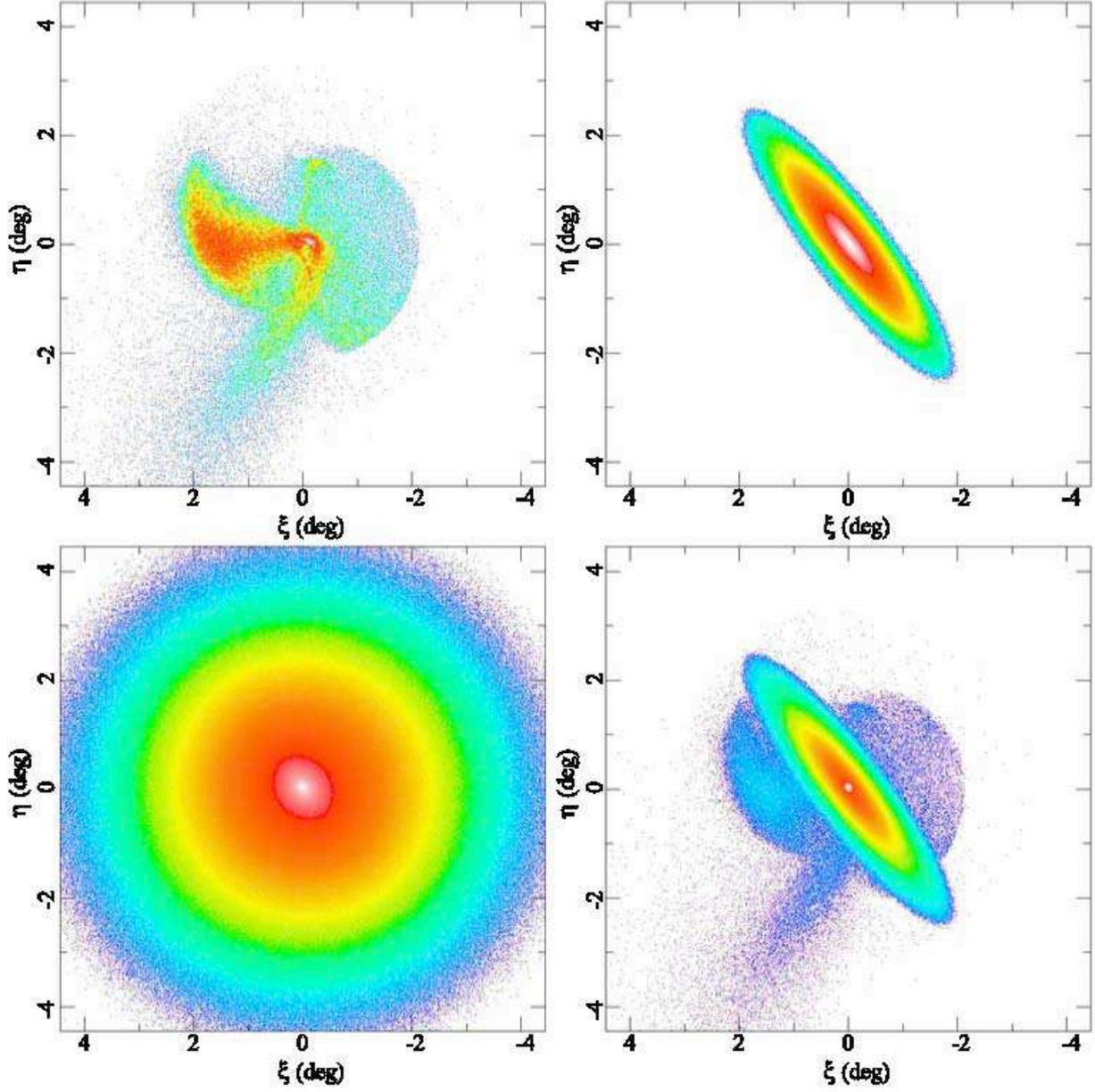}
\end{center}
\caption{Snapshot of the simulation data at 1\,Gyr in standard coordinates and  
colorcoded by the particle density (Mori \& Rich 2008). 
Each subpanel separately displays a different component: 
the accreted satellite only (top left), disk and bulge (top right), spherical halo (bottom left), 
and disk, bulge plus satellite (bottom right). }
\end{figure}
\begin{figure}
\begin{center}
\includegraphics[angle=0,width=0.7\hsize]{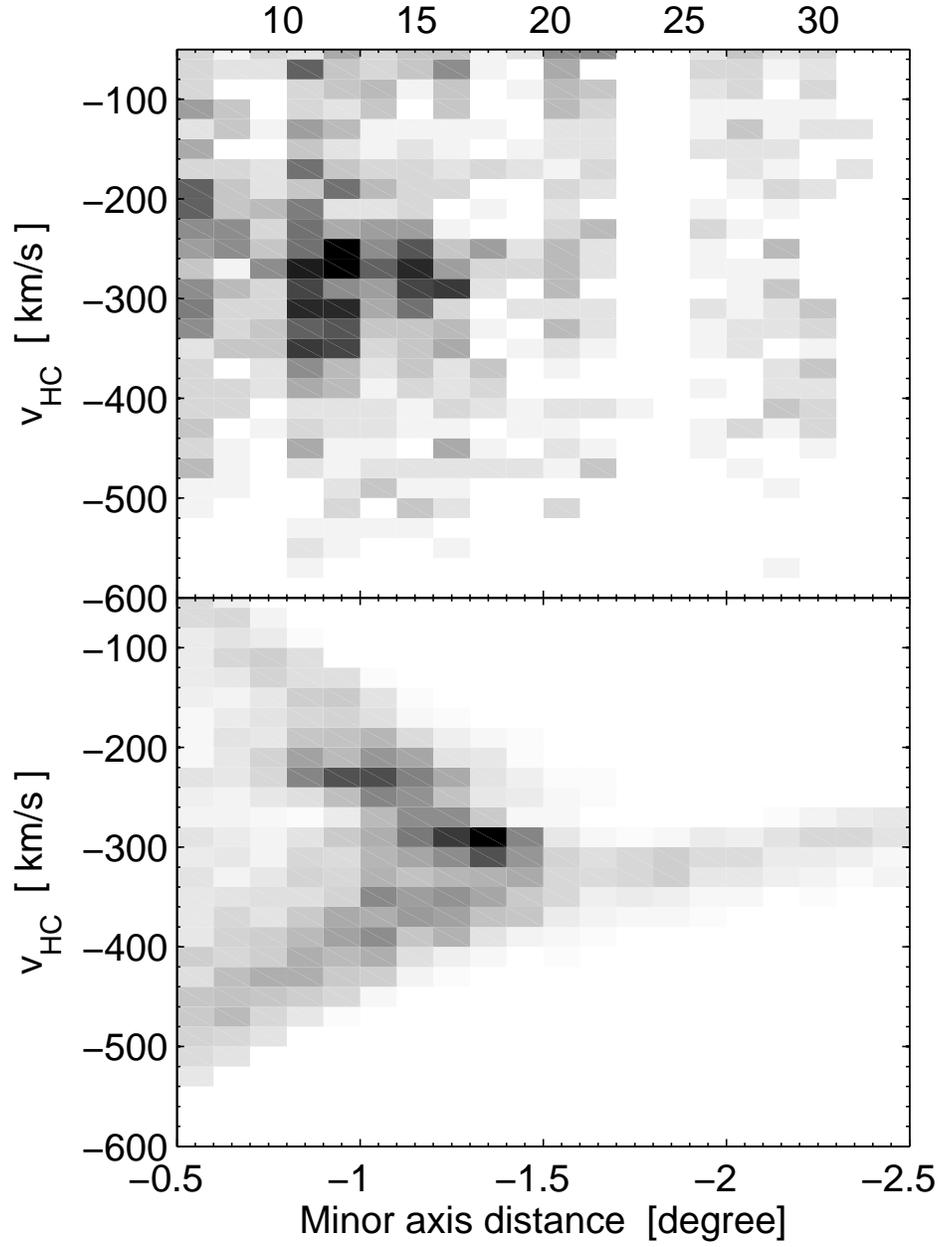}
\end{center}
\caption{Density distribution (with arbitrary scaling) of radial velocities along the minor axis. 
The top panel displays our observed minor axis 
data within 35 kpc, whereas the bottom panel shows the distribution of 
the satellite particles from our simulation. Numbers at the top indicate distances in kpc to guide the eye.}
\end{figure}
\begin{figure}
\begin{center}
\includegraphics[angle=0,width=0.95\hsize]{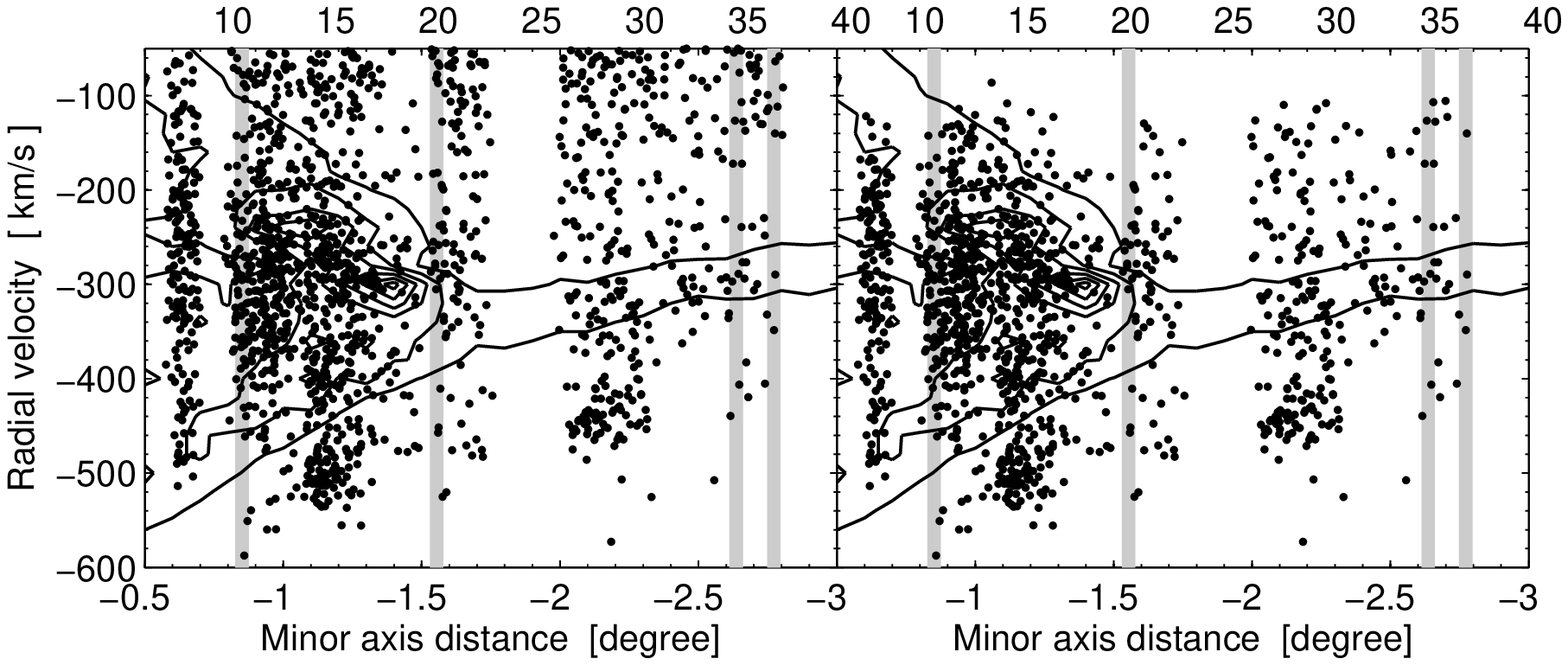}
\includegraphics[angle=0,width=0.45\hsize]{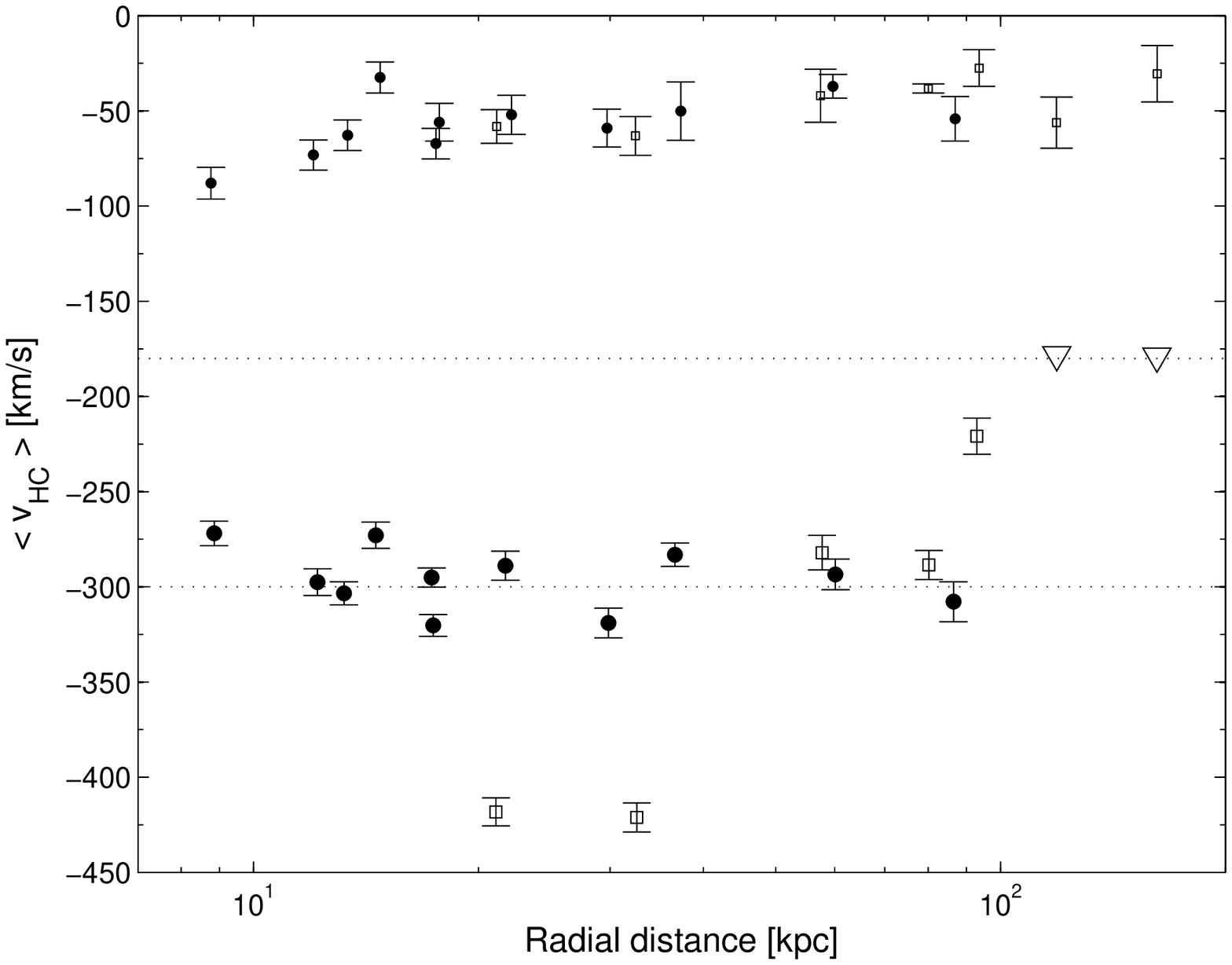}
\includegraphics[angle=0,width=0.45\hsize]{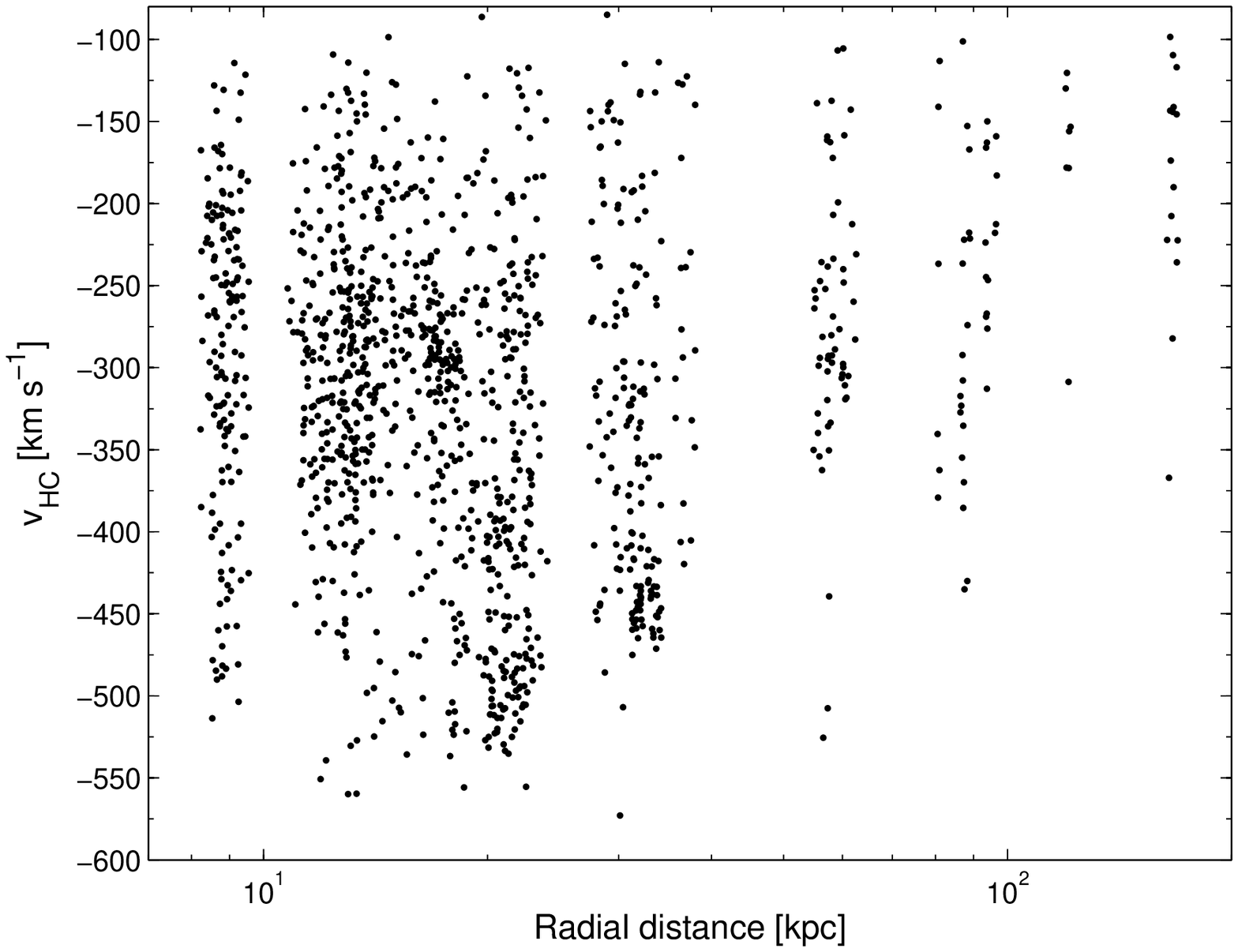}
\end{center}
\caption{Top panel: Radial velocities as a  function of radial distance along the minor axis. Shown left is the the complete spectroscopic sample, without the removal of foreground dwarfs,  
within the inner 40 kpc, while the data shown in the right panel have been cleared of the foreground contamination 
using the methods described in Sect.~4. Indicated as grey shaded bars are the HST fields of Brown 
et al. (2003; 2006; 2007). The black contours delineate the distribution from our simulations (see Fig.~11, bottom panel).  Numbers at the top indicate distances in kpc to guide the eye. 
 The bottom left panel shows the radial variation of  {\em mean} radial velocity 
for minor-axis (filled circles) and off-axis (open squares) fields. Small symbols at v$_{\rm HC}\sim-50$ 
km\,s$^{-1}$ 
are for the dwarf stars, while the larger symbols below $-$180 km\,s$^{-1}$ show the giant distributions.   
Comparison with the dotted lines at the systemic velocities of M31 ($-$300 km\,s$^{-1}$) and M33 
 ($-$180 km\,s$^{-1}$) suggests  that 
the giant samples in the outermost two fields appear to be suffering
from contamination by M33 members  (see also Fig.~19). The bottom right panel then shows the 
distribution of velocites for the full, dwarf-cleaned sample.  The stream fields (a3, H13s) reflect in the peaks below 
 $-$400 km\,s$^{-1}$ and the  peak at $-$520 km\,s$^{-1}$, both of which are 
 reproduced by the simple orbit model of Ibata et al. (2004).}
\end{figure}
\begin{figure}
\begin{center}
\includegraphics[angle=0,width=0.95\hsize]{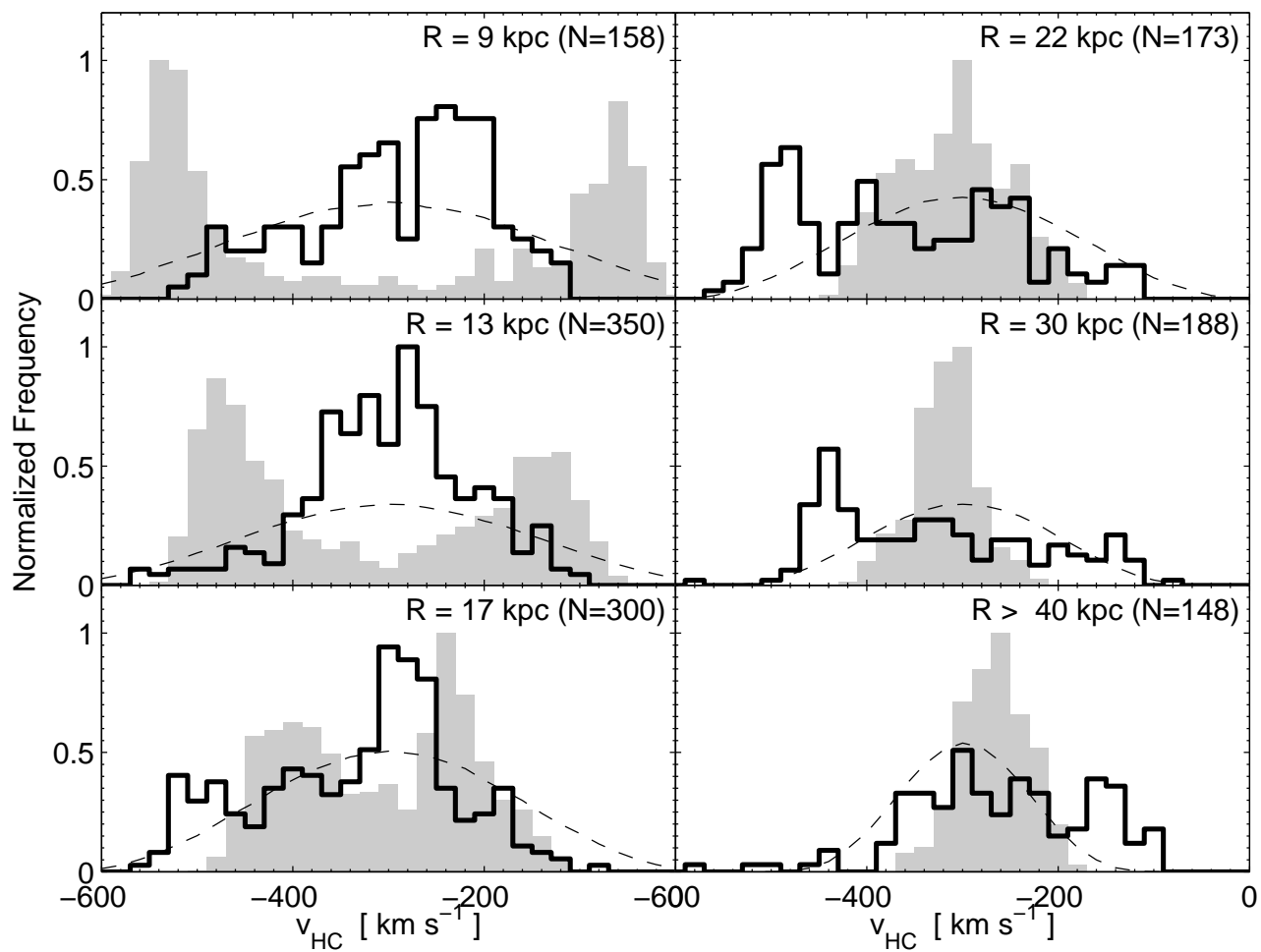}
\end{center}
\caption{Velocity histograms along different radial bins (with the numbers representing their mean locations). 
The solid line represents our observed, 
dwarf-cleaned velocity sample. Shown as a shaded histogram is the velocity distribution of 
the stream particles drawn from  our simulations at the same radial location as the observed fields, 
while the dashed thin line indicates the simulated contribution of  M31 halo particles to the velocity 
structures at these distances. Each distribution has been normalized to unity.}
\end{figure}
\begin{figure}
\includegraphics[angle=0,width=0.49\hsize]{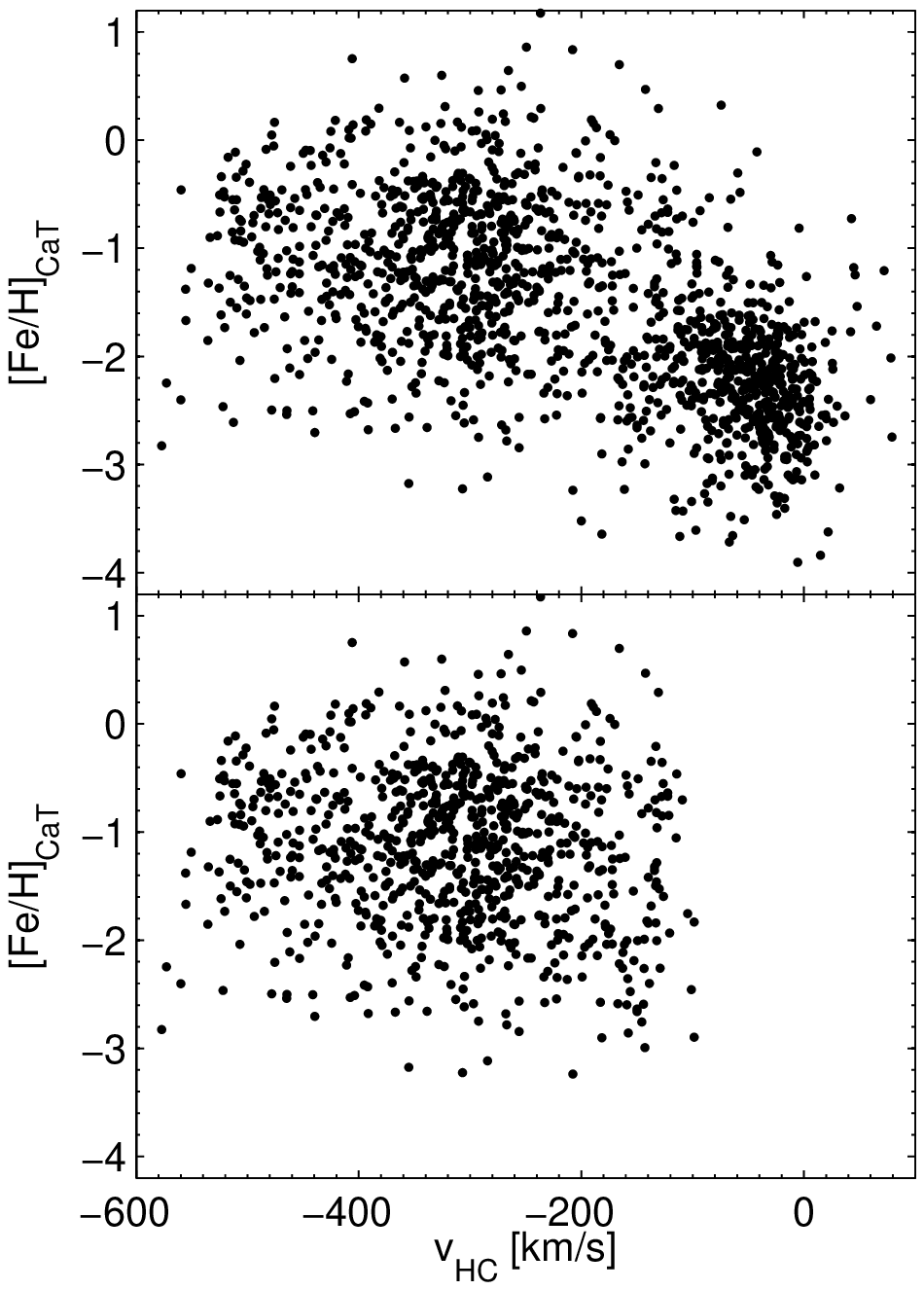}
\includegraphics[angle=0,width=0.5\hsize]{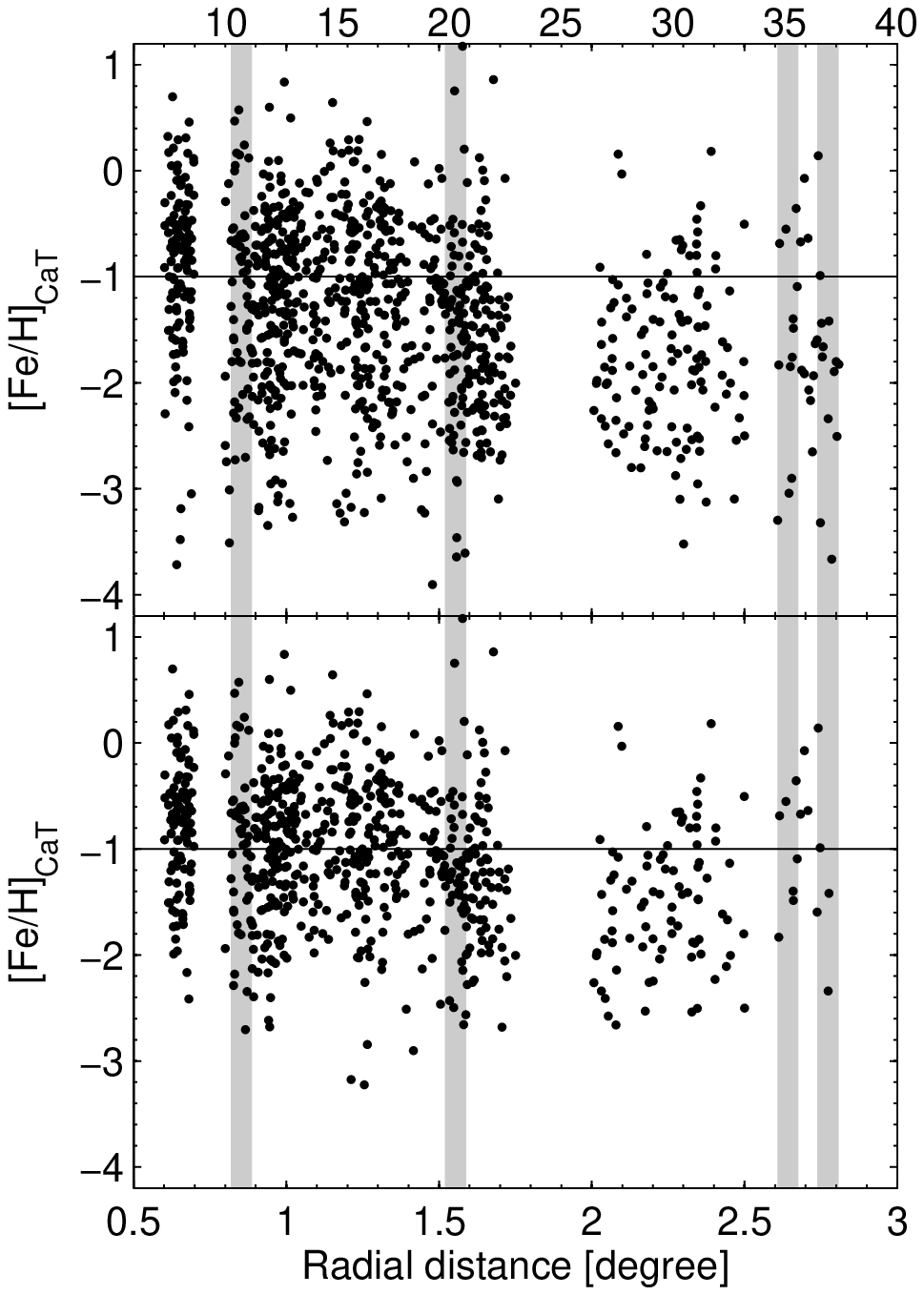}
\caption{Left panels: Spectroscopic metallicity  (on the scale of Carretta \& Gratton 1997) 
versus radial velocity. Right panels: 
the same metallicities within 40 kpc as a function of radial distance. 
The dwarf contamination clearly stands out as a clump around $-$2 dex above $\ga -150$ km\,s$^{-1}$. 
The top (bottom) panels each display the data set before (after) removal of this contamination. 
Shaded regions indicate the HST fields of Brown et al. (2003, 2006, 2007). For distributions of all metallicities out to 160 kpc see Fig.~15 (bottom panel). A solid line 
in the right panel has been added at [Fe/H]$_{\rm CaT}=-1$ for reference.}
\end{figure}
\begin{figure}
\begin{center}
\includegraphics[angle=0,width=0.5\hsize]{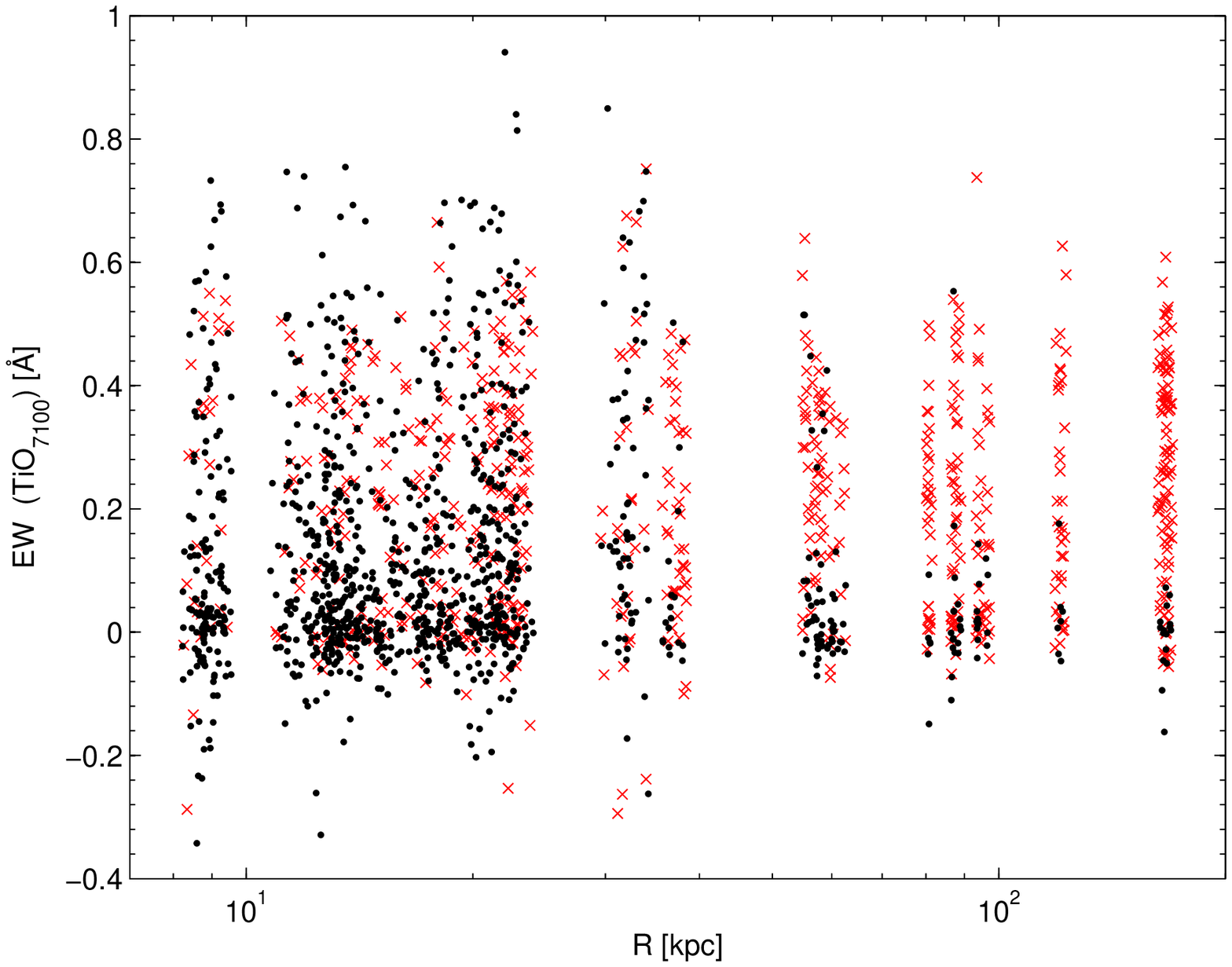}
\includegraphics[angle=0,width=0.4\hsize]{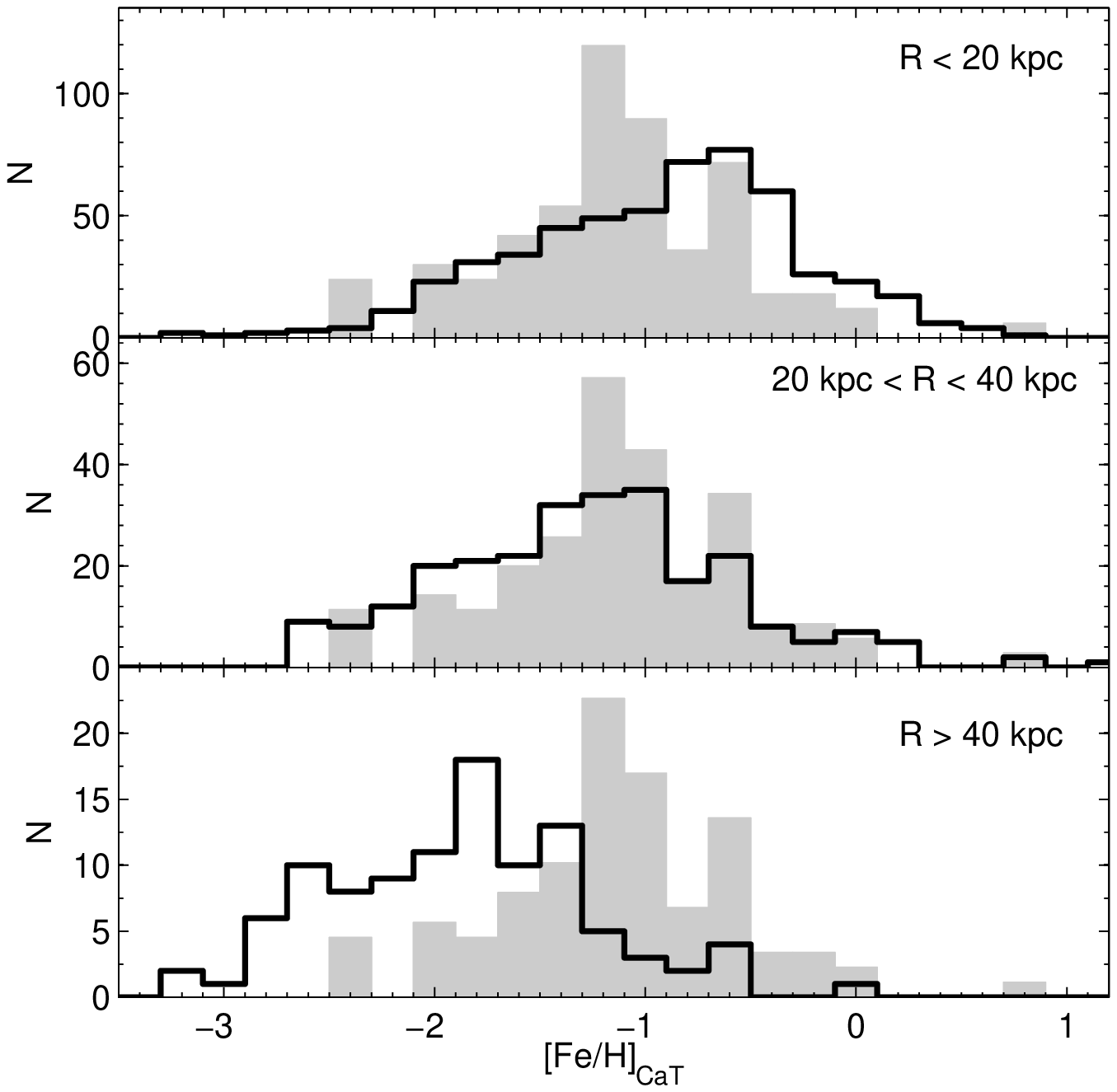}
\includegraphics[angle=0,width=0.9\hsize]{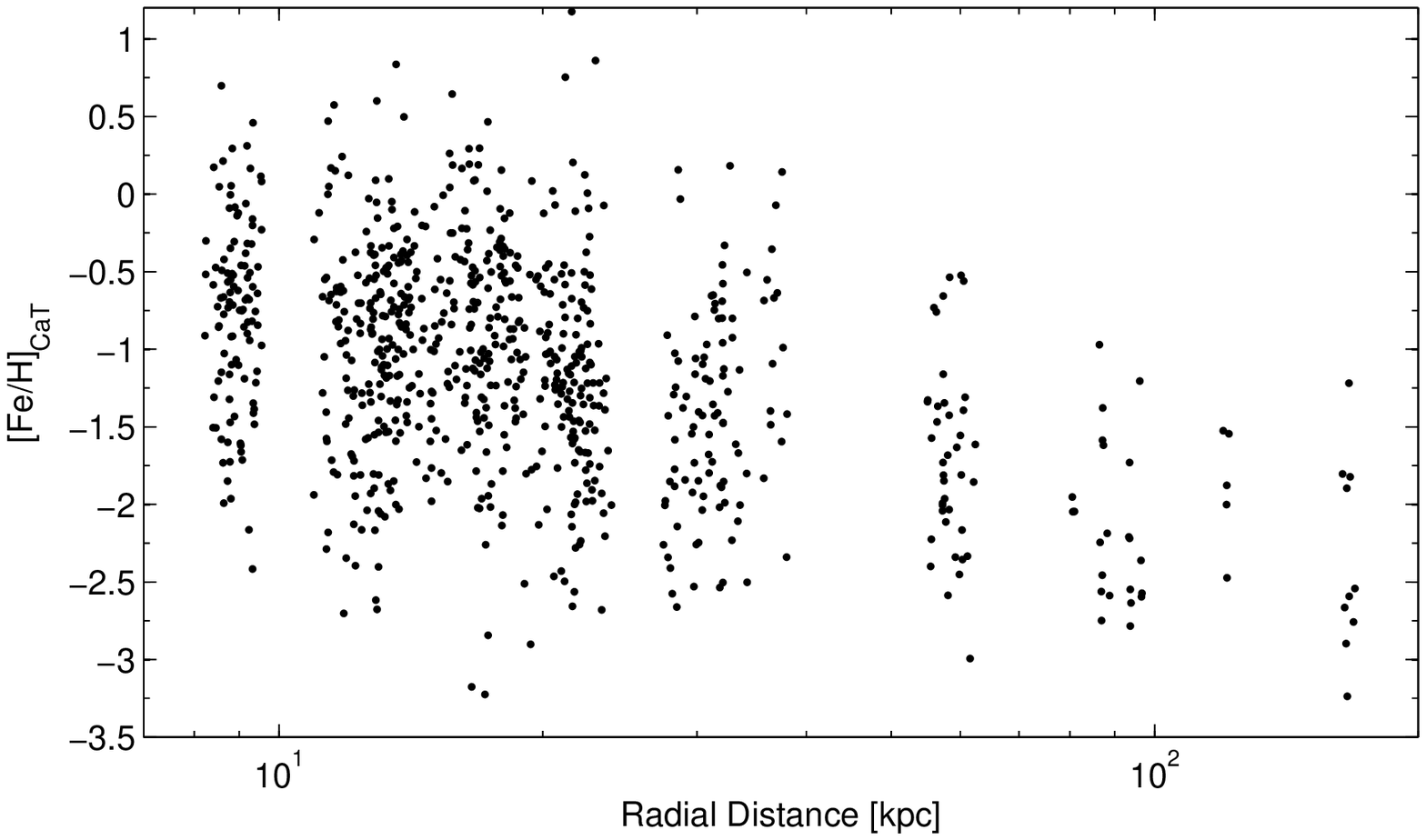}
\end{center}
\caption{A strong abundance gradient is 
qualitatively already visible  in the top left panel, which shows the strength of the TiO band at 7100\AA\ 
in our entire giant sample (black points) versus radial distance. Note that dwarfs (red crosses) cover the full range 
in TiO strengths at all radii. 
The radial metallicity gradient is then clearly present in the 
spectroscopic  MDFs  
in 3 different radial bins (Top right panel; solid lines) and  for our dwarf-cleaned CaT sample (bottom panel). 
However, none of the distributions fully resembles the stream component (shown as shaded histogram). 
Notice that there are no metal rich stars beyond $\sim$50 kpc. }
\end{figure}
\begin{figure}
\begin{center}
\includegraphics[angle=0,width=0.7\hsize]{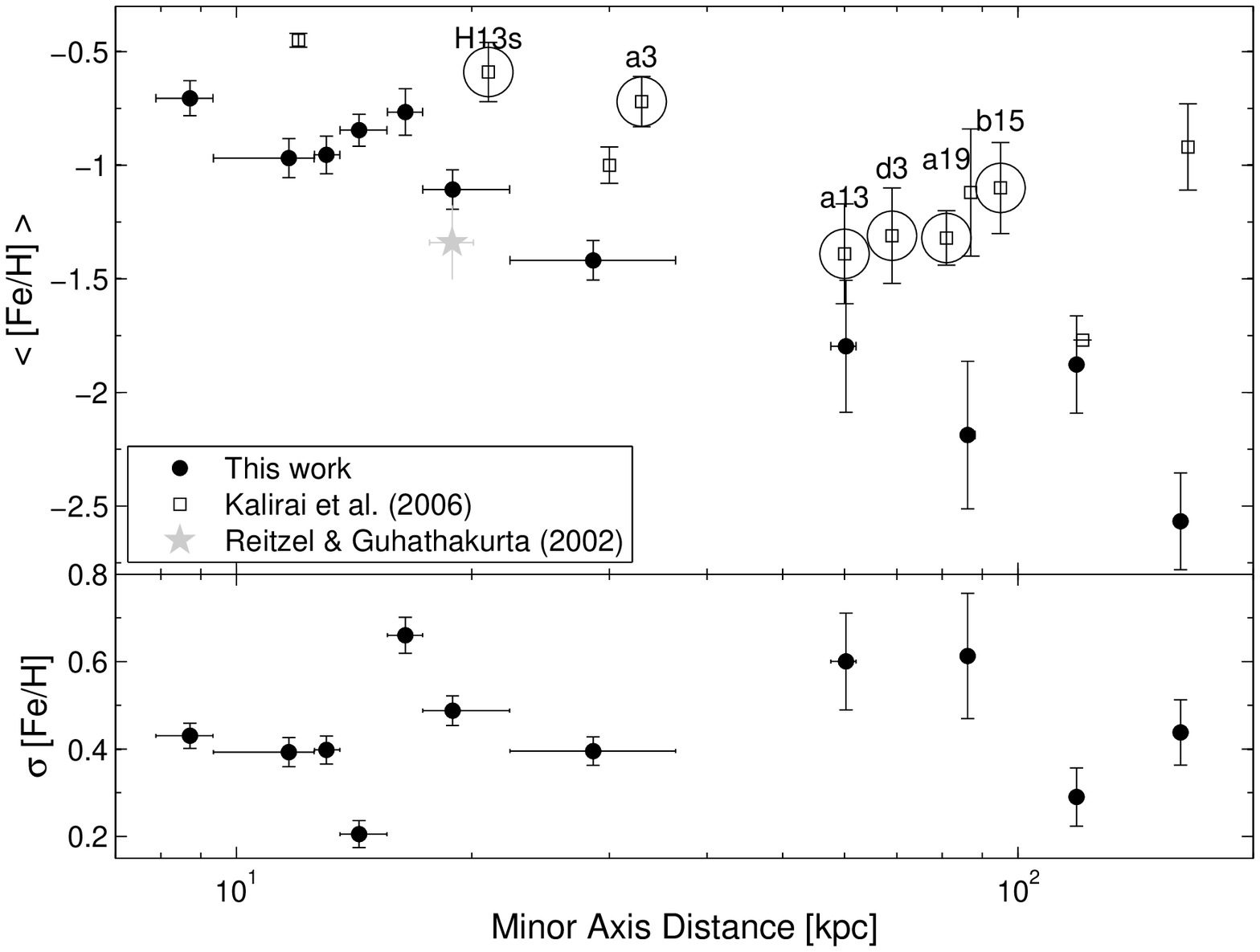}
\includegraphics[angle=0,width=0.7\hsize]{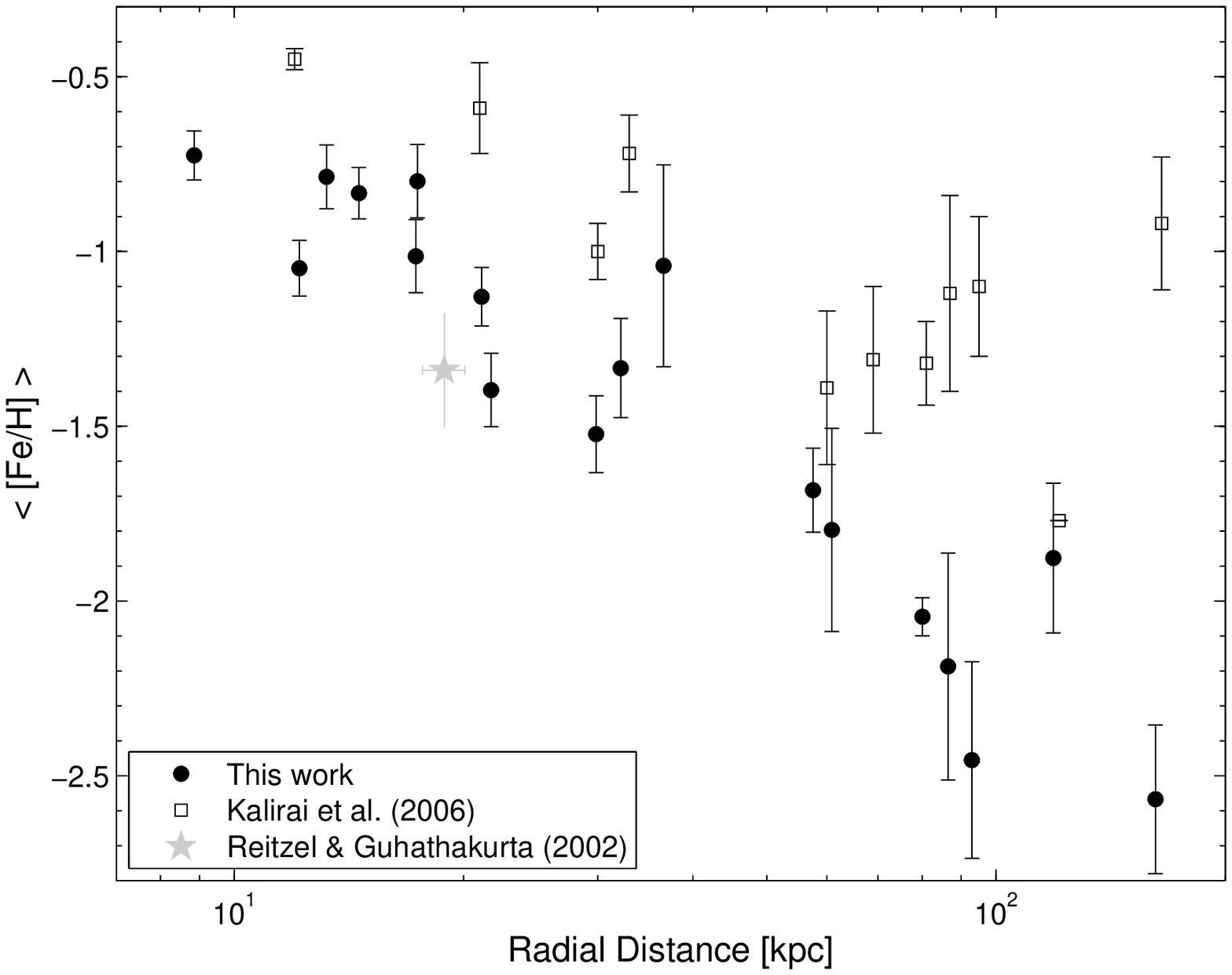}
\end{center}
\caption{Run of mean metallicity 
and metallicity dispersion as a function of radius. While both our data and those of Kalirai et al. (2006a) show the presence 
of a gradient, our measurements are more metal poor on average. 
The top panel only includes our {\em minor axis} fields with a radial binning to obtain the same number of stars per bin.  
The  points  encircled and labelled by their identifier are not located on the minor axis. 
Horizontal errorbars indicate the 
extent of our radial binning. The open star indicates the spectroscopic mean [Fe/H]  
measurement of Reitzel \& Guhathakurta (2002).
In the bottom panel, we also include the off-axis fields and compute the mean metallicities separately for 
each field. } 
\end{figure}
\begin{figure}
\begin{center}
\includegraphics[angle=0,width=0.37\hsize]{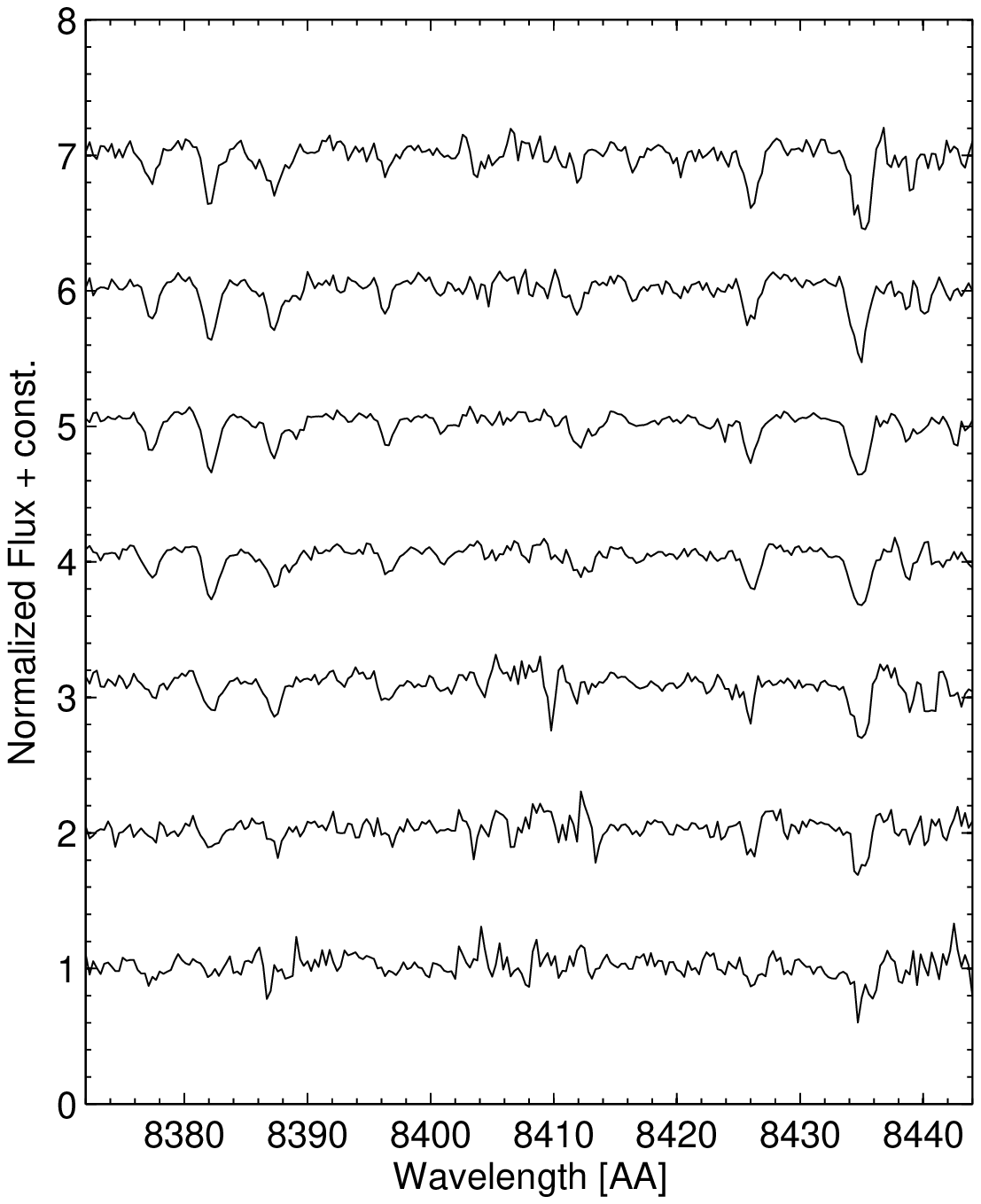}
\includegraphics[angle=0,width=0.57\hsize]{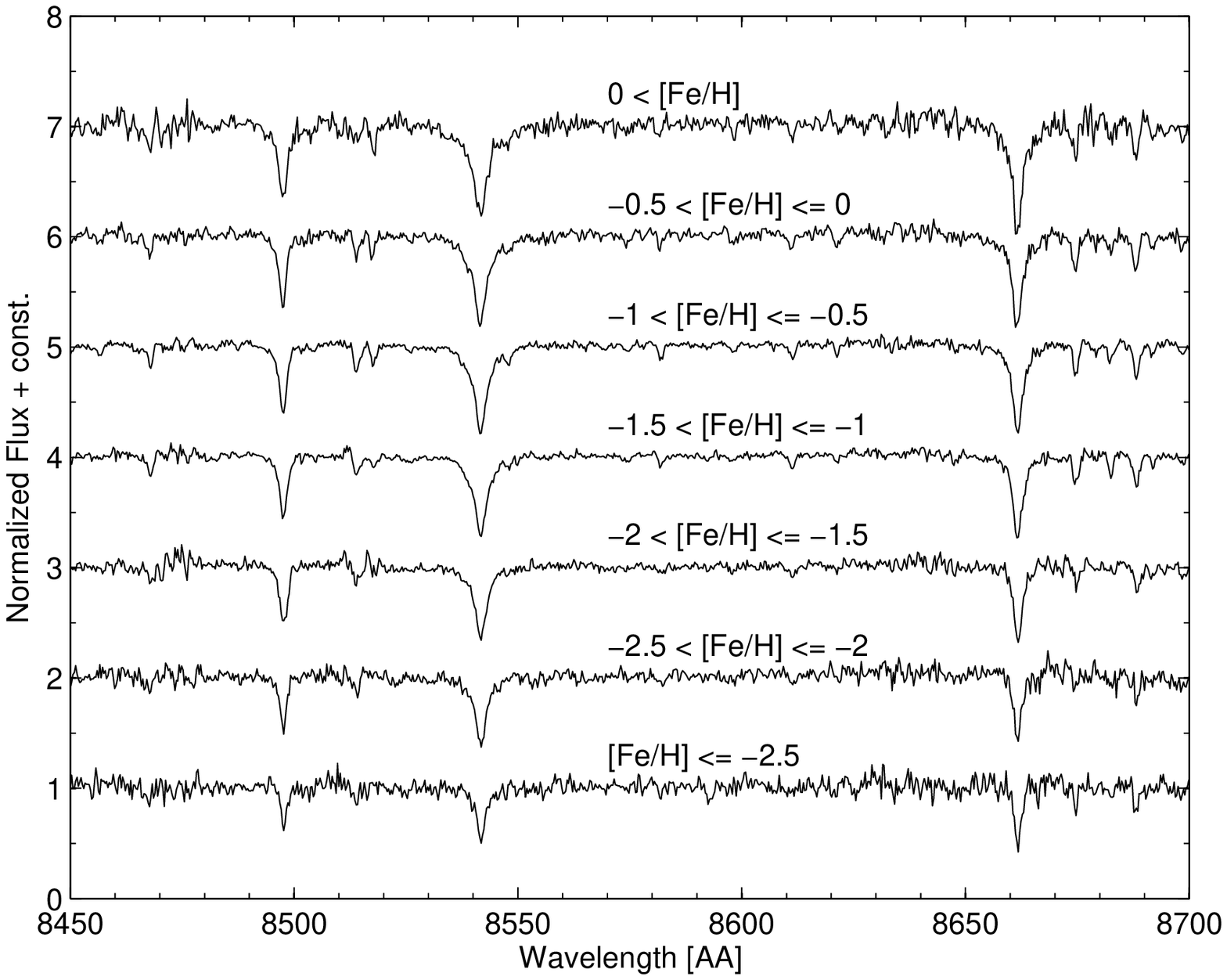}
\end{center}
\caption{Coadded spectra, grouped according to our derived CaT 
metallicity. Apart from the prominent CaT lines at 8498, 8542, 8662\AA (right panel), there 
are a number of weaker lines (mostly \ion{Fe}{1}), which  become progressively stronger in the more metal rich spectra (the CaT line strength $<\Sigma W>$ increases from 2.4\AA\ to 7.8\AA\ 
for the metallicity range covered in this figure). 
The coaddition also emphasizes a few $\alpha$-element lines that become weaker 
for the more metal poor stars: note for instance the strong \ion{Ti}{1} features at 8378,8426,8435\AA.}
\end{figure}
\begin{figure}
\begin{center}
\includegraphics[angle=0,width=0.7\hsize]{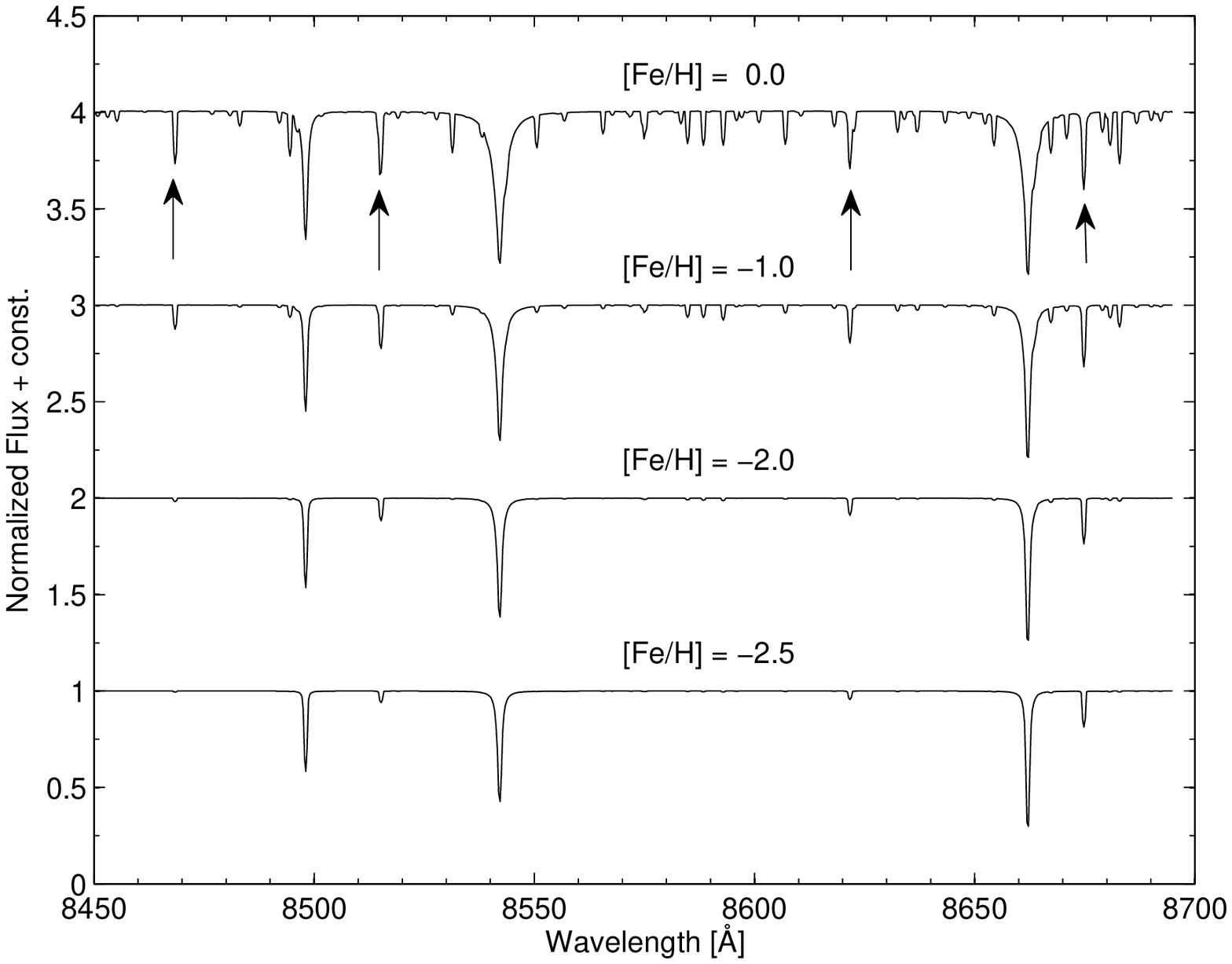}
\includegraphics[angle=0,width=0.7\hsize]{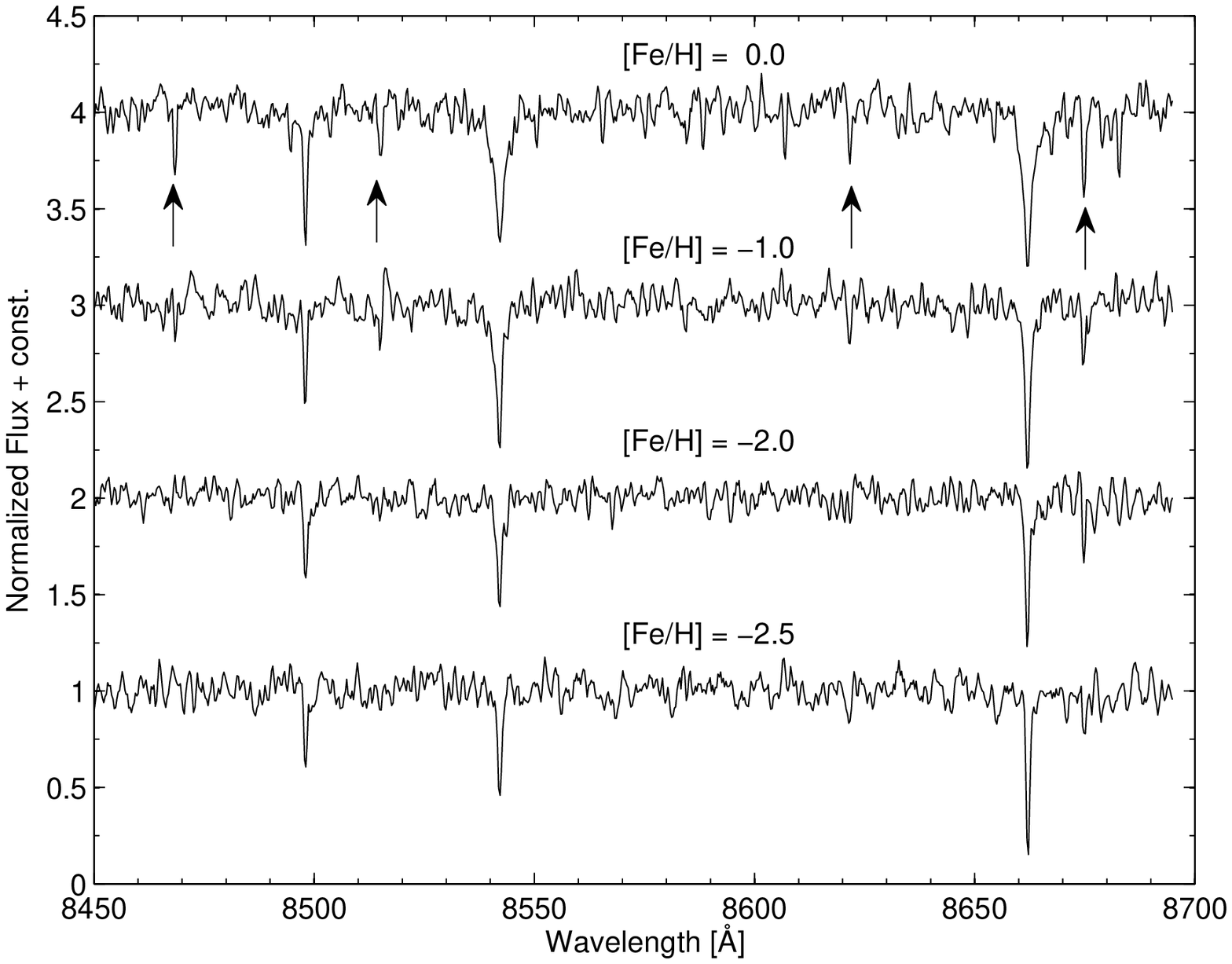}
\end{center}
\caption{Synthetic spectra for a typical red giant, using  Kurucz model atmospheres of different metallicities. The spectral resolution was reduced to  match that of DEIMOS. Additionaly, a noise component was added to the bottom panel to illustrate a 
S/N ratio of 10. Even at these low $S/N$, visual ranking by the strength of a number iron lines (indicated by arrows) is possible and observable in our spectra (see Fig.~17).}
\end{figure}
\begin{figure}
\begin{center}
\includegraphics[angle=0,width=0.45\hsize]{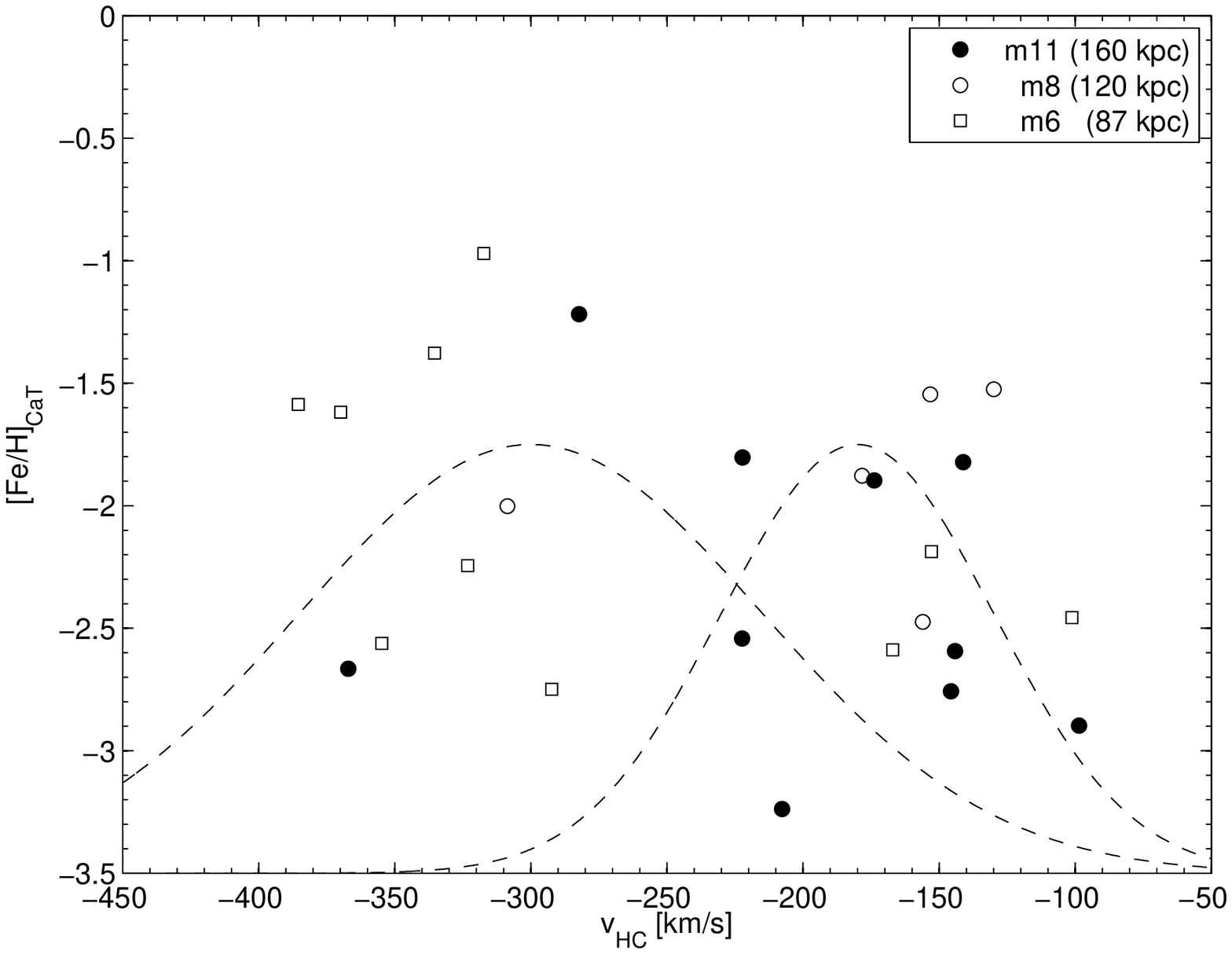}
\includegraphics[angle=0,width=0.45\hsize]{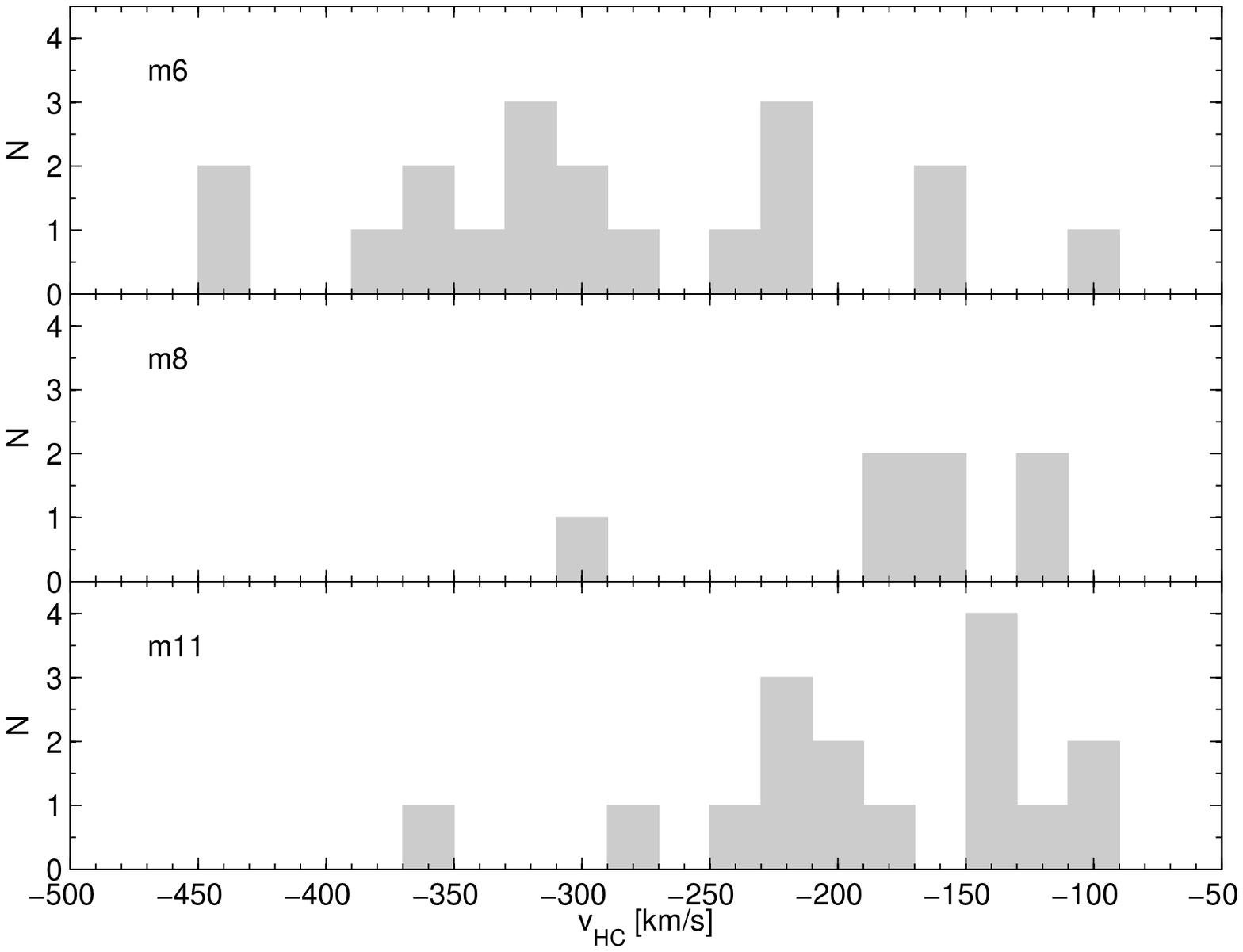}
\end{center}
\caption{Left panel: Metallicity and velocities for giant candidates in the outermost regions $>$40 kpc (small black dots). 
Different symbols highlight the distributions in the three outer fields. Also 
indicated  are the velocity number distributions of  the M31 halo (at a mean of $-300$ km\,s$^{-1}$) and 
that of M33 (around $-180$ km\,s$^{-1}$) with arbitrary scaling. The right panel shows the respective 
velocity histograms of these outer stars.}
\end{figure}
\begin{figure}
\begin{center}
\includegraphics[angle=0,width=0.7\hsize]{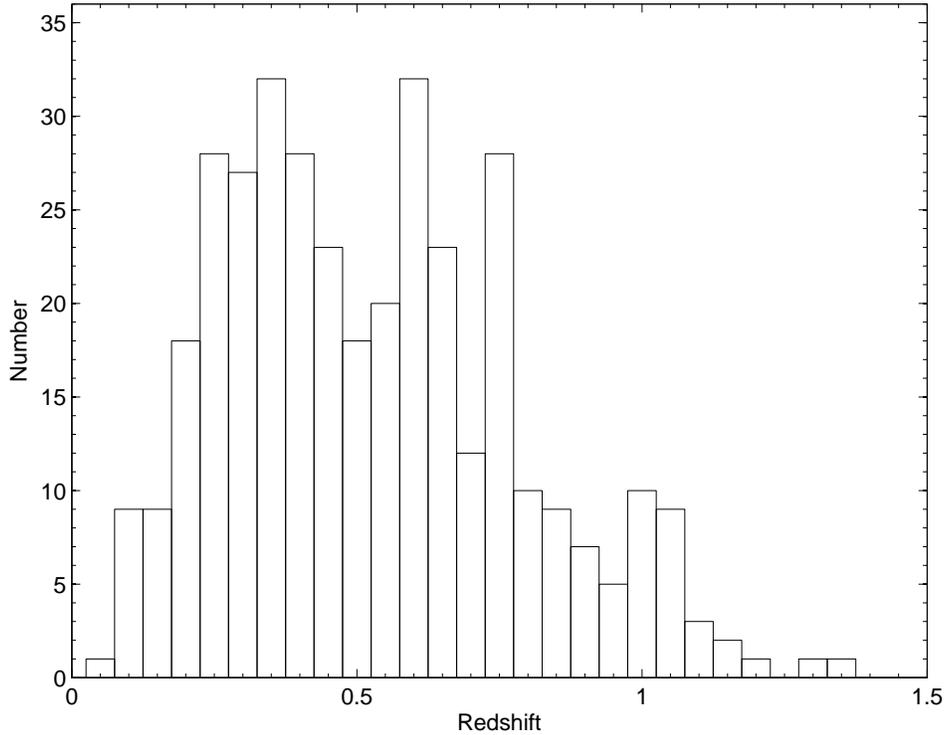}
\end{center}
\caption{Redshift distribution of background galaxies. Their respective color distribution is shown in 
Fig.~3.}
\end{figure}

\begin{thebibliography}{}
%
\bibitem[Armandroff \& Zinn(1988)]{1988AJ.....96...92A} Armandroff, T.~E., 
\& Zinn, R.\ 1988, \aj, 96, 92 
%
\bibitem[Armandroff \& Da Costa(1991)]{1991AJ....101.1329A} Armandroff, T.~E., 
\& Da Costa, G.~S.\ 1991, \aj, 101, 1329 
%
\bibitem[Ashman et al.(1994)]{1994AJ....108.2348A} Ashman, K.~M., Bird, 
C.~M., \& Zepf, S.~E.\ 1994, \aj, 108, 2348 
%
\bibitem[Battaglia et al. (2008)]{battaglia2008} Battaglia, G., Irwin, M., Tolstoy, E., 
Hill, V., Helmi, A., Letarte, B., \& Jablonka, P. 2008, MNRAS, 338, 183
%
\bibitem[Bellazzini et al.(2003)]{2003A&A...405..867B} Bellazzini, M., 
Cacciari, C., Federici, L., Fusi Pecci, F., \& Rich, M.\ 2003, \aap, 405, 
867 
%
%\bibitem[Bertelli et al.(1994)]{1994A&AS..106..275B} Bertelli, G., Bressan, 
%A., Chiosi, C., Fagotto, F., \& Nasi, E.\ 1994, \aaps, 106, 275 
%
\bibitem[Bosler et al.(2007)]{2007MNRAS.378..318B} Bosler, T.~L., 
Smecker-Hane, T.~A., \& Stetson, P.~B.\ 2007, \mnras, 378, 318 
%
\bibitem[Brown et al.(2003)]{2003ApJ...592L..17B} Brown, T.~M., Ferguson, 
H.~C., Smith, E., Kimble, R.~A., Sweigart, A.~V., Renzini, A., Rich, R.~M., 
\& VandenBerg, D.~A.\ 2003, \apjl, 592, L17 
%
\bibitem[Brown et al.(2006)]{2006ApJ...652..323B} Brown, T.~M., Smith, E., 
Ferguson, H.~C., Rich, R.~M., Guhathakurta, P., Renzini, A., Sweigart, 
A.~V., \& Kimble, R.~A.\ 2006, \apj, 652, 323 
%
\bibitem[Brown et al.(2007)]{2007ApJ...658L..95B} Brown, T.~M., et al.\ 
2007, \apjl, 658, L95 
%
\bibitem[Bullock \& Johnston(2005)]{2005ApJ...635..931B} Bullock, J.~S., \& 
Johnston, K.~V.\ 2005, \apj, 635, 931 
%
\bibitem[Carney et al.(1996)]{1996AJ....112..668C} Carney, B.~W., Laird, 
J.~B., Latham, D.~W., \& Aguilar, L.~A.\ 1996, \aj, 112, 668 
%
\bibitem[Carollo et al.(2007)]{2007Natur.450.1020C} Carollo, D., et al.\ 
2007, \nat, 450, 1020 
%
\bibitem[Carrera et al.(2007)]{2007AJ....134.1298C} Carrera, R., Gallart, 
C., Pancino, E., \& Zinn, R.\ 2007, \aj, 134, 1298 
%
\bibitem[Carretta & Gratton 1997]{Carretta1997} Carretta, E., \& Gratton, R. 1997, A\&AS, 121, 95
%
\bibitem[Cenarro et al.(2001)]{2001MNRAS.326..959C} Cenarro, A.~J., 
Cardiel, N., Gorgas, J., Peletier, R.~F., Vazdekis, A., \& Prada, F.\ 2001, 
\mnras, 326, 959 
%
\bibitem[Chapman et al.(2006)]{2006ApJ...653..255C} Chapman, S.~C., Ibata, 
R., Lewis, G.~F., Ferguson, A.~M.~N., Irwin, M., McConnachie, A., \& 
Tanvir, N.\ 2006, \apj, 653, 255 
%
\bibitem[Chapman et al.(2008)]{2008MNRASC} Chapman, S.~C., et al.  2008, MNRAS, 
submitted
%
\bibitem[Chiba \& Beers(2000)]{2000AJ....119.2843C} Chiba, M., \& Beers, 
T.~C.\ 2000, \aj, 119, 2843 
%
\bibitem[Cole et al.(2004)]{Cole04} Cole, A.~A., Smecker-Hane, 
T.~A., Tolstoy, E., Bosler, T.~L., \& Gallagher, J.~S.\ 2004, \mnras,
347, 367 
%
\bibitem[Durrell et al.(2001)]{2001AJ....121.2557D} Durrell, P.~R., Harris, 
W.~E., \& Pritchet, C.~J.\ 2001, \aj, 121, 2557 
%
\bibitem[Fardal et al.(2006)]{2006MNRAS.366.1012F} Fardal, M.~A., Babul, 
A., Geehan, J.~J., \& Guhathakurta, P.\ 2006, \mnras, 366, 1012 
%
\bibitem[Fardal et al.(2007)]{2007MNRAS.380...15F} Fardal, M.~A., 
Guhathakurta, P., Babul, A., \& McConnachie, A.~W.\ 2007, \mnras, 380, 15 
%
\bibitem[Ferguson et al.(2002)]{2002AJ....124.1452F} Ferguson, A.~M.~N., 
Irwin, M.~J., Ibata, R.~A., Lewis, G.~F., \& Tanvir, N.~R.\ 2002, \aj, 124, 
1452 
%
\bibitem[Ferguson et al.(2005)]{2005ApJ...622L.109F} Ferguson, A.~M.~N., 
Johnson, R.~A., Faria, D.~C., Irwin, M.~J., Ibata, R.~A., Johnston, K.~V., 
Lewis, G.~F., \& Tanvir, N.~R.\ 2005, \apjl, 622, L109 
%
\bibitem[Font et al.(2006)]{2006ApJ...646..886F} Font, A.~S., Johnston, 
K.~V., Bullock, J.~S., \& Robertson, B.~E.\ 2006, \apj, 646, 886 
%
\bibitem[Gallart et al.(2005)]{2005ARA&A..43..387G} Gallart, C., Zoccali, 
M., \& Aparicio, A.\ 2005, \araa, 43, 387 
%
\bibitem[Galleti et al.(2004)]{2004A&A...423..925G} Galleti, S., Bellazzini, M., \& Ferraro, F.~R.\ 2004, \aap, 423, 925 
%
\bibitem[Garnavich et al.(1994)]{1994AJ....107.1097G} Garnavich, P.~M., Vandenberg, D.~A., Zurek, 
D.~R., \& Hesser, J.~E.\ 1994, \aj, 107, 1097 
%
\bibitem[Gilbert et al.(2006)]{2006ApJ...652.1188G} Gilbert, K.~M., et al.\ 
2006, \apj, 652, 1188 
%
\bibitem[Gilbert et al.(2007)]{2007astro.ph..3029G} Gilbert, K.~M., et al.\ 
2007,  ApJ, 668, 245
%
%\bibitem[Girardi et al.(2002)]{2002A&A...391..195G} Girardi, L., Bertelli, 
%G., Bressan, A., Chiosi, C., Groenewegen, M.~A.~T., Marigo, P., Salasnich, 
%B., \& Weiss, A.\ 2002, \aap, 391, 195 
%
\bibitem[Guhathakurta et al.(2006)]{2006astro.ph..5172G} Guhathakurta, P., 
et al.\ 2006, AJ, submitted (astro-ph/0502366)
%
\bibitem[Guhathakurta et al.(2006)]{2006AJ....131.2497G} Guhathakurta, P., 
et al.\ 2006, \aj, 131, 2497 
%
\bibitem[Gwyn(2008)]{2008PASP..120..212G} Gwyn, S.~D.~J.\ 2008, \pasp, 120, 212 
%
\bibitem[Helmi et al.(2006)]{2006ApJ...651L.121H} Helmi, A., et al.\ 2006, 
\apjl, 651, L121 
%
\bibitem[Holland et al.(1996)]{1996AJ....112.1035H} Holland, S., Fahlman, 
G.~G., \& Richer, H.~B.\ 1996, \aj, 112, 1035 
%
\bibitem[Ibata et al.(2001)]{2001Natur.412...49I} Ibata, R., Irwin, M., 
Lewis, G., Ferguson, A.~M.~N., \& Tanvir, N.\ 2001, \nat, 412, 49 
%
\bibitem[Ibata et al.(2004)]{2004MNRAS.351..117I} Ibata, R., Chapman, S., 
Ferguson, A.~M.~N., Irwin, M., Lewis, G., 
\& McConnachie, A.\ 2004, \mnras, 351, 117 
%
\bibitem[Ibata et al.(2005)]{2005ApJ...634..287I} Ibata, R., Chapman, S., 
Ferguson, A.~M.~N., Lewis, G., Irwin, M., \& Tanvir, N.\ 2005, \apj, 634, 
287 
%
\bibitem[Ibata et al.(2007)]{2007arXiv0704.1318I} Ibata, R., Martin, N.~F., 
Irwin, M., Chapman, S., Ferguson, A.~M.~N., Lewis, G.~F., \& McConnachie, 
A.~W.\ 2007, ApJ, 671, 1591
%
\bibitem[Irwin et al.(2005)]{2005ApJ...628L.105I} Irwin, M.~J., Ferguson, 
A.~M.~N., Ibata, R.~A., Lewis, G.~F., \& Tanvir, N.~R.\ 2005, \apjl, 628, 
L105 
%
\bibitem[Jones et al.(1996)]{1996MNRAS.278..146J} Jones, J.~B., Gilmore, 
G., \& Wyse, R.~F.~G.\ 1996, \mnras, 278, 146 
%
\bibitem[Jorgensen et al.(1992)]{1992A&A...254..258J} J{\o}rgensen, U.~G., 
Carlsson, M., \& Johnson, H.~R.\ 1992, \aap, 254, 258 
%
\bibitem[Kalirai et al.(2006a)]{2006ApJ...648..389K} Kalirai, J.~S., et al.\ 
2006a, \apj, 648, 389 	% Metal poor halo -- gradient
%
\bibitem[Kalirai et al.(2006b)]{2006ApJ...641..268K} Kalirai, J.~S., 
Guhathakurta, P., Gilbert, K.~M., Reitzel, D.~B., Majewski, S.~R., Rich, 
R.~M., \& Cooper, M.~C.\ 2006b, \apj, 641, 268  % Stream
%
\bibitem[Kirby etal (2007)]{kirby07} Kirby, E.~N., Guhathakurta, P., Faber, S.~M., Koo, D.~C., 
Weiner, B.~J., \& Cooper, M.~C. 2007, ApJ, 660, 62
% 
\bibitem[Kirby et al.(2008)]{2008arXiv0804.3590K} Kirby, E.~N., 
Guhathakurta, P., \& Sneden, C.\ 2008, ApJ, in press (astro-ph/0804.3590) 
%
\bibitem[Koch \& Grebel(2006)]{2006AJ....131.1405K} Koch, A., \& Grebel, 
E.~K.\ 2006, \aj, 131, 1405 
%
\bibitem{koch06} Koch, A., Grebel, E.~K., Wyse, R.~F.~G., Kleyna, J.~T., Wilkinson, M.~I., Harbeck, D.~R.,  Gilmore, G.~F., \& Evans, N.~W.  2006, AJ, 131, 895
%
\bibitem[Koch et al.(2007)]{2007AJ....134..566K} Koch, A., Kleyna, J.~T., 
Wilkinson, M.~I., Grebel, E.~K., Gilmore, G.~F., Evans, N.~W., Wyse, 
R.~F.~G., \& Harbeck, D.~R.\ 2007, \aj, 134, 566 
%
\bibitem[Koch et al.(2008)]{2008AJ....135.1580K} Koch, A., Grebel, E.~K., 
Gilmore, G.~F., Wyse, R.~F.~G., Kleyna, J.~T., Harbeck, D.~R., Wilkinson, 
M.~I., \& Wyn Evans, N.\ 2008, \aj, 135, 1580 
%
\bibitem[Kuijken \& Dubinski(1995)]{1995MNRAS.277.1341K} Kuijken, K., \& 
Dubinski, J.\ 1995, \mnras, 277, 1341 
%
\bibitem[Majewski et al.(2000)]{2000AJ....119..760M} Majewski, S.~R., 
Ostheimer, J.~C., Patterson, R.~J., Kunkel, W.~E., Johnston, K.~V., \& 
Geisler, D.\ 2000, \aj, 119, 760 
%
\bibitem[Majewski et al.(2004)]{2004AJ....128..245M} Majewski, S.~R., et 
al.\ 2004, \aj, 128, 245 
%
\bibitem[Majewski et al.(2007)]{2007astro.ph..2635M} Majewski, S.~R., et 
al.\ 2007, ApJ,  670, L9
%
\bibitem[Marigo et al.(2008)]{2008A&A...482..883M} Marigo, P., Girardi, L., Bressan, A., Groenewegen, 
M.~A.~T., Silva, L., \& Granato, G.~L.\ 2008, \aap, 482, 883 
%
\bibitem[Martin et al.(2006)]{2006MNRAS.367L..69M} Martin, N.~F., Irwin, 
M.~J., Ibata, R.~A., Conn, B.~C., Lewis, G.~F., Bellazzini, M., Chapman, 
S., \& Tanvir, N.\ 2006a, \mnras, 367, L69 
%
\bibitem[Martin et al.(2006)]{2006MNRAS.371.1983M} Martin, N.~F., Ibata, 
R.~A., Irwin, M.~J., Chapman, S., Lewis, G.~F., Ferguson, A.~M.~N., Tanvir, 
N., \& McConnachie, A.~W.\ 2006b, \mnras, 371, 1983 
%
\bibitem[Martin et al.(2007)]{2007ApJ...668L.123M} Martin, N.~F., Ibata, 
R.~A., \& Irwin, M.\ 2007, \apjl, 668, L123 
%
\bibitem[McConnachie et al.(2005)]{2005MNRAS.356..979M} McConnachie, A.~W., 
Irwin, M.~J., Ferguson, A.~M.~N., Ibata, R.~A., Lewis, G.~F., \& Tanvir, 
N.\ 2005, \mnras, 356, 979 
%
\bibitem[McConnachie et al.(2006)]{2006ApJ...647L..25M} McConnachie, A.~W., 
Chapman, S.~C., Ibata, R.~A., Ferguson, A.~M.~N., Irwin, M.~J., Lewis, 
G.~F., Tanvir, N.~R., \& Martin, N.\ 2006, \apjl, 647, L25 
%
\bibitem[Merrifield \& Kuijken(1998)]{1998MNRAS.297.1292M} Merrifield, 
M.~R., \& Kuijken, K.\ 1998, \mnras, 297, 1292 
%
\bibitem[Mori \& Rich(2008)]{2008ApJ...674L..77M} Mori, M., \& Rich, R.~M.\ 2008, \apjl, 674, L77 
%
\bibitem[Mouhcine et al.(2005)]{2005ApJ...633..821M} Mouhcine, M., 
Ferguson, H.~C., Rich, R.~M., Brown, T.~M., \& Smith, T.~E.\ 2005, \apj, 
633, 821 
%
\bibitem[Mould \& Kristian(1986)]{1986ApJ...305..591M} Mould, J., \& 
Kristian, J.\ 1986, \apj, 305, 591 
%
\bibitem[O'Connell(1973)]{1973AJ.....78.1074O} O'Connell, R.~W.\ 1973, \aj, 
78, 1074 
%
\bibitem[Ostheimer(2003)]{2003PhDT.........4O} Ostheimer, J.~C.~J.\ 2003, 
Ph.D.~Thesis,  University of Virgina
%
\bibitem[Palma et al.(2003)]{2003AJ....125.1352P} Palma, C., Majewski, 
S.~R., Siegel, M.~H., Patterson, R.~J., Ostheimer, J.~C., \& Link, R.\ 
2003, \aj, 125, 1352 
%
\bibitem[Pasquini et al.(2002)]{2002Msngr.110....1P} Pasquini, L., et al.\ 
2002, The Messenger, 110, 1 
%
\bibitem[Pe{\~n}arrubia et al.(2006)]{2006ApJ...650L..33P} Pe{\~n}arrubia, 
J., McConnachie, A., \& Babul, A.\ 2006, \apjl, 650, L33 
%
\bibitem[Pritchet \& van den Bergh(1994)]{1994AJ....107.1730P} Pritchet, 
C.~J., \& van den Bergh, S.\ 1994, \aj, 107, 1730 
%
\bibitem[Reitzel \& Guhathakurta(2002)]{2002AJ....124..234R} Reitzel, 
D.~B., \& Guhathakurta, P.\ 2002, \aj, 124, 234 
%
\bibitem[Rich et al.(1996)]{1996AJ....111..768R} Rich, R.~M., Mighell, 
K.~J., Freedman, W.~L., \& Neill, J.~D.\ 1996, \aj, 111, 768 
%
\bibitem[Rich et al.(2005)]{2005AJ....129.2670R} Rich, R.~M., Corsi, C.~E., 
Cacciari, C., Federici, L., Fusi Pecci, F., Djorgovski, S.~G., \& Freedman, 
W.~L.\ 2005, \aj, 129, 2670 
%
\bibitem[Robin et al.(2003)]{2003A&A...409..523R} Robin, A.~C., Reyl{\'e}, 
C., Derri{\`e}re, S., \& Picaud, S.\ 2003, \aap, 409, 523 
%
\bibitem[Rocha-Pinto et al.(2004)]{2004ApJ...615..732R} Rocha-Pinto, H.~J., 
Majewski, S.~R., Skrutskie, M.~F., Crane, J.~D., 
\& Patterson, R.~J.\ 2004, \apj, 615, 732 
%
\bibitem[Rutledge1997a]{r3} Rutledge, G.~A., Hesser, J.~E., \& Stetson, P.~A., Mateo, M., 
Simard, L., Bolte, M., Friel, E.~D., \& Copin, Y. 1997a, 
PASP, 109, 883
%
\bibitem[Rutledge1997b]{r4} Rutledge, G.~A., Hesser, J.~E., \& Stetson, P.~A., 1997b
PASP, 109, 907 
%
\bibitem[Schiavon et al.(1997)]{1997ApJ...479..902S} Schiavon, R.~P., 
Barbuy, B., Rossi, S.~C.~F., \& Milone, A.\ 1997, \apj, 479, 902 
%
\bibitem[Schlegel et al.(1998)]{1998ApJ...500..525S} Schlegel, D.~J., 
Finkbeiner, D.~P., \& Davis, M.\ 1998, \apj, 500, 525 
%
\bibitem[Sherwin et al.(2008)]{2007arXiv0709.1156S} Sherwin, B.~D., Loeb, 
A., \& O'Leary, R.~M.\ 2008, MNRAS, 386, 1179
%
\bibitem[Simon \& Geha(2007)]{2007arXiv0706.0516S} Simon, J.~D., \& Geha, 
M.\ 2007, ApJ, 669, 327
%
\bibitem[Smith \& Drake(1990)]{1990A&A...231..125S} Smith, G., \& Drake, 
J.~J.\ 1990, \aap, 231, 125 
%
\bibitem[Stanek \& Garnavich(1998)]{1998ApJ...503L.131S} Stanek, K.~Z., \& 
Garnavich, P.~M.\ 1998, \apjl, 503, L131 
%
\bibitem[tonry]{t0} Tonry, J.~L., \& Davis, M. 1979, \aj, 84, 1511
%
\bibitem[Widrow et al.(2003)]{2003ApJ...588..311W} Widrow, L.~M., Perrett, 
K.~M., \& Suyu, S.~H.\ 2003, \apj, 588, 311 
%
\end{thebibliography}
\end{document}